\documentclass[12pt,letterpaper]{article}
\usepackage[a4paper, total={7in, 10in}]{geometry}

\usepackage{graphicx}
\usepackage{helvet}
\usepackage{authblk}
\usepackage[hidelinks]{hyperref}
\usepackage{amsmath} 
\usepackage{amssymb} 
\usepackage{xcolor}
\usepackage{float}
\usepackage[super,comma,sort&compress]  
   {natbib}
\usepackage[right]{lineno} \linenumbers
\usepackage{tabularx}
\usepackage{makecell}
\usepackage[normalem]{ulem}
\usepackage{longtable}
\usepackage{csvsimple}
\usepackage{booktabs}
\usepackage[normalem]{ulem}
\newcommand{\angstrom}{\mbox{\AA}}

\newcommand*{\mfigrefs}[2][]{%
  \hyperref[{fig:#2}]{%
    Figures~\ref*{fig:#2}%
    \ifx\\#1\\%
    \else
      #1%
    \fi
  }%
}

\newcommand*{\figref}[2][]{%
  \hyperref[{fig:#2}]{%
    Figure~\ref*{fig:#2}%
    \ifx\\#1\\%
    \else
      #1%
    \fi
  }%
}

\makeatletter
\renewcommand{\maketitle}{\bgroup\setlength{\parindent}{0pt}
\begin{flushleft}
  \textbf{\@title}
  
  \@author
\end{flushleft}\egroup}
\makeatother

\title{ANTIPASTI: interpretable prediction of antibody binding affinity exploiting Normal Modes and Deep Learning}
\date{}

\author[1,*]{Kevin Michalewicz}
\author[1]{Mauricio Barahona}
\author[1,2,**]{Barbara Bravi}

\affil[1]{Department of Mathematics, Imperial College London, London SW7 2AZ, United Kingdom}
\affil[2]{Lead contact}

\affil[*]{Correspondence: k.michalewicz22@imperial.ac.uk}
\affil[**]{Correspondence: b.bravi21@imperial.ac.uk}

\begin{document}
\nolinenumbers
\maketitle

\section*{SUMMARY}

The high binding affinity of antibodies towards their cognate targets is key to eliciting effective immune responses, as well as to the use of antibodies as research and therapeutic tools. Here, we propose ANTIPASTI, a Convolutional Neural Network model that achieves state-of-the-art performance in the prediction of antibody binding affinity using as input a representation of antibody-antigen structures in terms of Normal Mode correlation maps derived from Elastic Network Models. This representation  captures not only structural features but energetic patterns of local and global residue fluctuations. The learnt representations are interpretable: they reveal similarities of binding patterns among antibodies targeting the same antigen type, and can be used to quantify the importance of antibody regions contributing to binding affinity. Our results show the importance of the antigen imprint in the Normal Mode landscape, and the dominance of cooperative effects and long-range correlations between antibody regions to determine binding affinity.

\section*{KEYWORDS}


Antibody, Binding Affinity, Deep Learning, Interpretability, Normal Mode Analysis, Protein Structures

\section*{INTRODUCTION}

Antibodies are proteins that play a major role in the  immune response by binding to harmful foreign agents, \textit{i.e.}, antigens. Antibodies are both highly effective and specific in their binding capabilities. These properties, which are crucial when targeting antigens responsible for infections, have also led to the development of engineered antibodies for the treatment of other diseases, \textit{e.g.}, by targeting cancer cells~\cite{Strohl2017, Lu2020, Jin2022Rev}.

Structurally, antibodies have a prototypical Y-shape: the stem of the `Y' is invariant and responsible for the communication with the rest of the immune system, whereas each of the two identical tips of the `Y' comprises variable regions that are tailored to specific antigens~\cite{Chiu2019}. Each tip is composed of a light and a heavy chain~\cite{sormanni_third_2018}, which contain the entire \textit{paratope}, \textit{i.e.}, the collection of antibody residues that interact with the antigen's \textit{epitope}. Most of the paratope residues are distributed in CDRs (complementarity-determining regions), three on each chain. In general, CDRs involve between 6 and 20 amino acids~\cite{CDRlength}, with the CDR-H3 (third complementarity-determining region of the heavy chain) typically harbouring more epitope binding residues than other regions~\cite{Sela2013,akbar_silico_2022,Reis2022}. Antibodies that are not yet fully developed, and hence not specialised for a particular epitope, are called \textit{germline} or \textit{naïve}. Upon exposure of the immune system to an antigen, a process entailing somatic mutations leads to \textit{matured} antibodies capable of binding with increased affinity to the specific target.

There is an increasing number of datasets of antibodies and their corresponding targets, which include usually amino acid sequences~\cite{ABCDdatabase,OAS} and sometimes three-dimensional structures~\cite{SAbDab2014,SAbDab2022,pdb}. As such, they have enabled the training of Deep Learning models for function prediction and characterisation even from datasets of relatively modest size~\cite{Chinery2023}. In general, Deep Learning addresses different tasks, such as regression or classification, by means of multiple non-linear layers organised according to particular architectures. In the case of antibody modelling, Convolutional Neural Networks (CNNs) have been applied to structural data to predict contacts~\cite{ambrosetti_proabc-2_2020} and antibody-antigen binding interfaces~\cite{Pittala2020}, and to find potential binders for epitopes of interest~\cite{Schneider2021}. Convolutions are powerful when dealing with protein structures due to their robustness in detecting features, regardless of their exact position~\cite{Alzubaidi2021} through a consistent set of learnable parameters.
This characteristic makes it possible to find common structural motifs, which can provide insights into protein function~\cite{Singh2003}. Furthermore, given the multi-scale nature of proteins, the hierarchical architecture of CNNs is ideal for capturing features at different levels of abstraction~\cite{Soleymani2018}.

While less available as of yet, structural data are often more informative about the specific configuration of binding, and hence affinity. Indeed, sequence-based methods have proven to be less effective, with a reduced ability to capture spatial synergies, as shown by the modest reported accuracies in Refs.~\cite{Kang2021SequencebasedDL, Sirin2015}. To date, only a few methods have used structural information to predict antibody binding affinity~\cite{Kurumida2020, Myung2021, YANG2023108364}. However, these approaches do not use the entire variable region of the antibody, relying instead on a handcrafted engineering of a few features as paratope contact-based and area-based descriptors. Yet, given that protein structures are specified by hundreds to thousands of atomic spatial coordinates, methods trained on structures tend to be computationally expensive~\cite{Durairaj2023}. Ideally, one would want to find an embedding that captures the structural properties of all the amino acids in the antibody variable region as well as those of the antigen, while also taking into account their relationships. The choice of a parsimonious yet informative input data representation becomes particularly crucial to developing any such structure-based approaches.  As described below, our proposed method uses CNNs applied to residue-level descriptions of antibodies to circumvent these problems.

The need to include broader structural information stems from the fact that the molecular determinants of binding affinity are not fully understood, thus affecting our ability to identify sites and regions to target for antibody engineering. For instance, although the CDR-H3 is generally acknowledged to be the main driver of high-affinity binding~\cite{leem_deciphering_2022}, several studies indicate that the other CDRs also play an important role~\cite{dangelo2018, Phillips2021}.  
Furthermore, interactions across antibody regions are observed to favour high binding affinity, including bonds between the FR-L2 (second framework region of the light chain) and heavy chain CDR loops, which help stabilise and orient them into the antigen-bound conformation~\cite{Phillips2021}.

In a different line of research, delivering generalisable insights on how structural rigidity affects binding affinity has also proved difficult. On one hand, a positive correlation between rigidity and  high affinity has been reported~\cite{Schmidt2013, Xu2015, sormanni_rational_2015, Mishra2018, fernandez-quintero_local_2020} and increased localised rigidity with lower residue fluctuations (B-factors) has been found in the CDR-H3 of affinity-matured antibodies relative to their \textit{naïve} counterparts~\cite{laffy_proarticleuous_2017}. On the other hand, it was found that maturation reduces flexibility only if there is initially a strong binding to the conserved antigen sites~\cite{Ovchinnikov2018}, and that mutations that stabilise CDR loops in a binding-compatible conformation can confer high binding affinity without CDR-H3 rigidification~\cite{Phillips2023}. Therefore simple metrics of residue flexibility like B factors are unlikely to be sufficient for the accurate prediction of antibody binding affinity in general.

The computational study of such structural fluctuations via molecular dynamics is hampered by the long-time scales involved, with only a limited number of case studies available~\cite{Ovchinnikov2018, Zimmermann2010,Tomar2022}. Alternatively, Normal Mode Analysis (NMA)~\cite{Dykeman2010} allows the study of correlated molecular fluctuations in protein structures, and is especially efficient when applied to coarse-grained descriptions of proteins such as Elastic Network Models (ENMs)~\cite{Tirion1996,bahar_global_2010}, which model the protein as a network of residues interacting via empirically calibrated harmonic potentials~\cite{Yang2009}. The Normal Modes of such networks capture fluctuations around equilibrium with different frequencies and spatial ranges. Although NMA is well established, with several software tools widely used by the community for proteins in general~\cite{Dobbins2008,nmaBio3D,LopezBlanco2011, Bakan2011}, it has not been exploited to characterise the structural fluctuations of antibody-antigen complexes and how they contribute to binding affinity. In this context, priority has been given to modelling the binding interface alone~\cite{Madsen2024}, a problem for which NMA is not best suited.

In this work, we develop ANTIPASTI (ANTIbody Predictor of Affinity from STructural Information), an interpretable Deep Learning approach that harnesses the information encoded in the molecular structure and its associated interactions for the numerical prediction of binding affinity. Specifically,  ANTIPASTI is based on a CNN architecture that learns a representation of antibody structures from binding affinity data and residue-level Normal Mode correlation maps derived from Elastic Network Models. The learnt representation can then be used to predict the binding affinity of any given structure and to extract interpretable structural features connected to antibody regions that are key contributors to increase and decrease binding affinity. 

\section*{RESULTS}

\begin{figure}[H]
    \centering
    \includegraphics[clip, trim=0 4cm 0.3cm 0, width=\textwidth]{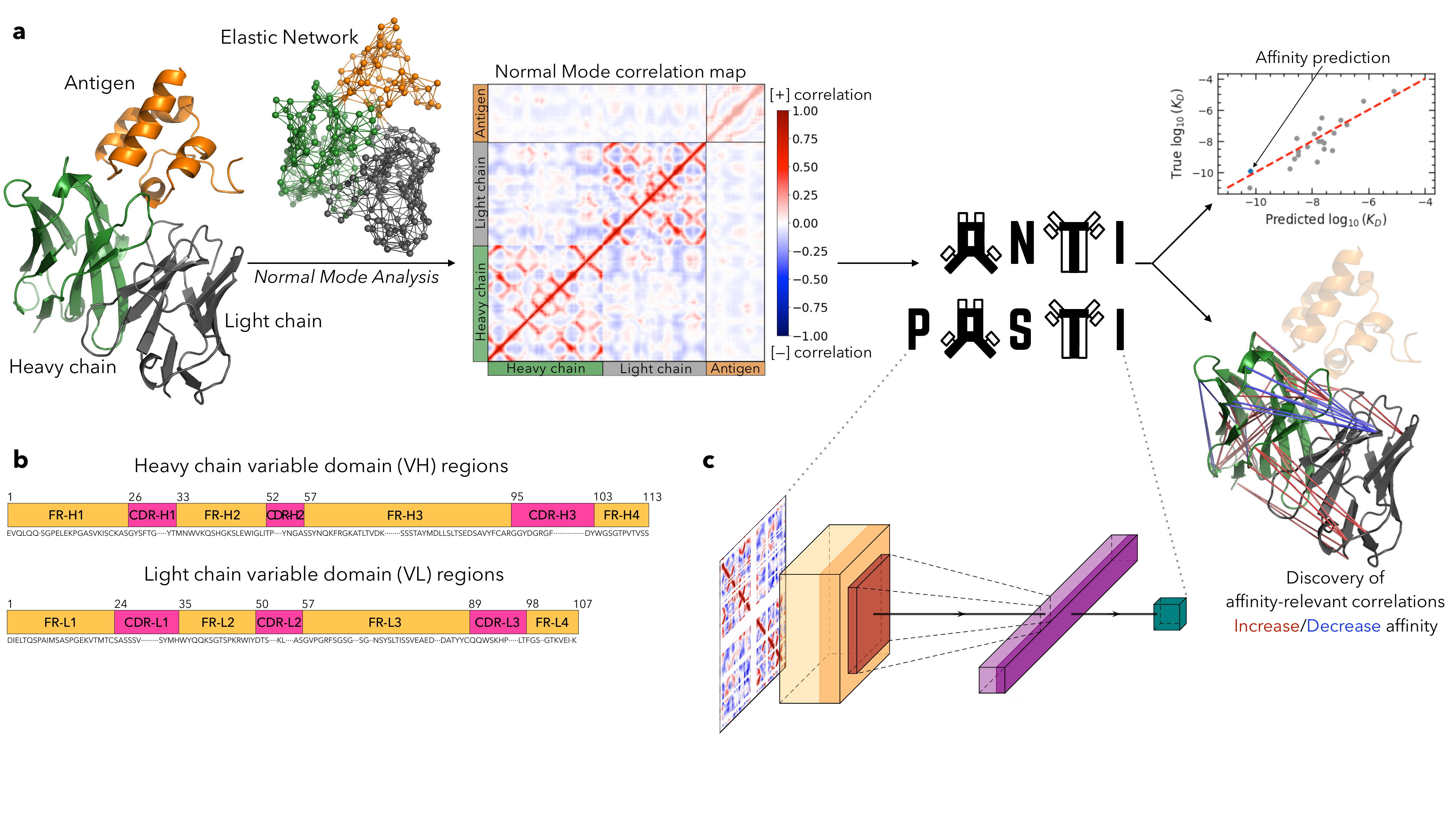}
    \caption{\textbf{Overview of ANTIPASTI.} (\textbf{a}) From the PDB structure of the antibody-antigen complex, an Elastic Network Model (ENM) representation is created. The correlation between residues is then calculated from the Normal Modes, which carry the antigen's imprint on the antibody residues. Gaps are then added in place of absent residues through an alignment (see \textbf{b}), and the antigen residues are removed from the correlation map only after this point to produce the input image to our CNN model (see \textbf{c}). ANTIPASTI processes this input image and  produces a binding affinity prediction and a map indicating the relevant correlations. (\textbf{b}) Regions are defined with an alignment to the Chothia position numbering scheme~\cite{Chothia1987}. (\textbf{c}) Architecture of the ANTIPASTI Deep Learning model (see \hyperref[methods]{STAR Methods}).}
    \label{fig:1}
\end{figure}

\subsection*{The ANTIPASTI Deep Learning architecture}

ANTIPASTI uses structural data of antibody-antigen complexes together with experimentally measured values of their binding dissociation constant ($K_D$) to train a Deep Learning model based on a CNN architecture in a supervised fashion. Our training dataset consists of $634$ antibody-antigen structures, retrieved from SAbDab~\cite{SAbDab2014}, annotated with an experimental $K_D$ value (see \hyperref[methods]{STAR Methods}). The input to our model is a residue-residue correlation map. This map is calculated from the Normal Modes of an Elastic Network Model (ENM) of the antibody-antigen complex, a residue-level representation of the structure containing energy interactions (\hyperref[methods]{STAR Methods}).
The Normal Mode Analysis of the ENM antibody-antigen structural complex encodes quantitative  information on regions that undergo correlated or anti-correlated fluctuations (\figref[a]{1}). Since binding affinity is determined by the specific conformation of the antibody and its cognate antigen, we compute the input correlation map of the antibody variable domains and the \emph{full} antigen, retaining in this way the antigen's imprint on the antibody residues (\hyperref[methods]{STAR Methods}). We then extract the portion of the Normal Mode map corresponding to the antibody residues to obtain our input, which is set to have the same size by adopting a uniform numbering scheme for antibody residue positions (Chothia scheme~\cite{Chothia1987}, see \figref[b]{1}). 

The ANTIPASTI neural network architecture is composed of a CNN with $n_\text{f}$ filters of shape $k\times k$, a non-linearity and a $p$-pooling layer, followed by a fully-connected layer, which outputs the $\log_{10}(K_D)$ prediction (\figref[c]{1}, Methods). The representation of antibody structures via Normal Mode correlation maps provides image-like inputs, for which CNNs, firstly developed for applications in computer vision~\cite{Alzubaidi2021}, are particularly well-suited. During training, the model learns new representations of antibody structures guided by the supervised task of mapping residue-residue correlations in the antigen-bound conformation to binding affinity. The representations are kept interpretable as they can be mapped back to the original antibody structure. To ensure reproducibility, our pipeline and data are publicly available with instructions and tutorials provided on GitHub (see \hyperref[code]{Code availability}).

Given structural data for an antibody-antigen complex, ANTIPASTI first computes the ENM representation and its corresponding Normal Mode correlation map for the antibody (under the imprint of the antigen). This correlation map is then used as the input to the CNN model to predict the binding affinity constant. In addition, the representations can be inspected to discover which correlations and antibody regions are most relevant to binding affinity (\figref[c]{1}).

\subsection*{ANTIPASTI accurately predicts binding affinity}

We conducted hyperparameter optimisation by means of $K$-fold cross-validation ($K=10$) and identified a CNN with pooling as the best ANTIPASTI architecture, which we will use in the following sections for various analyses (Figure S1a). The second best model was a CNN without pooling, which we included in our accuracy tests. We evaluated these two best model candidates on the test set for five random seeded train-test splits (see~\hyperref[methods]{STAR Methods} for more details) and we measured their performance via the $R$ correlation score between the true $K_D$ and the predicted values by ANTIPASTI. As a baseline, we also considered a basic linear regression to predict binding affinity from Normal Mode correlation maps (\hyperref[methods]{STAR Methods}).

We then investigated the accuracy of ANTIPASTI for prediction of the binding affinity constant $K_D$. 
\figref[a]{2} shows the predicted $K_D$ values of the CNN with pooling on an unseen test set for five training/test splits compared to the true values, with a Pearson correlation of $R=0.86$ when combining the data points. The splits were checked to ensure there is less than 10\% redundancy and that the antigens are different between the training and test data (\hyperref[methods]{STAR Methods}). The mean $R$ score computed over 5 test sets is $R=0.85$. The other architecture (CNN without pooling) exhibits comparable performance, with mean $R$ score of $0.86$ (\figref[c]{2}). Such a performance compares favourably with existing sequence-based and structure-based predictors of antibody binding affinity~\cite{Sirin2015,Kang2021SequencebasedDL,Kurumida2020, YANG2023108364}, which have $R$ scores ranging from $0.45-0.79$, albeit a full comparison cannot be carried out consistently (Table S1). This is due to differences in method design and data selection, which prevent us from re-training and evaluating their models on our more general data. 
For comparison, the linear regression on Normal Modes achieves a mean $R=0.64$, substantially lower than ANTIPASTI (\figref[c]{2}), indicating that is crucial to include the non-linearities in the network layers with robust feature detection properties, such as convolutional layers. Furthermore, we note that the ANTIPASTI architectures are computationally efficient: the training time of all the architectures implemented here remained below $10$ minutes (Apple M1, macOS Big Sur).

\begin{figure}[H]
\centering

  \includegraphics[clip, trim=1cm 3.7cm 1.2cm 0, width=\linewidth]{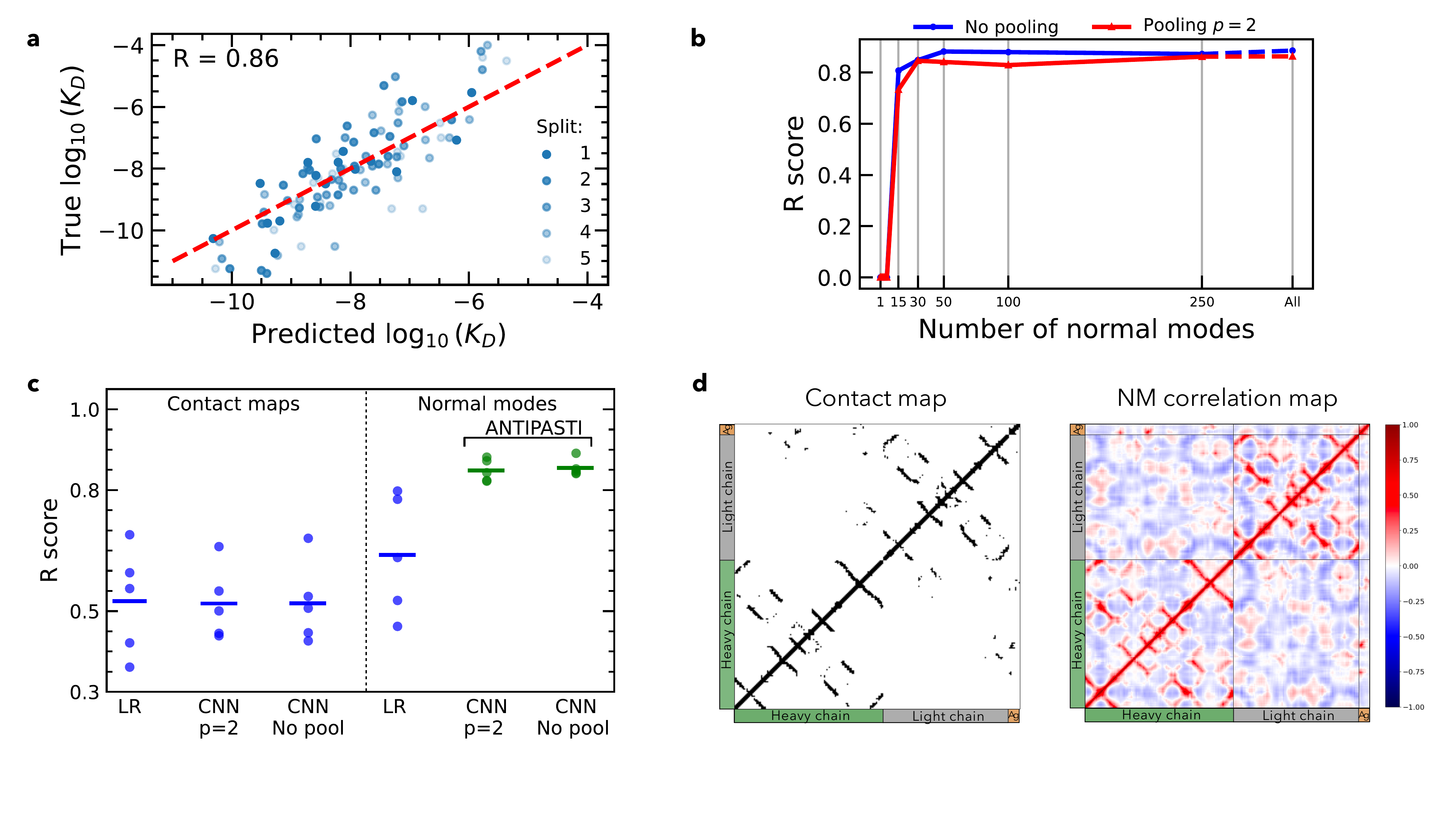}

\caption{\textbf{ANTIPASTI performance.} (\textbf{a}) ANTIPASTI predictions on the test set for five training/test splits (see Methods).
(\textbf{b}) Test performance as a function of the number of Normal Modes used. Note that the first six are trivial and that they are ordered from the lowest to highest associated eigenvalue (\hyperref[methods]{STAR Methods}).
(\textbf{c}) Test performance for five distinct train-test random splits comparing the two best ANTIPASTI architectures - with $p=2$ pooling and no pooling - to simpler approaches: linear regression (LR), and CNNs and LR trained on contact maps. Mean $R$ scores are indicated by a horizontal line. (\textbf{d}) Example of a contact map and Normal Mode correlation map (PDB entry: \texttt{3fn0}).
}
\label{fig:2}
\end{figure}

We next analysed the robustness of our results to the number of Normal Modes used to build the input map. Normal modes are ordered by frequency (from low to high), hence capturing structural fluctuations from global to increasingly local length scales relative to the protein size. A natural question is if the number of modes could be reduced while maintaining the predictive performance or, in other words, what spatial scales of correlated fluctuations matter for binding affinity. To address this, we compute the R score as a function of the number of Normal Modes until we observe an elbow, signaling a stabilisation in performance. We observed that the prediction accuracy remained high when excluding high frequency Normal Modes: as shown in \figref[b]{2}, the performance of ANTIPASTI is unaffected as long as the first $30$ modes (\textit{i.e.}, those with lowest frequency and largest spatial scales) are kept ($R = 0.845$). As a general trend, this indicates that binding affinity is predominantly affected by long- and medium-range spatial correlations, while very short-range correlations can be discarded with little loss of prediction power for binding affinity.

\subsection*{Normal Mode correlation maps encode richer information than contact maps}

We also explored whether residue-residue contact maps, where a contact is recorded if the physical distance between $\alpha$-Carbon atoms is less than a threshold (typically $6-12\angstrom$~\cite{Baldi2003}), are
sufficient to estimate $K_D$ instead of opting for Normal Modes. To tackle this question, we carried out a hyperparameter optimisation for this type of input and then re-trained ANTIPASTI using contact maps as input data (\hyperref[methods]{STAR Methods}).

Using the same protocol as for our Normal Mode computations, we found that CNN models applied to contact map inputs lead to test set $R$ scores ranging between $0.36$ and $0.68$ (\figref[c]{2}). This confirms that the antibody structural connections mediated through 3D space are informative, yet substantially less so than the structural fluctuations (\emph{dynamical} as much as structural information) captured by Normal Mode maps.  Note also that linear regression and CNN models perform indistinguishably on contact map inputs (\figref[c]{2}), a fact that can be explained due to the sparsity of contact maps (\figref[d]{2}), \textit{i.e.}, the large amount of zero pixels makes the hierarchical convolutional coarse-graining less useful. 

\subsection*{ANTIPASTI representations capture distinctive binding modes for different antigen types}

The internal representations learnt by the model during training (specifically the output layer) can be used to reveal pairwise residue correlations that are key to binding affinity. Note that the sum of the elements of the output layer gives $\log_{10}{(K_D)}$ (see \hyperref[methods]{STAR Methods}). Yet the individual elements contain information about correlations that produce an increase or a decrease of the predicted binding affinity.  To visualise this, we reshape the output layer into a residue-residue \textit{affinity-relevant correlation map} (\figref[a]{3}, \mfigrefs[d-e]{3}, \hyperref[methods]{STAR Methods}). 

To further understand the patterns of affinity-relevant correlations, we applied UMAP (Uniform Manifold Approximation and Projection) dimensionality reduction~\cite{McInnes2018} to the output layer representations of all available antibodies in our dataset. We found that antibodies cluster according to the type of antigen in the UMAP projection  (\figref[b]{3}) even though the inputs are restricted only to antibody residues. This finding confirms that the model leverages similarities in the pattern of correlated fluctuations induced by the imprint of antigens. In particular, antibodies binding to small size linear targets (\emph{i.e.}, haptens and peptides) cluster closely, hinting that target size and the presence of a secondary structure in the antigen impose  constraints on binding. On the other hand, nanobodies stand out as a separate cluster, in line with the biologically sensible expectation that a distinctive binding conformation characterises antibodies with a single domain (see also~Figures S2a-c). 
We have checked that even when re-training ANTIPASTI on the heavy chain alone, nanobodies stay separable from the heavy chains of paired-HL antibodies with the same target type (Figure S2d). Similarly, there is a tight cluster of anti-HIV antibodies comprising $14$ structures, all of which correspond to the same species (Homo sapiens); have the same type of light chain ($\kappa$); and belong to the same heavy chain V gene family (IGHV1). Another tight cluster is seen for $6$ antibodies targeting the Amyloid Beta peptide (Alzheimer's disease), also sharing these properties in most cases (Mus musculus/$\kappa$/Other) (\figref[b]{3}, Figures S2a-c).

\begin{figure}[H]
\centering
  \includegraphics[clip, trim=0cm 0.8cm 2.5cm 0, width=0.95\linewidth]{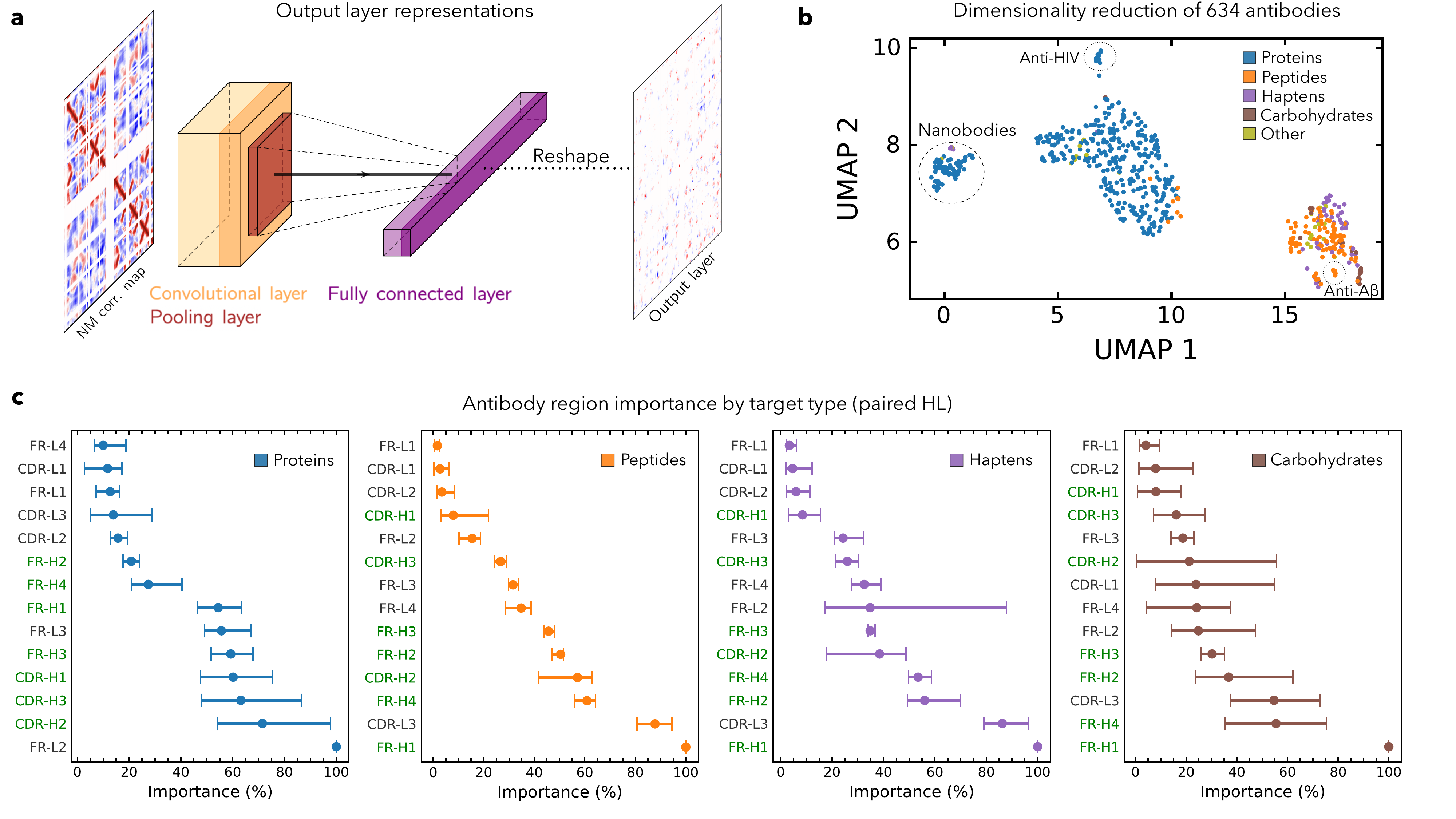}
  \vspace{0.4cm}
  
  \includegraphics[clip, trim=0.1cm 0cm .5cm 0, width=0.95\linewidth]{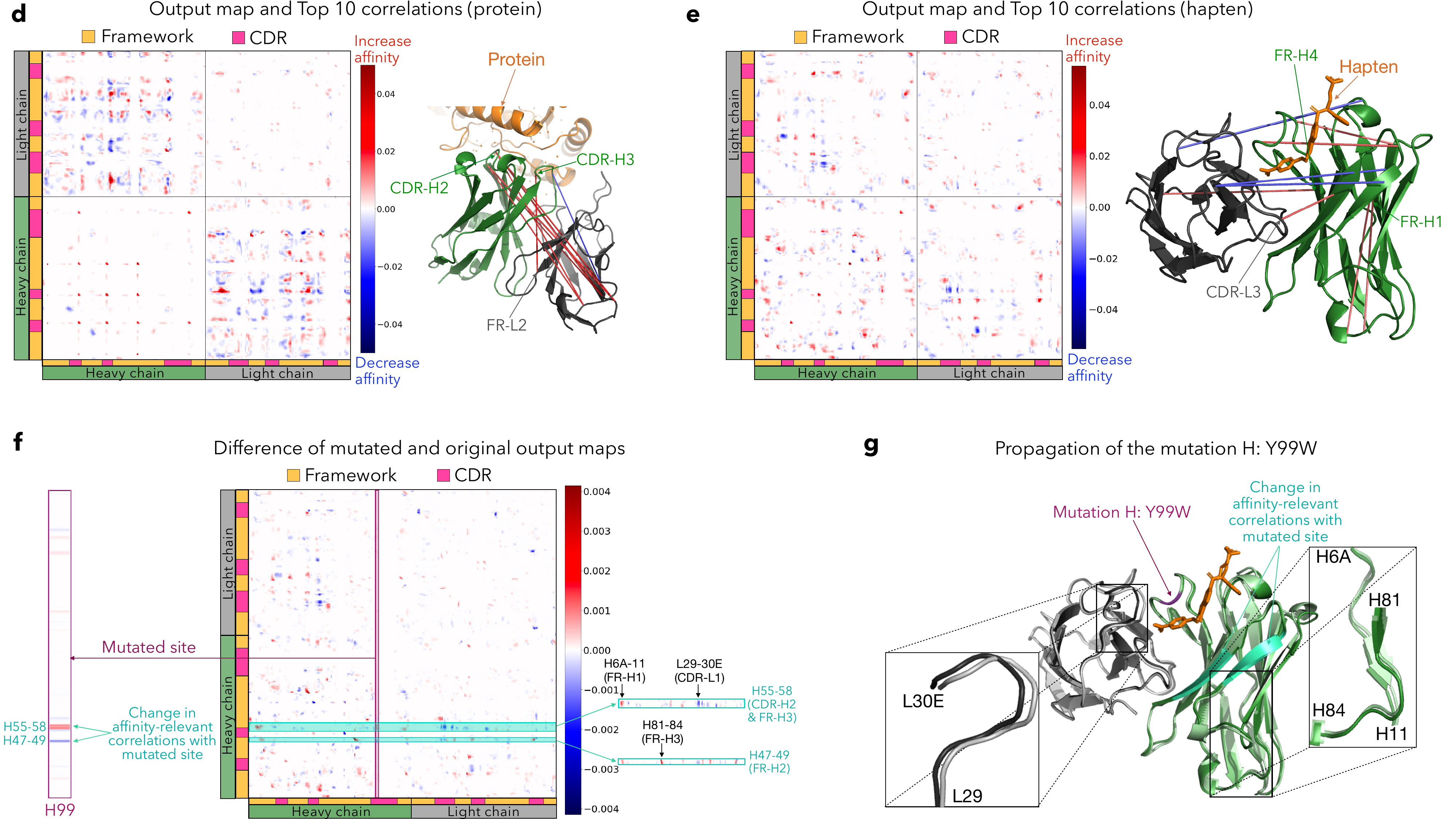}
\caption{\textbf{Residue-residue correlations relevant to binding affinity.} (\textbf{a}) ANTIPASTI output representations indicating which correlations increase or decrease binding affinity. (\textbf{b}) Dimensionality reduction (UMAP) of the output layer representations groups antibodies by their binding target type. (\textbf{c}) Rankings of antibody regions' importance, 
expressed as a percentage of the importance of the best region, for 4 of the groups of antibodies with different target type highlighted in b (antibodies with paired heavy and light chains only). We report the average (dot) and extreme values (error bars) over 5 training/test splits. (\textbf{d}) Relevant correlations for binding affinity for an antibody with a protein target (Mntc with Mab 305-78-7 complex, PDB entry: \texttt{5hdq}). (\textbf{e}) Relevant correlations for binding affinity in the complex of an antibody targeting the hapten dabigatran (PDB entry: \texttt{4yhi}). (\textbf{f}) Difference in affinity-relevant correlations between a mutated antibody (\texttt{4yho}) and its original structure (\texttt{4yhi}), whose variable regions differ only at site H99 (CDR-H3). The mutated antibody presents a higher affinity. (\textbf{g}) \texttt{4yhi} and \texttt{4yho} structure overlay. The greatest variations in affinity-relevant correlations with the mutated site identify two $\beta$-strands (green-cyan). These are in turn correlated with numerous sites, three of which correspond to the most substantial conformational shifts between \texttt{4yhi} and \texttt{4yho}.
}
\label{fig:3}
\end{figure}

\subsection*{ANTIPASTI representations identify important regions for binding affinity}

We next used the ANTIPASTI output layer representations to define a measure of antibody region importance towards high binding affinity. Our measure quantifies the loss in accuracy due to the exclusion of the correlations of the residues in a  particular region with all other sites, normalised to be a percentage of the top region (\hyperref[methods]{STAR Methods}). The results of the region importance measure are presented in \figref[c]{3} for the antibodies grouped by antigen type. 

As a general trend, antibodies that have haptens and peptides as antigens exhibit a similar ranking of important regions, different to antibodies targeting carbohydrates and especially proteins. This is consistent with the separation of protein-targeting antibodies in the UMAP projection (\figref[b]{3}).  In addition, 
heavy chain regions have high importance: $5$ ($4$ for carbohydrates) of the $7$ most relevant regions belong to the heavy chain. In particular, the CDR loops of the heavy chain  are among the first $4$ regions for antibodies with protein targets. Indeed, heavy chain CDR loops are known to harbour the majority of epitope-binding residues~\cite{Koenig2017,akbar_silico_2022} some of which have prominent effects on affinity to protein targets when mutated~\cite{Koenig2017, Phillips2023}. CDR loops are surpassed only by the second framework of the light chain (FR-L2). Previous experimental work has shown that mutations in FR-L2 can lead to improved thermostability, a pre-requisite to binding~\cite{Koenig2017}, and residues within FR-L2 are part of the heavy-light chain interface involved in highly stabilising interactions with the CDR-H3, thus modulating its rigidity~\cite{Xu2015, Phillips2023} (see \figref[d]{3} for an example showing highly important correlations linking CDR-H3 and FR-L2). Framework regions, in particular FR-H1,2,4, also stand out in the importance ranking for smaller targets (\figref[c]{3}), consistent with an active role of framework sites in affinity and stability by influencing conformational dynamics during antigen binding~\cite{Koenig2017}. 

Altogether, these observations suggest synergistic effects between regions in determining high-affinity binding. To quantify this further, we determined the proportion of importance due to the inter-region \textit{vs.} intra-region correlations. We found that higher importance of inter-region correlations, underlining the role of synergy between regions in antigen binding. Repeating this computation to quantify intra-chain \textit{vs.} inter-chain correlations corroborated that the importance of light chain framework regions stems to a large extent from correlations with the heavy chain regions (Figure S3a).

\subsection*{ANTIPASTI representations reveal long-range correlations as key to affinity}

To further validate the insight that affinity-relevant correlations span large distances in the structure, we considered the $10$ most important correlations for each structure in the dataset. 
We found that most of them are long-range: $80\%$ of the top interactions are between residues more than $10\angstrom$ apart. The distribution also has a peak at short-distance, indicating that the model detects  important short-range physical interactions (Figure S3d).

To better understand affinity-relevant long-range correlations predicted by ANTIPASTI, we sought experimental data on single site mutations with an effect on antibody affinity. We could only find one mutational study~\cite{Schiele2015} reporting both affinity values and the corresponding antigen-antibody complex structure, corresponding to antibody fragment idarucizumab targeting the hapten dabigatran. The original complex (PDB code: \texttt{4yhi}, \figref[e]{3}) had $K_D=180$ $pM$, and a single site mutation within the CDR-H3, from tyrosine to tryptophan (mutation H:Y99W),  led to a $10$-fold improvement in binding affinity in the mutated complex (PDB code: \texttt{4yho}). Although globally similar, there are a few residues whose changes in the corresponding affinity-relevant correlations contain most of the information relevant to the change in binding affinity following the mutation (Figure S3e). By inspecting the top 25 changes in affinity-relevant correlations, we found that these residues are concentrated around two $\beta$-strands (H47-49 and H55-58,~\figref[f]{3}), which partly lie at the interface between the heavy and light chains (\figref[g]{3}). In particular, H47-49 belongs to the binding pocket of dabigatran. Via structural analyses, Ref.~\cite{Schiele2015} attributed the increased binding affinity to an optimisation of the steric fit to the paratope due to the incorporated tryptophan in the CDR-H3 pushing dabigatran towards the side of the binding pocket (residue H29 and $\beta$-strand residue H49), forming a new $\pi$-stacking bond. Indeed, the two $\beta$-strands (H47-49 and H55-58) undergo conformational shifts following the mutation (see zoomed insets on the overlaid structures in \figref[g]{3}) and the loss of a hydrogen bond between L30C and H99 upon the mutation might lead to a tighter fit of dabigatran towards H29/H49~\cite{Schiele2015}. These changes lead to large differences in the ANTIPASTI-predicted affinity-relevant correlations with two neighbouring $\beta$-strands, which amplify and mediate the spread of the mutational effect to more flexible regions in the antibody (\figref[f]{3}). Note that these two $\beta$-strands already appear in the top $10$ correlations predicted by ANTIPASTI for the original (unmutated) structure (\figref[e]{3}), as they present correlations with the CDR-L1 and CDR-L3, thus hinting at their potentially central role for modifying binding affinity.

\subsection*{Evaluation tests with AlphaFold-predicted structures}

Finally, we assessed the accuracy of ANTIPASTI affinity predictions for AlphaFold-predicted structures. Predicted structures are increasingly used as tools to understand protein functions, for homology detection~\cite{Monzon2022} and for target prediction by inverse docking~\cite{Wang2022}. To determine whether the performance of ANTIPASTI on AlphaFold-predicted structures is comparable to that on experimental PDB structures, we generated AlphaFold structures~\cite{Jumper2021, Evans2022} from the sequences of $21$ recently published antibody-antigen complexes that were not available at the time of the training of AlphaFold, and are therefore unseen by AlphaFold (\hyperref[methods]{STAR Methods}). We then computed the Normal Mode correlation maps of these predicted structures and used them as inputs to ANTIPASTI to predict their binding affinity, $\hat{K}_D$. We found that the affinity prediction is comparable to that of the original PDB structures (Pearson's correlation coefficient $R=0.9$, \figref[a]{4}; for the predictions using the original structures, which yielded $R=0.78$, see Figure S4a).

To quantify further the effect of the uncertainty in  AlphaFold predictions, we used the two AlphaFold-provided confidence metrics: the predicted Local Distance Difference Test (pLDDT) for each residue, and the Predicted Alignment Error (PAE) matrix. \figref[b]{4} shows these metrics for the predicted complex with PDB entry \texttt{4f3f}.

As expected, the larger the uncertainty in the AlphaFold prediction the larger the difference between the predicted affinities for experimental and AlphaFold structures. This can be seen in the negative correlation  when plotting the average pLDDT as a function of $\Delta\!\log_{10}(\hat{K}_D)$ (\figref[c]{4}, Spearman's correlation coefficient $\rho = -0.8$, p-value $=1.2\times10^{-5}$). Although there is also a positive correlation between the maximum PAE versus $\Delta\!\log_{10}(\hat{K}_D)$ when considering the entire PAE matrix (\figref[c]{4}, Spearman's correlation coefficient $\rho=0.74$, p-value = $1.4\times10^{-4}$), most of these maxima are reached at residues at the interface between antibody and antigen (Figure S4b, Spearman's coefficient $\rho=0.68$, p-value = $8\times10^{-4}$). This result indicates the need for accurate AlphaFold prediction of the antibody-antigen interface to obtain a consistent affinity estimation with ANTIPASTI. We assessed also the Interface Root Mean Square Deviation (IRMSD) between the original and AlphaFold-predicted structures (Figure S4c), and we obtained a weaker correlation with $\Delta\!\log_{10}(\hat{K}_D)$ ($\rho=0.49$, p-value = $0.02$). The results of this analysis, reported in Figure S4c, suggest that small $\Delta\!\log_{10}(\hat{K}_D)$ tends to correspond to low IRMSD, but the reverse association does not necessarily hold.

\begin{figure}[H]
\centering
  \includegraphics[clip, trim=1.25cm 0 3.25cm 0.4cm, width=\linewidth]{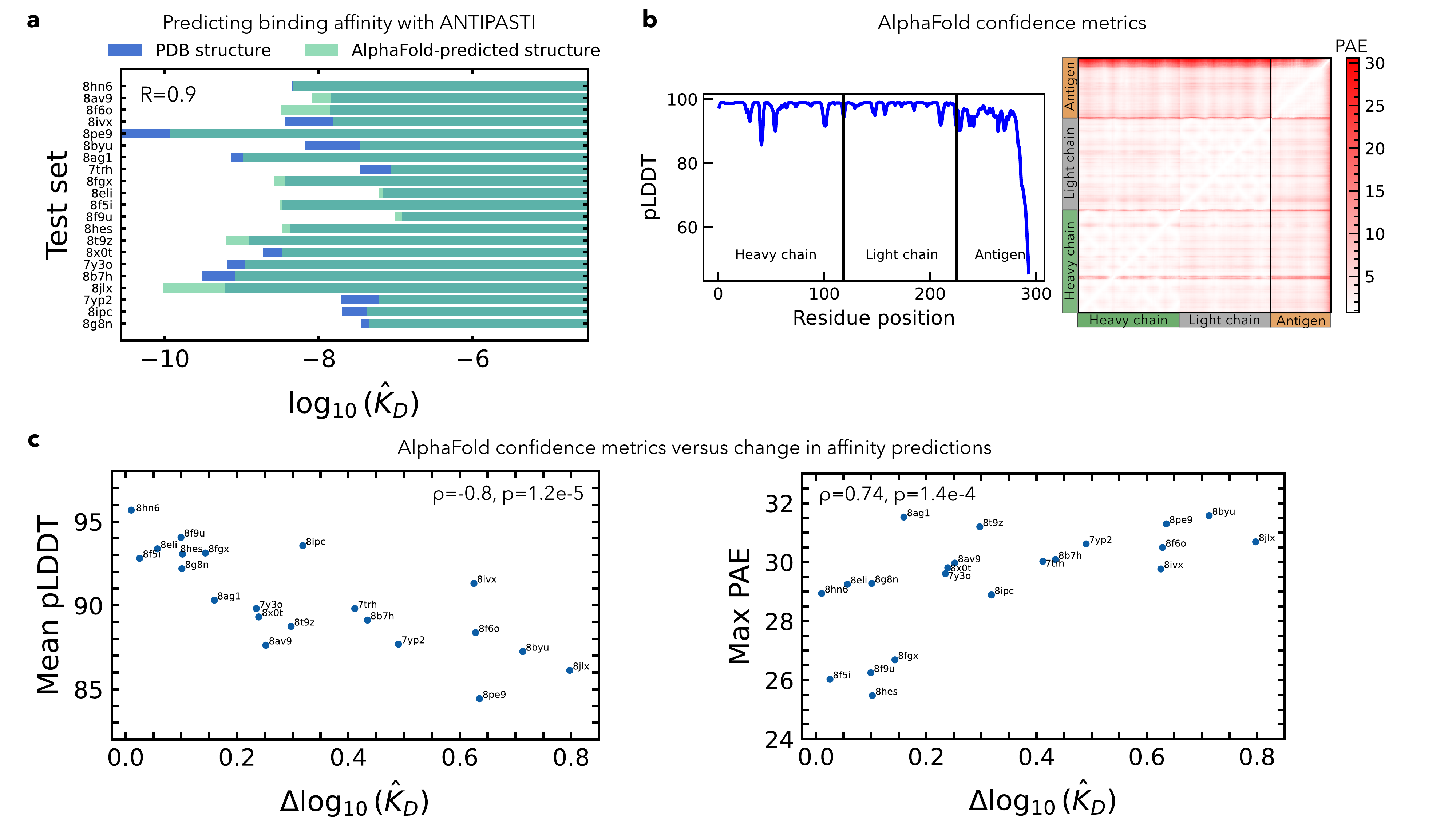}
\caption{\textbf{Evaluation of ANTIPASTI on AlphaFold-predicted structures.} (\textbf{a}) Comparison of $\log_{10}(\hat{K}_D)$ between the PDB structures and those predicted by AlphaFold for the same test set as in \figref[a]{2}. (\textbf{b}) Alphafold confidence in folding prediction for the Msln7-64 Morab-009 Fab complex (PDB entry \texttt{4f3f}). The predicted Local Distance Difference Test (pLDDT) for each residue and the Predicted Alignment Error (PAE) matrix are displayed. (\textbf{c}) Mean of the per-residue confidence, assessed by the pLDDT metric (Spearman $\rho=-0.8$), and the maximum PAE (Spearman $\rho=0.74$), as functions of $\Delta\!\log_{10}(\hat{K}_D)$.}
\label{fig:4}
\end{figure}

\section*{DISCUSSION}

In this work, we have developed ANTIPASTI, a new Machine Learning tool that leverages structural fluctuation information (obtained from Normal Mode correlation maps) to predict the binding affinity of antibodies using a CNN with a single filter bank and a fully-connected layer. We showed that ANTIPASTI can predict $K_D$ from Normal Mode correlation maps with high generalisation power and achieves greater accuracy than existing methods (test set $R=0.86$). Furthermore, training and inference are possible with low computational cost and using only a CPU. 

A main ingredient of our approach is the use of residue-residue correlation maps obtained through Normal Mode Analysis of Elastic Network Model representations of antibody-antigen complexes as inputs that capture structural correlations without the need for long molecular dynamics simulations. These inputs include the antigen and the entire variable region of antibodies, which is acknowledged to be central to antigen binding~\cite{Janeway2001, Igawa2011}. Although the method could be seamlessly extended to structures comprising also the constant region of the antibody, its application is challenging with the current data availability. In particular, many published antibody-antigen structures do not contain an experimentally resolved constant region, hence reducing the number of samples for training. Furthermore, the inclusion of the constant region would require to learn a substantial number of additional parameters from a training dataset of limited size. Assessing the performance of ANTIPASTI when trained on antibody full structures would be an interesting direction for future research. Although the antigen is part of the NMA analysis, so as to retain the antigen imprint in the Normal Mode landscape of the antibody, the model is then trained on correlation maps of the imprinted antibody section alone. This choice allows ANTIPASTI to focus on the properties and regions of antibodies determining optimal binding, and can also potentially widen its scope of application to data that lack the cognate antigen in the resolved structure. To examine the importance of the antigen imprint for our method, we also tested an antigen-agnostic version (where the NMA analysis is carried out without the antigen section), and we found reduced predictive power, but still with information on biological characteristics in the output layer maps (Figure S2e-h).

The use of NMA correlation maps computed from ENMs as inputs exhibits a clear advantage compared to contact maps, albeit the latter encodes structural constraints that can be linked to structural fluctuations through an associated Gaussian Network Model~\cite{Rader2005}. The fact that our NMA correlations maps are based on ENM interaction potentials that have been calibrated on protein simulation studies, and the fact that we use of systematically derived dynamic residue-residue correlations as inputs (rather than the mere structural constraints in contact maps) could explain why a CNN trained on NMA correlation maps performs better than on contact maps alone.

The slow and large-scale fluctuations linked to low frequency modes are found to be most relevant to the prediction of binding affinity, suggesting long-range effects of antigen-binding conformational changes, possibly mediated through allosteric mechanisms~\cite{amor_prediction_2016,Tang2020}. 
The detailed characterisation of the conformational changes contributing to binding affinity in antibodies is a direction for potential research in the future, which will benefit from a combination of approaches including structural studies and molecular dynamics simulations~\cite{Barozet2018,Liu2024}. Through the use of spectral properties in the Normal Modes, ANTIPASTI processes fluctuations about equilibrium at different scales, which are linked to structural changes in the binding process. Such information, integrated nonlinearly across scales, has been shown to be predictive of structural~\cite{peach_unsupervised_2019} and allosteric~\cite{amor_prediction_2016,Wu2022} processes, and the predictive performance of ANTIPASTI clearly highlights their relevance for our regression task here.

In this work, coarse-graining from the atomistic to the residue level makes data preprocessing and training more computationally affordable, and it allows us to work with a manageable number of learnt parameters based on the available data. Yet, the relative advantages of coarse-grained \textit{vs.} atomistic protein structure representations~\cite{amor_prediction_2016, Wu2022} are worth further investigation.

We have also assessed the impact on the estimated binding affinity of AlphaFold-predicted, as opposed to experimentally resolved, structures. Our results on a small test set showed that evaluating a model trained on solved structures performs equally well on both types of structures giving support to the application of ANTIPASTI to predict the binding affinity of antibodies whose structure has not yet been experimentally obtained. We also found that the the discrepancy with the affinity estimated from AlphaFold structures (relative to the one estimated from the original structure) increases with the uncertainty of the structural prediction. Hence the AlphaFold-provided measure of uncertainty can be used as an indication on the level of certainty of the prediction. On a more conceptual level, our analysis indicated that the accurate prediction of the antibody-antigen contact interface is crucial to improve AlphaFold-predicted structures for the purpose of affinity prediction. We remark, that we have not used AlphaFold-predicted structures to augment the ANTIPASTI training set, as this can pose problems of data quality because at present AlphaFold is not guaranteed to predict accurately antibody-antigen complex structures~\cite{Evans2022, tubiana_scannet_2022,Pak2023}.

The learnt CNN filters can be used to establish which antibody regions are important for binding affinity. We find differences according to the type of antigen, especially between proteins and other small targets that presumably reflect patterns of binding configuration induced by the size of the target. Our model also allows us to establish an intrinsic measure of region importance based on to their impact on affinity prediction. Further work to establish model-agnostic approaches to quantifying importance through input perturbation~\cite{LiShapley2020}, or occlusion sensitivity~\cite{Li2021} would also be worth of further research.

The study of the affinity-relevant correlations obtained with ANTIPASTI reveals long-range effects that are leveraged by the model to predict binding affinity, consistent with antigen-distal mutations being beneficial to affinity~\cite{Koenig2017}. Similarly, non-additive dependencies (epistasis) between distant sites have been involved in changes in binding affinity upon joint mutations~\cite{Adams2019}, with these dependencies stretching along all the CDRs and FRs of the heavy chain~\cite{Phillips2021} and across the heavy and light chain~\cite{Phillips2023}. These observations are also in line with the extended inter-region and inter-chain synergy in binding affinity suggested by our model. Formulating the prediction of binding affinity in terms of pairwise residue correlations through ANTIPASTI also offers a natural avenue for quantifying the synergy between regions, while avoiding the computationally costly evaluation of pairwise interactions through randomisation approaches. The synergistic patterns detectable in this way could be further explored in future as a help towards rational processes of acquisition of high binding affinity and specificity in antibodies. However, more data on the systematic assessment of the roles of residues in binding affinity by point or pairwise mutations with accompanying structures would be needed to clarify the mechanisms underlying affinity-relevant correlations uncovered by the ANTIPASTI learnt representations. We hope that the tools we provide here can contribute towards the goal of antibody design in conjunction with laboratory-controlled mutagenesis aimed at antibody affinity improvement. In this regard, a main challenge for the future would be to build upon these insights to develop a strategy to propose mutations that can lead to viable antibody structures with enhanced binding affinity. Additionally, docking analyses and the integration of docked structures into our framework will be an important direction for future research. We plan to further investigate the robustness and applicability of ANTIPASTI in such contexts.

\section*{Resource availability}


\subsection*{Lead contact}


Requests for further information and resources should be directed to and will be fulfilled by the lead contact, Barbara Bravi (b.bravi21@imperial.ac.uk).

\subsection*{Materials availability}


This study did not generate new materials.

\subsection*{Data and code availability}





\begin{itemize}
    \item Antibody-antigen PDB files and their corresponding binding affinity values, sourced from SAbDab, have been deposited at Mendeley Data and are publicly available as of the date of publication. The DOI is listed in the key resources table.
    \item All original code has been deposited at \href{https://github.com/kevinmicha/ANTIPASTI}{https://github.com/kevinmicha/ANTIPASTI} and Zenodo, and is publicly available as of the date of publication. The Zenodo DOI is listed in the key resources table.
    \item Any additional information required to reanalyze the data reported in this paper is available from the lead contact upon request.
\end{itemize}

\section*{Acknowledgments}


K.M. acknowledges support from the President's Scholarship at Imperial College London. M.B. acknowledges support by the Engineering and Physical Sciences Research Council (EPSRC) under grant EP/N014529/1 funding the EPSRC Centre for Mathematics of Precision Healthcare at Imperial College London. All authors are grateful to Francesco A. Aprile, Camila M. Clemente and Clément Nizak for their valuable feedback and suggestions.

\section*{Author contributions}


Study concept and design: K.M., M.B., and B.B.; Development of source
code: K.M.; Analysis and interpretation of data: K.M., M.B., B.B.; Writing and revision of the manuscript: K.M., M.B., and B.B.;
Study supervision: M.B. and B.B.

\section*{Declaration of interests}


The authors declare no competing interests.

\newpage

\section*{STAR METHODS}\label{methods}


\subsection*{Key resources table}

\subsection*{Method details}


\subsubsection*{Data collection and pre-processing}

The Structural Antibody Database (SAbDab)~\cite{SAbDab2014,SAbDab2022} contains all the antibody structures available in the Protein Data Bank archive (PDB)~\cite{pdb}. We downloaded all bound antibody structures with affinity data from SAbDab as of 23 June 2023 in PDB format using the Chothia numbering~\cite{Dondelinger2018}.

Entries without affinity or antigen data, VL single-domain antibodies and single-chain variable fragments (scFvs) were discarded. Structures with less than three atoms in any residue (incomplete amino acids), with a resolution worse than $4 \angstrom$~\cite{ruffolo_antibody_2022} or with very few residues in the heavy chain (less than $30$) were excluded as well. We applied no filters based on the antibody isotype. Just $80$ out of the remaining $634$ antibodies (available in the GitHub repository) are nanobodies consisting of a heavy chain only, meaning that most of them have two paired chains. For both chains, we considered all amino acids in the variable region, corresponding, in Chothia numbering, to the range of positions from 1 to $113$ for the heavy chain and from 1 to $107$ for the light chain (\figref[b]{1}).

\subsubsection*{Normal Mode Analysis and the computation of correlation maps}

The main goal of Normal Mode Analysis (NMA) is the characterisation of dynamical fluctuations around equilibria in physical systems~\cite{Dykeman2010}. This analysis is only valid near the equilibrium (\textit{i.e.}, for small deformations) and for short times. Still, Normal Modes have been useful for protein structures beyond this strict regime of validity, since they capture global information on the expected dynamical behaviour upon a perturbation (like the binding of a ligand at a certain site) that is a consequence of the connectivity of the protein’s network of residues. The key step of NMA is to simplify the dynamics using a set of generalised coordinates known as Normal Modes, which are derived from the spectral decomposition of the Hessian of the potential describing the system, as follows.

Let $\mathbf{r^{*}}\in \mathbb{R}^{3N}$ be the coordinates of a system of $N$ particles in 3D space at equilibrium. Then, its potential energy under a small perturbation can be written (to second order) as:
\[U(\mathbf{r}) = U(\mathbf{r}^{*}) + (\mathbf{r} - \mathbf{r}^{*})^\top \left. \nabla U(\mathbf{r})\right | _{\mathbf{r}^{*}} + \frac{1}{2} (\mathbf{r} - \mathbf{r}^{*})^\top 
\left. \nabla^2 U(\mathbf{r}) \right 
|_{\mathbf{r}^{*}} 
(\mathbf{r} - \mathbf{r}^{*}) + \dots
\]
Here, $U(\mathbf{r}^{*})$ is arbitrary and can be set to zero~\cite{nmaBauer}, while the gradient term is zero at a local minimum. Therefore, the potential energy can be expressed as a function of components of the Hessian matrix $\nabla^2 U(\mathbf{r})$ evaluated at the minimum $\mathbf{r^{*}}$. 

The solutions to the equations of motion under a quadratic potential as given by $U$ take the form:
$$\mathbf{r}-\mathbf{r}^{*}\propto\sum_{l=0}^{3N-1}C^{(l)}\mathbf{a}^{(l)}\cos(\omega^{(l)}t+\phi^{(l)})$$
The fluctuations of the particles can be thus expressed as linear combinations of Normal Modes, indexed by $l$. Hence $C^{(l)}$ and $\phi^{(l)}$ are, respectively, the amplitude and the phase of the $l^{th}$ Normal Mode, $\mathbf{a}^{(l)}\in \mathbb{R}^{3N}$ is the $l^{th}$ eigenvector of the Hessian matrix $\left(\nabla^2 U(\mathbf{r})\right)^{*}$ and $\omega^{(l)}$ is its associated eigenvalue~\cite{Tama2001}. It is important to note that the first six NM are trivial, as they are associated to rotational and translational invariances~\cite{Dubanevics2022}.

A potential that assumes that each particle behaves as a harmonic oscillator~\cite{VanWynsberghe2006} was proposed in Ref.~\cite{Hinsen2005} and underpins the so-called Elastic Network Models (ENMs):
$$U(\mathbf{r}) = \sum_{\substack{i,j \\ r_{ij}^{*}<R_c}}k(r_{ij}^{*})(r_{ij}-r_{ij}^{*})^2$$
with $r_{ij}$ the distance between particles $i$ and $j$, $R_c$ a cut-off radius and $k(r_{ij}^{*})$ the pairwise force terms, here specified by the $\alpha$-Carbon force field. The latter, derived from the fitting to the Amber94~\cite{Cornell1995} potential, is given by:
$$
  k(r) =
    \begin{cases}
      8.6.10^2r-2.39.10^3 \text{ for } r<4\angstrom \\
      128.10^4r^{-6} \text{ otherwise}
    \end{cases}       
$$
We use several Bio3D~\cite{nmaBio3D} functions and an in-house script to implement this formalism at the residue level. Note that non-protein antigens are not explicitly included by Bio3D when computing the Normal Modes. However, we have checked that including non-protein antigens using a more computationally expensive atomistic approach for the computation of NM correlation maps does not affect our conclusions on clustering by binding target type and on region importance for non-protein antigens (see Figures S2e and S3c). We compute the Normal Modes of antibody-antigen complexes, and the associated Normal Mode correlation maps:
$$X_{ij} \propto\sum_{l=0}^{L-1} \frac{{\mathbf{a}^{(l)}_i}\cdot\mathbf{a}^{(l)}_j}{{\omega^{(l)}}^2} $$
where $\cdot$ is the dot product in $\mathbb{R}^3$, $\mathbf{a}^{(l)}_i$ represents the
Cartesian coordinates of the $l^{th}$ eigenvector of the Hessian matrix $\left(\nabla^2 U(\mathbf{r})\right)^{*}$ corresponding to residue $i$, and similarly for $\mathbf{a}^{(l)}_j$. $L$ is the number of modes considered, which can be at most $3N$. The maps are then normalised to the $[-1,1]$ range.

In our work, we calculate these maps in the presence of the antigens to capture the imprint of the antigen correlations with the antibodies; once computed, the antigen residues are eliminated from the input maps for our algorithm. Missing residues (gaps) are encoded as zeros in the Normal Mode maps, denoting absence of correlation of their Normal Modes. The final value for $N$ is $292$ and it corresponds to the length of heavy and light chains with alignment to Chothia. This is performed identically for the nanobodies, but as expected the input maps are zero-valued in all pixels of the light chain and of the blocks shared by the heavy and light chains. To construct the NM correlation maps, we rely on the software~\cite{nmaBio3D}, a package to compute the Normal Modes of a given protein structure, including fluctuations, deformation, and residue cross-correlation analysis.

\subsubsection*{ANTIPASTI Neural Network architecture}

ANTIPASTI is based on a convolutional neural network which receives as input Normal Mode correlation maps $X\in\mathbb{R}^{N\times N}$ (see \figref[c]{1}). The model is PyTorch-based~\cite{PyTorchPaper} and comprises a bank of $n_{\text{f}}$ $k\times k$ convolutional filters, followed by a ReLU activation function and a $p\times p$ max pooling layer with stride $s$. 
The latter is followed by a bias-free fully-connected layer from which the value of $\log_{10}(K_D)$ is computed as follows. Let $\mathbf{z}\in\mathbb{R}^M$ be the flattened input of the ANTIPASTI fully-connected layer, produced by the network's convolutional layer (with $n_{\text{f}}$ filters and the non-linear activation) and the max pooling layer applied to a Normal Mode correlation map $X\in\mathbb{R}^{N\times N}$.

Let $\mathbf{w}\in\mathbb{R}^M$ be the weights of the fully-connected layer. Then, since there is no bias, the prediction of $K_D$ is simply
$$\log_{10}(\hat{K}_D) = \mathbf{w}^\top\mathbf{z}=\sum_{i=0}^{M-1}w_i z_i$$
where for the described architecture
$$
M=n_{\text{f}}\left(\left\lfloor\frac{N-k+1-p}{s}+1\right\rfloor\right)^2$$
as every convolution operation yields a matrix of shape $(N-k+1)\times(N-k+1)$ which is subtracted by $p$ and downsized by $s$ through max pooling with stride $s$ and kernel shape $p\times p$. In this work, we take $s$ always equal to {$1$}.

\subsubsection*{Training and evaluation}

We split $634$ available correlation maps with their corresponding $\log_{10}(K_D)$ values into a training set and a test set in a $95-5$ ratio, repeating the split 5 times.

For each split, the training set contains 602 complexes and the test set 32 complexes. To avoid redundancy, for every pair of antigen-antibody complexes in the training and test set respectively, we check that the antibody sequence identity is below a threshold of 90\%~\cite{Davila2022} and that the antigen sequences are not identical. Specifically, we proceed as follows: a complex is randomly selected and added to the test set. If the selected complex satisfies the non-redundancy condition with respect to all the complexes in the training set, it is kept; otherwise it is returned to the training set. This process is repeated until the required test size of 32 complexes (5\% of the dataset) is achieved.

We ensure \textit{a posteriori} that the antibody-antigen complexes are sufficiently representative by making sure that different types of antigens are present and that the range of binding affinity values is sufficiently large (see \figref[a]{2}).

Training is performed in a supervised fashion with the AdaBelief optimiser and a learning rate $l_r$. AdaBelief~\cite{AdaBelief} guarantees the fast convergence found in adaptive methods such as Adam, good generalisation as in accelerated schemes such as Stochastic Gradient Descent (SGD) and non-exploding gradients during training. The parameters were set to $\beta_1 = 0.9$, $\beta_2 = 0.999$ and $\epsilon=10^{-8}$. The latter is a typical configuration for a regression task~\cite{zhuang2021acprop}.

Given a Lagrange multiplier $\lambda\in\mathbb{R}_{>0}$, a batch of size $B_s$ of input data points $X^{(i)}$, and a function $f_{\theta}:\mathbb{R}^{N\times N} \rightarrow \mathbb{R}$, the loss function used to learn $\theta$ is given by:

$$\mathcal{L}(\theta)=\frac{1}{B_s}\sum_{i=0}^{B_s-1}\left( \log_{10}(K_{D}^{(i)})-f_{\theta}(X^{(i)})\right)^2+\lambda\left\lVert \theta\right\rVert_1$$

It consists of a data fidelity term and a regularisation term. The first one quantifies the mean squared error (MSE) between the ground truth $\log_{10}(K_D)$ and the prediction $\log_{10}(\hat{K}_D)=f_{\theta}(X)$, and its expected value is computed as the empirical mean over a batch ($B_s=32$). To favour sparsity in the weights of the neural network while avoiding overfitting, the $\ell_1$ regularisation~\cite{Ma2019} is employed in the second term with $\lambda=2\times10^{-3}$.

Hyperparametric search was performed with Optuna~\cite{Akiba2019Optuna} and 10-fold cross-validation on the training set for a given training/test split (approximately 602 structures) by considering all the possible combinations of $n_\text{f}\in\lbrace{2,4,8\rbrace}$, $k\in\lbrace{3,4,5\rbrace}$, $p\in\lbrace{1,2,3\rbrace}$ and $l_r\in\lbrace{5\times10^{-5}, 1\times10^{-4},5\times10^{-4},}$
${1\times10^{-3}\rbrace}$. With comparable accuracy, the top two candidates were (Figure S1a, Table S2):
\begin{itemize}
    \item $n_\text{f}=4$, $k=4$, $p=2$, and $l_r=1\times10^{-4}$ (83000 parameters);
    \item $n_\text{f}=4$, $k=4$, $p=1$, and $l_r=5\times10^{-5}$ (334000 parameters).
\end{itemize}

These methodological choices (cross-validation and introduction of an $\ell_1$ regularisation) are made to keep overfitting under control, given the limited dataset size. The training was conducted for $5$ different random seeded train-test splits using these optimal hyperparameters. The performance of ANTIPASTI was then evaluated on the test sets through the Pearson's correlation coefficient $R$ between predicted and ground truth binding affinity values (figref[a,c]{2}).

A linear regression model with no bias term and the same loss function was also considered. In essence, this model maps the input pixels to $\log_{10}{(K_D)}$ using $N^2$ weights.

Contact maps (with $\alpha$-Carbon cutoff distances of $8\angstrom$~\cite{Guven2023}) were also examined for comparison with the Normal Mode correlation maps. 
The optimal hyperparameters were the same as for NMA, except that for the case without pooling the optimal learning rate $l_r$ turned out to be $1\times10^{-4}$ instead of $5\times10^{-5}$.

\subsubsection*{Interpretation of the learnt output layer representations}

We construct a $\sqrt{\frac{M}{n_{\text{f}}}}\times\sqrt{\frac{M}{n_{\text{f}}}}$ array $F$ using the weights of the fully-connected layer $\mathbf{w}\in\mathbb{R}^M$ and its input $\mathbf{z}\in\mathbb{R}^M$. We refer to $F$ as the \textit{output layer representation} and it is obtained by reshaping the $M$ output pixels of the fully-connected layer:

$$F_{jk}:= -\sum_{l=0}^{n_\text{f}-1}w_{i}z_{i}=F_{kj};\;i=j+\sqrt{\frac{M}{n_{\text{f}}}}k+l\frac{M}{n_{\text{f}}}$$
with $j=0,\dots,\sqrt{\frac{M}{n_{\text{f}}}}-1; k=0,\dots,\sqrt{\frac{M}{n_{\text{f}}}}-1$, then
$$\log_{10}(\hat{K}_D) = -\sum_{j,k}F_{jk}$$
The positive (respectively, negative) pixels of $F$ can be interpreted as the Normal Mode correlations in $X$ responsible for its high (respectively, low) binding affinity.

For the sake of better visualisation and inspection, the $F$ arrays are post-processed with UMAP~\cite{McInnes2018} using a minimum Euclidean distance of $0.1$ and a minimum of $90$ neighbours. We coloured the data points according to sequence properties extracted directly from SAbDab, such as antigen type (\figref[b]{3}), V gene family, light chain type and antibody species (Figures 3a-c, Figures S2b-d). For the latter two, ellipses comprising $75\%$ of the data points were added to show the separation of the different classes.

\subsubsection*{Antigen-agnostic case}

We tested the scenario in which we train ANTIPASTI without taking antigen information into account when calculating the Normal Modes. We call this scenario the \textit{antigen-agnostic} case, which requires a bound antibody-antigen conformation but not the antigen sequence or its atom coordinates. Although the accuracy (Figure S1b) is lower than that of the standard case (with \textit{antigen imprint}), we found, when following the same procedure as for Figure 3b, noteworthy separations according to various biological properties in the UMAP maps of the output layer representations (Figures S2b–d). 

\subsubsection*{Antibody regions importance}

To study which heavy and light chain regions have a greater impact on the binding affinity, we computed the importance $I$ for each region $s^*$ with a mean number of residues across the entire dataset $\bar{N}(s^*)$ as follows:

$$I(s^*)=100\frac{\bar{N}(s_{\text{Best}})}{\bar{N}(s^*)}\frac{\bigg\rvert\text{MSE}\big\rvert_{s \neq s^*}-\text{MSE}\bigg\lvert}{\bigg\rvert\text{MSE}\big\rvert_{s \neq s_{\text{Best}}}-\text{MSE}\bigg\lvert}$$

We estimated the importance $I(s^*)$ separately for each group of antibodies binding to different antigen types (\figref[c]{3}). The Mean Squared Error (MSE) was computed as:
\begin{equation*}
\label{eq:mse}
\text{MSE} = \frac{1}{A_t}\sum_{i=0}^{A_t-1}\left( \log_{10}(K_{D}^{(i)})-f_{\theta^{*}}(X^{(i)})\right)^2
\end{equation*}
where $f_{\theta^{*}}$ denotes the binding affinity prediction $\log_{10}(\hat{K}_D)$ by the final ANTIPASTI architecture with learnt parameters $\theta^{*}$. $A_t$ is the number of antibodies in the training set binding to a given antigen type $t$, the types being proteins, peptides, haptens and carbohydrates ($A_t = 347$, $113$, $53$ and $24$ respectively). Nanobodies were excluded as they would introduce a bias in the importance in favour of the heavy chain regions.

Excluding a region means that all output layer pixels in the region and their correlations with other regions are removed for the predictions. $s_\text{Best}$ is the region that, when excluded, causes the MSE to deviate the most from that obtained with all regions. Hence $s_\text{Best}$ represents the region that contributes the most to an accurate prediction of $K_D$. Regions importance is defined in such a way as to be $100$ for $s_\text{Best}$, thus for each region it can be expressed as a percentage of $s_\text{Best}$ importance.

We estimated the regions importance for the entire Framework and CDR regions of the heavy and light chains (\figref[c]{3}). The mean importance for each region was determined by averaging the importance scores across the different splits. To compute the limits of the error bars, we considered the extreme cases (\textit{i.e.}, the minimum and maximum) in terms of MSE deviation occurring in any of the splits. To build Figure S3a, we examined for each region $s^*$ the proportion of MSE deviation attributable to correlations between residues within the same region (intra-region) and that resulting from correlations with residues in other regions (inter-region). We proceeded with the same logic for Figure S3b, this time considering correlations within the chain (heavy or light) to which $s^*$ belongs, as opposed to those established between $s^*$ and residues from the other chain.

\subsubsection*{AlphaFold prediction}

We used the ColabFold v1.5.5~\cite{Mirdita2022} implementation of AlphaFold2. The antigen and antibody were chained together in one sequence and, for each prediction, $3$ recycles were used with an early stop tolerance set to \texttt{auto}. We gave the possibility of detecting templates from a $100\%$ clustered PDB, \textit{i.e.}, the option \texttt{template\_mode} was set to \texttt{pdb100}. While retaining the original model training weights, ColabFold features a faster search via MMseqs2~\cite{Steinegger2017} instead of using the time-consuming MSA generation step~\cite{Ruffolo2023}.

We used $21$ samples of a test set extracted manually from the PDB. It consists of the latest published antibody-antigen entries that satisfy our data acceptance conditions (see Data collection and pre-processing), limiting them to one per study to avoid redundancy. We then compared the $\log_{10}(K_D)$ prediction of the original structure and that of the AlphaFold-estimated structure (\figref[a]{4}). We refer to the subtraction of these two quantities as $\Delta\!\log_{10}(\hat{K}_D)$ (\figref[c]{4}). Specifically, the resulting PDB codes were: \texttt{7trh}~\cite{7TRH}, \texttt{7y3o}~\cite{7Y3O}, \texttt{7yp2}~\cite{7YP2}, \texttt{8ag1}~\cite{8AG1}, \texttt{8av9}~\cite{8AV9}, \texttt{8b7h}~\cite{8B7H}, \texttt{8byu}~\cite{8BYU}, \texttt{8eli}~\cite{8ELI}, \texttt{8f5i}~\cite{8F5I}, \texttt{8f6o}~\cite{8F6O}, \texttt{8f9u}~\cite{8F9U}, \texttt{8fgx}~\cite{8FGX}, \texttt{8g8n}~\cite{8G8N}, \texttt{8hes}~\cite{8HES}, \texttt{8hn6}~\cite{8HN6}, \texttt{8ipc}~\cite{8IPC}, \texttt{8ivx}~\cite{8IVX}, \texttt{8jlx}~\cite{8JLX}, \texttt{8pe9}~\cite{8PE9}, \texttt{8t9z}~\cite{8T9Z} and \texttt{8x0t}~\cite{8X0T}, the least recent of them being published in May 2023. Their labelled $-\log_{10}(K_D)$ values range between $7$ and $11$, hence are representative of a wide range of binding affinities.

The Interface Root Mean Square Deviation (IRMSD) between the original and AlphaFold-predicted structures (Figure S4c) was computed using Rosetta~\cite{Fleishman2011} IRmsd filter with default options.

\subsubsection*{Python package}\label{code}

All the code is available in a GitHub repository
as a Python package at~\url{github.com/kevinmicha/ANTIPASTI}. It includes installation and dependency management instructions, ready-to-run environments and four tutorials in the notebook format. We also provide unit tests with 98.5\% coverage of the total lines of code.

We wrote the code in a modular way, in such a way that it is possible to use NMA or contact maps, a CNN or LR and customise both models by just modifying the settings. This makes it immediately adaptable to architectures with more trainable parameters in the future if additional data becomes available.

\subsection*{Quantification and Statistical Analysis}

In Figures 1b, 2a, 2c, 4a, S1b and S4a, the R score corresponds to the Pearson correlation between predictions $\lbrace\hat{y}_i\rbrace_{i=1}^{n}$ and ground truth values $\lbrace y_i\rbrace_{i=1}^{n}$:
$$-1 \leq \,  R:=\frac{\sum_{i=1}^{n} (\hat{y}_i - m_{\hat{y}})(y_i - m_{y})}{\sqrt{\sum_{i=1}^{n} (\hat{y}_i - m_{\hat{y}})^2 \sum_{i=1}^{n} (y_i - m_{y})^2}} \,  \leq 1.$$
Here, $m_{\hat{y}}$ and $m_{y}$ denote the means of the predicted values and the ground truth values, respectively, and $n$ is the number of samples in the test set ($n=32$ in this work, except in Figure 4a, where $n=21$). R scores of $-1$, $0$ and $1$ indicate perfect negative correlation, no correlation and perfect positive correlation, respectively, between predictions and ground truth values.

In Figures 4c-d and S6b-c, the Spearman correlation coefficient $\rho$ was computed between $\Delta\!\log_{10}(\hat{K}_D)$ and various metrics, such as pLDDT, PAE and IRMSD from AlphaFold, defined as: 
$$ -1 \leq \, \rho := 1 - 6\frac{\sum_{i=1}^n d_i^2}{n(n^2-1)} \, \leq 1,$$
where $n$ is the number of samples ($n=21$ in this work), and $d_i$ represents the difference between the ranks of the $i^{\text{th}}$ pair of values. Again, $\rho=1$ and $\rho=-1$ indicate  perfect monotonic increasing and decreasing relationships, respectively, and $\rho =0$ indicates no ordinal relationship. 
The p-values correspond to a hypothesis test where the null hypothesis postulates no ordinal correlation between samples, and $p<0.05$ is typically considered as significant.

Both Pearson and Spearman correlation coefficients were computed using SciPy functions~\cite{Virtanen2020}. Information on estimated statistics for each result shown is provided in the caption of the corresponding figures.

\section*{Supplemental information index}




\begin{description}
  \item Document S1. Figures S1–S4 and Tables S1 and S2
\end{description}

\newpage


\bibliography{paper}

\begin{thebibliography}{113}
\providecommand{\natexlab}[1]{#1}
\providecommand{\href}[2]{#2}
\providecommand{\path}[1]{#1}
\providecommand{\DOIprefix}{}
\providecommand{\ArXivprefix}{arXiv:}
\providecommand{\URLprefix}{}
\providecommand{\doi}[1]{\href{http://dx.doi.org/#1}{\path{#1}}}
\providecommand{\BIBand}{and}
\providecommand{\bibinfo}[2]{#2}
\ifx\xfnm\undefined \def\xfnm[#1]{\unskip,\space#1}\fi
\makeatletter\def\@biblabel#1{#1.}\makeatother
\bibitem[{Strohl(2017)}]{Strohl2017}
\bibinfo{author}{Strohl, W.~R.} (\bibinfo{year}{2017}). \bibinfo{title}{Current progress in innovative engineered antibodies}.
\newblock \bibinfo{journal}{Protein \& Cell} \emph{\bibinfo{volume}{9}}, \bibinfo{pages}{86--120}. \DOIprefix\doi{10.1007/s13238-017-0457-8}.
\bibitem[{Lu et~al.(2020)Lu, Hwang, Liu, Lee, Tsai, Li and Wu}]{Lu2020}
\bibinfo{author}{Lu, R.-M.}, \bibinfo{author}{Hwang, Y.-C.}, \bibinfo{author}{Liu, I.-J.}, \bibinfo{author}{Lee, C.-C.}, \bibinfo{author}{Tsai, H.-Z.}, \bibinfo{author}{Li, H.-J.}, and \bibinfo{author}{Wu, H.-C.} (\bibinfo{year}{2020}). \bibinfo{title}{Development of therapeutic antibodies for the treatment of diseases}.
\newblock \bibinfo{journal}{Journal of Biomedical Science} \emph{\bibinfo{volume}{27}}, \bibinfo{pages}{1}. \DOIprefix\doi{10.1186/s12929-019-0592-z}.
\bibitem[{Jin et~al.(2022)Jin, Sun, Liang, Gu, Ning, Xu, Chen and Pan}]{Jin2022Rev}
\bibinfo{author}{Jin, S.}, \bibinfo{author}{Sun, Y.}, \bibinfo{author}{Liang, X.}, \bibinfo{author}{Gu, X.}, \bibinfo{author}{Ning, J.}, \bibinfo{author}{Xu, Y.}, \bibinfo{author}{Chen, S.}, and \bibinfo{author}{Pan, L.} (\bibinfo{year}{2022}). \bibinfo{title}{Emerging new therapeutic antibody derivatives for cancer treatment}.
\newblock \bibinfo{journal}{Signal Transduction and Targeted Therapy} \emph{\bibinfo{volume}{7}}, \bibinfo{pages}{39}. \DOIprefix\doi{10.1038/s41392-021-00868-x}.
\bibitem[{Chiu et~al.(2019)Chiu, Goulet, Teplyakov and Gilliland}]{Chiu2019}
\bibinfo{author}{Chiu, M.~L.}, \bibinfo{author}{Goulet, D.~R.}, \bibinfo{author}{Teplyakov, A.}, and \bibinfo{author}{Gilliland, G.~L.} (\bibinfo{year}{2019}). \bibinfo{title}{Antibody structure and function: The basis for engineering therapeutics}.
\newblock \bibinfo{journal}{Antibodies} \emph{\bibinfo{volume}{8}}. \DOIprefix\doi{10.3390/antib8040055}.
\bibitem[{Sormanni et~al.(2018)Sormanni, Aprile and Vendruscolo}]{sormanni_third_2018}
\bibinfo{author}{Sormanni, P.}, \bibinfo{author}{Aprile, F.~A.}, and \bibinfo{author}{Vendruscolo, M.} (\bibinfo{year}{2018}). \bibinfo{title}{Third generation antibody discovery methods: \textit{in silico} rational design}.
\newblock \bibinfo{journal}{Chemical Society Reviews} \emph{\bibinfo{volume}{47}}, \bibinfo{pages}{9137--9157}. \DOIprefix\doi{10.1039/C8CS00523K}.
\bibitem[{Svilenov et~al.(2021)Svilenov, Sacherl, Protzer, Zacharias and Buchner}]{CDRlength}
\bibinfo{author}{Svilenov, H.}, \bibinfo{author}{Sacherl, J.}, \bibinfo{author}{Protzer, U.}, \bibinfo{author}{Zacharias, M.}, and \bibinfo{author}{Buchner, J.} (\bibinfo{year}{2021}). \bibinfo{title}{Mechanistic principles of an ultra-long bovine cdr reveal strategies for antibody design}.
\newblock \bibinfo{journal}{Nature Communications} \emph{\bibinfo{volume}{12}}. \DOIprefix\doi{10.1038/s41467-021-27103-z}.
\bibitem[{Sela-Culang et~al.(2013)Sela-Culang, Kunik and Ofran}]{Sela2013}
\bibinfo{author}{Sela-Culang, I.}, \bibinfo{author}{Kunik, V.}, and \bibinfo{author}{Ofran, Y.} (\bibinfo{year}{2013}). \bibinfo{title}{The structural basis of antibody-antigen recognition}.
\newblock \bibinfo{journal}{Frontiers in Immunology} \emph{\bibinfo{volume}{4}}. \DOIprefix\doi{10.3389/fimmu.2013.00302}.
\bibitem[{Akbar et~al.(2022)Akbar, Robert, Weber, Widrich, Frank, Pavlović, Scheffer, Chernigovskaya, Snapkov, Slabodkin, Mehta, Miho, Lund-Johansen, Andersen, Hochreiter, Haff, Klambauer, Sandve and Greiff}]{akbar_silico_2022}
\bibinfo{author}{Akbar, R.}, \bibinfo{author}{Robert, P.~A.}, \bibinfo{author}{Weber, C.~R.}, \bibinfo{author}{Widrich, M.}, \bibinfo{author}{Frank, R.}, \bibinfo{author}{Pavlović, M.}, \bibinfo{author}{Scheffer, L.}, \bibinfo{author}{Chernigovskaya, M.}, \bibinfo{author}{Snapkov, I.}, \bibinfo{author}{Slabodkin, A.}, \bibinfo{author}{Mehta, B.~B.}, \bibinfo{author}{Miho, E.}, \bibinfo{author}{Lund-Johansen, F.}, \bibinfo{author}{Andersen, J.~T.}, \bibinfo{author}{Hochreiter, S.}, \bibinfo{author}{Haff, I.~H.}, \bibinfo{author}{Klambauer, G.}, \bibinfo{author}{Sandve, G.~K.}, and \bibinfo{author}{Greiff, V.} (\bibinfo{year}{2022}). \bibinfo{title}{In silico proof of principle of machine learning-based antibody design at unconstrained scale}.
\newblock \bibinfo{journal}{mAbs} \emph{\bibinfo{volume}{14}}. \DOIprefix\doi{10.1080/19420862.2022.2031482}.
\bibitem[{Reis et~al.(2022)Reis, Barletta, Gagliardi, Fortuna, Soler and Rocchia}]{Reis2022}
\bibinfo{author}{Reis, P. B. P.~S.}, \bibinfo{author}{Barletta, G.~P.}, \bibinfo{author}{Gagliardi, L.}, \bibinfo{author}{Fortuna, S.}, \bibinfo{author}{Soler, M.~A.}, and \bibinfo{author}{Rocchia, W.} (\bibinfo{year}{2022}). \bibinfo{title}{Antibody-antigen binding interface analysis in the big data era}.
\newblock \bibinfo{journal}{Frontiers in Molecular Biosciences} \emph{\bibinfo{volume}{9}}. \DOIprefix\doi{10.3389/fmolb.2022.945808}.
\bibitem[{Lima et~al.(2020)Lima, Gasteiger, Marcatili, Duek, Bairoch and Cosson}]{ABCDdatabase}
\bibinfo{author}{Lima, W.}, \bibinfo{author}{Gasteiger, E.}, \bibinfo{author}{Marcatili, P.}, \bibinfo{author}{Duek, P.}, \bibinfo{author}{Bairoch, A.}, and \bibinfo{author}{Cosson, P.} (\bibinfo{year}{2020}). \bibinfo{title}{The abcd database: a repository for chemically defined antibodies}.
\newblock \bibinfo{journal}{Nucleic Acids Research} \emph{\bibinfo{volume}{48}}, \bibinfo{pages}{D261–D264}. \DOIprefix\doi{10.1093/nar/gkz714}.
\bibitem[{Olsen et~al.(2022)Olsen, Boyles and Deane}]{OAS}
\bibinfo{author}{Olsen, T.~H.}, \bibinfo{author}{Boyles, F.}, and \bibinfo{author}{Deane, C.~M.} (\bibinfo{year}{2022}). \bibinfo{title}{Observed antibody space: A diverse database of cleaned, annotated, and translated unpaired and paired antibody sequences}.
\newblock \bibinfo{journal}{Protein Science} \emph{\bibinfo{volume}{31}}, \bibinfo{pages}{141--146}. \DOIprefix\doi{10.1002/pro.4205}.
\bibitem[{Dunbar et~al.(2013)Dunbar, Krawczyk, Leem, Baker, Fuchs, Georges, Shi and Deane}]{SAbDab2014}
\bibinfo{author}{Dunbar, J.}, \bibinfo{author}{Krawczyk, K.}, \bibinfo{author}{Leem, J.}, \bibinfo{author}{Baker, T.}, \bibinfo{author}{Fuchs, A.}, \bibinfo{author}{Georges, G.}, \bibinfo{author}{Shi, J.}, and \bibinfo{author}{Deane, C.~M.} (\bibinfo{year}{2013}). \bibinfo{title}{{SAbDab: the structural antibody database}}.
\newblock \bibinfo{journal}{Nucleic Acids Research} \emph{\bibinfo{volume}{42}}, \bibinfo{pages}{D1140--D1146}. \DOIprefix\doi{10.1093/nar/gkt1043}.
\bibitem[{Schneider et~al.(2021{\natexlab{a}})Schneider, Raybould and Deane}]{SAbDab2022}
\bibinfo{author}{Schneider, C.}, \bibinfo{author}{Raybould, M. I.~J.}, and \bibinfo{author}{Deane, C.~M.} (\bibinfo{year}{2021}{\natexlab{a}}). \bibinfo{title}{{SAbDab in the age of biotherapeutics: updates including SAbDab-nano, the nanobody structure tracker}}.
\newblock \bibinfo{journal}{Nucleic Acids Research} \emph{\bibinfo{volume}{50}}, \bibinfo{pages}{D1368--D1372}. \DOIprefix\doi{10.1093/nar/gkab1050}.
\bibitem[{Berman et~al.(2000)Berman, Westbrook, Feng, Gilliland, Bhat, Weissig, Shindyalov and Bourne}]{pdb}
\bibinfo{author}{Berman, H.~M.}, \bibinfo{author}{Westbrook, J.}, \bibinfo{author}{Feng, Z.}, \bibinfo{author}{Gilliland, G.}, \bibinfo{author}{Bhat, T.~N.}, \bibinfo{author}{Weissig, H.}, \bibinfo{author}{Shindyalov, I.~N.}, and \bibinfo{author}{Bourne, P.~E.} (\bibinfo{year}{2000}). \bibinfo{title}{{The Protein Data Bank}}.
\newblock \bibinfo{journal}{Nucleic Acids Research} \emph{\bibinfo{volume}{28}}, \bibinfo{pages}{235--242}. \DOIprefix\doi{10.1093/nar/28.1.235}.
\bibitem[{Chinery et~al.(2022)Chinery, Wahome, Moal and Deane}]{Chinery2023}
\bibinfo{author}{Chinery, L.}, \bibinfo{author}{Wahome, N.}, \bibinfo{author}{Moal, I.}, and \bibinfo{author}{Deane, C.~M.} (\bibinfo{year}{2022}). \bibinfo{title}{{Paragraph—antibody paratope prediction using graph neural networks with minimal feature vectors}}.
\newblock \bibinfo{journal}{Bioinformatics} \emph{\bibinfo{volume}{39}}. \DOIprefix\doi{10.1093/bioinformatics/btac732}.
\bibitem[{Ambrosetti et~al.(2020)Ambrosetti, Olsen, Olimpieri, Jiménez-García, Milanetti, Marcatilli and Bonvin}]{ambrosetti_proabc-2_2020}
\bibinfo{author}{Ambrosetti, F.}, \bibinfo{author}{Olsen, T.~H.}, \bibinfo{author}{Olimpieri, P.~P.}, \bibinfo{author}{Jiménez-García, B.}, \bibinfo{author}{Milanetti, E.}, \bibinfo{author}{Marcatilli, P.}, and \bibinfo{author}{Bonvin, A. M. J.~J.} (\bibinfo{year}{2020}). \bibinfo{title}{{proABC}-2: {PRediction} of {AntiBody} contacts v2 and its application to information-driven docking}.
\newblock \bibinfo{journal}{Bioinformatics} \emph{\bibinfo{volume}{36}}, \bibinfo{pages}{5107--5108}. \DOIprefix\doi{10.1093/bioinformatics/btaa644}.
\bibitem[{Pittala and Bailey-Kellogg(2020)}]{Pittala2020}
\bibinfo{author}{Pittala, S.}, and \bibinfo{author}{Bailey-Kellogg, C.} (\bibinfo{year}{2020}). \bibinfo{title}{{Learning context-aware structural representations to predict antigen and antibody binding interfaces}}.
\newblock \bibinfo{journal}{Bioinformatics} \emph{\bibinfo{volume}{36}}, \bibinfo{pages}{3996--4003}. \DOIprefix\doi{10.1093/bioinformatics/btaa263}.
\bibitem[{Schneider et~al.(2021{\natexlab{b}})Schneider, Buchanan, Taddese and Deane}]{Schneider2021}
\bibinfo{author}{Schneider, C.}, \bibinfo{author}{Buchanan, A.}, \bibinfo{author}{Taddese, B.}, and \bibinfo{author}{Deane, C.~M.} (\bibinfo{year}{2021}{\natexlab{b}}). \bibinfo{title}{{DLAB: deep learning methods for structure-based virtual screening of antibodies}}.
\newblock \bibinfo{journal}{Bioinformatics} \emph{\bibinfo{volume}{38}}, \bibinfo{pages}{377--383}. \DOIprefix\doi{10.1093/bioinformatics/btab660}.
\bibitem[{Alzubaidi et~al.(2021)Alzubaidi, Zhang, Humaidi, Al-dujaili, Duan, Al-Shamma, Santamar{\'i}a, Fadhel, Al-Amidie and Farhan}]{Alzubaidi2021}
\bibinfo{author}{Alzubaidi, L.}, \bibinfo{author}{Zhang, J.}, \bibinfo{author}{Humaidi, A.~J.}, \bibinfo{author}{Al-dujaili, A.}, \bibinfo{author}{Duan, Y.}, \bibinfo{author}{Al-Shamma, O.}, \bibinfo{author}{Santamar{\'i}a, J.~I.}, \bibinfo{author}{Fadhel, M.~A.}, \bibinfo{author}{Al-Amidie, M.}, and \bibinfo{author}{Farhan, L.} (\bibinfo{year}{2021}). \bibinfo{title}{Review of deep learning: concepts, cnn architectures, challenges, applications, future directions}.
\newblock \bibinfo{journal}{Journal of Big Data} \emph{\bibinfo{volume}{8}}. \DOIprefix\doi{10.1186/s40537-021-00444-8}.
\bibitem[{Singh and Saha(2003)}]{Singh2003}
\bibinfo{author}{Singh, R.}, and \bibinfo{author}{Saha, M.} (\bibinfo{year}{2003}). \bibinfo{title}{Identifying structural motifs in proteins}.
\newblock \bibinfo{journal}{Pacific Symposium on Biocomputing}, pp.  \bibinfo{pages}{228--239}. \DOIprefix\doi{10.1142/9789812776303_0022}.
\bibitem[{Soleymani et~al.(2018)Soleymani, Dabouei, Kazemi, Dawson and Nasrabadi}]{Soleymani2018}
\bibinfo{author}{Soleymani, S.}, \bibinfo{author}{Dabouei, A.}, \bibinfo{author}{Kazemi, H.}, \bibinfo{author}{Dawson, J.}, and \bibinfo{author}{Nasrabadi, N.~M.} (\bibinfo{year}{2018}).
\newblock \bibinfo{title}{Multi-level feature abstraction from convolutional neural networks for multimodal biometric identification}.
\newblock \DOIprefix\doi{10.48550/arXiv.1807.01332}.
\bibitem[{Kang et~al.(2021)Kang, Leng, Guo and Pan}]{Kang2021SequencebasedDL}
\bibinfo{author}{Kang, Y.}, \bibinfo{author}{Leng, D.}, \bibinfo{author}{Guo, J.}, and \bibinfo{author}{Pan, L.} (\bibinfo{year}{2021}).
\newblock \bibinfo{title}{Sequence-based deep learning antibody design for in silico antibody affinity maturation}.
\newblock \DOIprefix\doi{10.48550/arXiv.2103.03724}.
\bibitem[{Sirin et~al.(2015)Sirin, Apgar, Bennett and Keating}]{Sirin2015}
\bibinfo{author}{Sirin, S.}, \bibinfo{author}{Apgar, J.}, \bibinfo{author}{Bennett, E.}, and \bibinfo{author}{Keating, A.} (\bibinfo{year}{2015}). \bibinfo{title}{Ab-bind: Antibody binding mutational database for computational affinity predictions}.
\newblock \bibinfo{journal}{Protein Science} \emph{\bibinfo{volume}{25}}. \DOIprefix\doi{10.1002/pro.2829}.
\bibitem[{Kurumida et~al.(2020)Kurumida, Saito and Kameda}]{Kurumida2020}
\bibinfo{author}{Kurumida, Y.}, \bibinfo{author}{Saito, Y.}, and \bibinfo{author}{Kameda, T.} (\bibinfo{year}{2020}). \bibinfo{title}{Predicting antibody affinity changes upon mutations by combining multiple predictors}.
\newblock \bibinfo{journal}{Scientific Reports} \emph{\bibinfo{volume}{10}}. \DOIprefix\doi{10.1038/s41598-020-76369-8}.
\bibitem[{Myung et~al.(2021)Myung, Pires and Ascher}]{Myung2021}
\bibinfo{author}{Myung, Y.}, \bibinfo{author}{Pires, D. E.~V.}, and \bibinfo{author}{Ascher, D.~B.} (\bibinfo{year}{2021}). \bibinfo{title}{{CSM-AB: graph-based antibody–antigen binding affinity prediction and docking scoring function}}.
\newblock \bibinfo{journal}{Bioinformatics} \emph{\bibinfo{volume}{38}}, \bibinfo{pages}{1141--1143}. \DOIprefix\doi{10.1093/bioinformatics/btab762}.
\bibitem[{Yang et~al.(2023)Yang, Wang and Zhu}]{YANG2023108364}
\bibinfo{author}{Yang, Y.~X.}, \bibinfo{author}{Wang, P.}, and \bibinfo{author}{Zhu, B.~T.} (\bibinfo{year}{2023}). \bibinfo{title}{Binding affinity prediction for antibody–protein antigen complexes: A machine learning analysis based on interface and surface areas}.
\newblock \bibinfo{journal}{Journal of Molecular Graphics and Modelling} \emph{\bibinfo{volume}{118}}, \bibinfo{pages}{108364}. \DOIprefix\doi{10.1016/j.jmgm.2022.108364}.
\bibitem[{Durairaj et~al.(2023)Durairaj, {de Ridder} and {van Dijk}}]{Durairaj2023}
\bibinfo{author}{Durairaj, J.}, \bibinfo{author}{{de Ridder}, D.}, and \bibinfo{author}{{van Dijk}, A.~D.} (\bibinfo{year}{2023}). \bibinfo{title}{Beyond sequence: Structure-based machine learning}.
\newblock \bibinfo{journal}{Computational and Structural Biotechnology Journal} \emph{\bibinfo{volume}{21}}, \bibinfo{pages}{630--643}. \DOIprefix\doi{10.1016/j.csbj.2022.12.039}.
\bibitem[{Leem et~al.(2022)Leem, Mitchell, Farmery, Barton and Galson}]{leem_deciphering_2022}
\bibinfo{author}{Leem, J.}, \bibinfo{author}{Mitchell, L.~S.}, \bibinfo{author}{Farmery, J. H.~R.}, \bibinfo{author}{Barton, J.}, and \bibinfo{author}{Galson, J.~D.} (\bibinfo{year}{2022}). \bibinfo{title}{Deciphering the language of antibodies using self-supervised learning}.
\newblock \bibinfo{journal}{Patterns} \emph{\bibinfo{volume}{3}}, \bibinfo{pages}{100513}. \DOIprefix\doi{10.1016/j.patter.2022.100513}.
\bibitem[{D’Angelo et~al.(2018)D’Angelo, Ferrara, Naranjo, Erasmus, Hraber and Bradbury}]{dangelo2018}
\bibinfo{author}{D’Angelo, S.}, \bibinfo{author}{Ferrara, F.}, \bibinfo{author}{Naranjo, L.}, \bibinfo{author}{Erasmus, M.~F.}, \bibinfo{author}{Hraber, P.}, and \bibinfo{author}{Bradbury, A. R.~M.} (\bibinfo{year}{2018}). \bibinfo{title}{Many routes to an antibody heavy-chain cdr3: Necessary, yet insufficient, for specific binding}.
\newblock \bibinfo{journal}{Frontiers in Immunology} \emph{\bibinfo{volume}{9}}. \DOIprefix\doi{10.3389/fimmu.2018.00395}.
\bibitem[{Phillips et~al.(2021)Phillips, Lawrence, Moulana, Dupic, Chang, Johnson, Cvijovic, Mora, Walczak and Desai}]{Phillips2021}
\bibinfo{author}{Phillips, A.~M.}, \bibinfo{author}{Lawrence, K.~R.}, \bibinfo{author}{Moulana, A.}, \bibinfo{author}{Dupic, T.}, \bibinfo{author}{Chang, J.}, \bibinfo{author}{Johnson, M.~S.}, \bibinfo{author}{Cvijovic, I.}, \bibinfo{author}{Mora, T.}, \bibinfo{author}{Walczak, A.~M.}, and \bibinfo{author}{Desai, M.~M.} (\bibinfo{year}{2021}). \bibinfo{title}{Binding affinity landscapes constrain the evolution of broadly neutralizing anti-influenza antibodies}.
\newblock \bibinfo{journal}{eLife} \emph{\bibinfo{volume}{10}}. \DOIprefix\doi{10.7554/eLife.71393}.
\bibitem[{Schmidt et~al.(2013)Schmidt, Xu, Khan, O’Donnell, Khurana, King, Manischewitz, Golding, Suphaphiphat, Carfi, Settembre, Dormitzer, Kepler, Zhang, Moody, Haynes, Liao, Shaw and Harrison}]{Schmidt2013}
\bibinfo{author}{Schmidt, A.~G.}, \bibinfo{author}{Xu, H.}, \bibinfo{author}{Khan, A.~R.}, \bibinfo{author}{O’Donnell, T.}, \bibinfo{author}{Khurana, S.}, \bibinfo{author}{King, L.~R.}, \bibinfo{author}{Manischewitz, J.}, \bibinfo{author}{Golding, H.}, \bibinfo{author}{Suphaphiphat, P.}, \bibinfo{author}{Carfi, A.}, \bibinfo{author}{Settembre, E.~C.}, \bibinfo{author}{Dormitzer, P.~R.}, \bibinfo{author}{Kepler, T.~B.}, \bibinfo{author}{Zhang, R.}, \bibinfo{author}{Moody, M.~A.}, \bibinfo{author}{Haynes, B.~F.}, \bibinfo{author}{Liao, H.-X.}, \bibinfo{author}{Shaw, D.~E.}, and \bibinfo{author}{Harrison, S.~C.} (\bibinfo{year}{2013}). \bibinfo{title}{Preconfiguration of the antigen-binding site during affinity maturation of a broadly neutralizing influenza virus antibody}.
\newblock \bibinfo{journal}{Proceedings of the National Academy of Sciences} \emph{\bibinfo{volume}{110}}, \bibinfo{pages}{264--269}. \DOIprefix\doi{10.1073/pnas.1218256109}.
\bibitem[{Xu et~al.(2014)Xu, Schmidt, O'donnell, Therkelsen, Kepler, Moody, Haynes, Liao, Harrison and Shaw}]{Xu2015}
\bibinfo{author}{Xu, H.}, \bibinfo{author}{Schmidt, A.}, \bibinfo{author}{O'donnell, T.}, \bibinfo{author}{Therkelsen, M.}, \bibinfo{author}{Kepler, T.}, \bibinfo{author}{Moody, M.}, \bibinfo{author}{Haynes, B.}, \bibinfo{author}{Liao, H.}, \bibinfo{author}{Harrison, S.}, and \bibinfo{author}{Shaw, D.} (\bibinfo{year}{2014}). \bibinfo{title}{Key mutations stabilize antigen-binding conformation during affinity maturation of a broadly neutralizing influenza antibody lineage}.
\newblock \bibinfo{journal}{Proteins: Structure, Function, and Bioinformatics} \emph{\bibinfo{volume}{83}}. \DOIprefix\doi{10.1002/prot.24745}.
\bibitem[{Sormanni et~al.(2015)Sormanni, Aprile and Vendruscolo}]{sormanni_rational_2015}
\bibinfo{author}{Sormanni, P.}, \bibinfo{author}{Aprile, F.~A.}, and \bibinfo{author}{Vendruscolo, M.} (\bibinfo{year}{2015}). \bibinfo{title}{Rational design of antibodies targeting specific epitopes within intrinsically disordered proteins}.
\newblock \bibinfo{journal}{Proceedings of the National Academy of Sciences} \emph{\bibinfo{volume}{112}}, \bibinfo{pages}{9902--9907}. \DOIprefix\doi{10.1073/pnas.1422401112}.
\bibitem[{Mishra and Mariuzza(2018)}]{Mishra2018}
\bibinfo{author}{Mishra, A.~K.}, and \bibinfo{author}{Mariuzza, R.~A.} (\bibinfo{year}{2018}). \bibinfo{title}{Insights into the structural basis of antibody affinity maturation from next-generation sequencing}.
\newblock \bibinfo{journal}{Frontiers in Immunology} \emph{\bibinfo{volume}{9}}. \DOIprefix\doi{10.3389/fimmu.2018.00117}.
\bibitem[{Fernández-Quintero et~al.(2020)Fernández-Quintero, Loeffler, Bacher, Waibl, Seidler and Liedl}]{fernandez-quintero_local_2020}
\bibinfo{author}{Fernández-Quintero, M.~L.}, \bibinfo{author}{Loeffler, J.~R.}, \bibinfo{author}{Bacher, L.~M.}, \bibinfo{author}{Waibl, F.}, \bibinfo{author}{Seidler, C.~A.}, and \bibinfo{author}{Liedl, K.~R.} (\bibinfo{year}{2020}). \bibinfo{title}{Local and {Global} {Rigidification} {Upon} {Antibody} {Affinity} {Maturation}}.
\newblock \bibinfo{journal}{Frontiers in Molecular Biosciences} \emph{\bibinfo{volume}{7}}. \DOIprefix\doi{10.3389/fmolb.2020.00182}.
\bibitem[{Laffy et~al.(2017)Laffy, Dodev, Macpherson, Townsend, Lu, Dunn-Walters and Fraternali}]{laffy_proarticleuous_2017}
\bibinfo{author}{Laffy, J. M.~J.}, \bibinfo{author}{Dodev, T.}, \bibinfo{author}{Macpherson, J.~A.}, \bibinfo{author}{Townsend, C.}, \bibinfo{author}{Lu, H.~C.}, \bibinfo{author}{Dunn-Walters, D.}, and \bibinfo{author}{Fraternali, F.} (\bibinfo{year}{2017}). \bibinfo{title}{Proarticleuous antibodies characterised by their physico-chemical properties: {From} sequence to structure and back}.
\newblock \bibinfo{journal}{Progress in Biophysics and Molecular Biology} \emph{\bibinfo{volume}{128}}, \bibinfo{pages}{47--56}. \DOIprefix\doi{10.1016/j.pbiomolbio.2016.09.002}.
\bibitem[{Ovchinnikov et~al.(2018)Ovchinnikov, Louveau, Barton, Karplus and Chakraborty}]{Ovchinnikov2018}
\bibinfo{author}{Ovchinnikov, V.}, \bibinfo{author}{Louveau, J.~E.}, \bibinfo{author}{Barton, J.~P.}, \bibinfo{author}{Karplus, M.}, and \bibinfo{author}{Chakraborty, A.~K.} (\bibinfo{year}{2018}). \bibinfo{title}{Role of framework mutations and antibody flexibility in the evolution of broadly neutralizing antibodies}.
\newblock \bibinfo{journal}{eLife} \emph{\bibinfo{volume}{7}}. \DOIprefix\doi{10.7554/eLife.33038}.
\bibitem[{Phillips et~al.(2023)Phillips, Maurer, Brooks, Dupic, Schmidt and Desai}]{Phillips2023}
\bibinfo{author}{Phillips, A.~M.}, \bibinfo{author}{Maurer, D.~P.}, \bibinfo{author}{Brooks, C.}, \bibinfo{author}{Dupic, T.}, \bibinfo{author}{Schmidt, A.~G.}, and \bibinfo{author}{Desai, M.~M.} (\bibinfo{year}{2023}). \bibinfo{title}{Hierarchical sequence-affinity landscapes shape the evolution of breadth in an anti-influenza receptor binding site antibody}.
\newblock \bibinfo{journal}{eLife} \emph{\bibinfo{volume}{12}}. \DOIprefix\doi{10.7554/eLife.83628}.
\bibitem[{Zimmermann et~al.(2010)Zimmermann, Romesberg, Brooks and Thorpe}]{Zimmermann2010}
\bibinfo{author}{Zimmermann, J.}, \bibinfo{author}{Romesberg, F.~E.}, \bibinfo{author}{Brooks, C. L.~I.}, and \bibinfo{author}{Thorpe, I.~F.} (\bibinfo{year}{2010}). \bibinfo{title}{Molecular description of flexibility in an antibody combining site}.
\newblock \bibinfo{journal}{The Journal of Physical Chemistry B} \emph{\bibinfo{volume}{114}}, \bibinfo{pages}{7359--7370}. \DOIprefix\doi{10.1021/jp906421v}.
\bibitem[{Tomar et~al.(2022)Tomar, Licari, Bauer, Singh, Li and Kumar}]{Tomar2022}
\bibinfo{author}{Tomar, D.~S.}, \bibinfo{author}{Licari, G.}, \bibinfo{author}{Bauer, J.}, \bibinfo{author}{Singh, S.~K.}, \bibinfo{author}{Li, L.}, and \bibinfo{author}{Kumar, S.} (\bibinfo{year}{2022}). \bibinfo{title}{Stress-dependent flexibility of a full-length human monoclonal antibody: Insights from molecular dynamics to support biopharmaceutical development}.
\newblock \bibinfo{journal}{Journal of Pharmaceutical Sciences} \emph{\bibinfo{volume}{111}}, \bibinfo{pages}{628--637}. \DOIprefix\doi{10.1016/j.xphs.2021.10.039}.
\bibitem[{Dykeman and Sankey(2010)}]{Dykeman2010}
\bibinfo{author}{Dykeman, E.~C.}, and \bibinfo{author}{Sankey, O.~F.} (\bibinfo{year}{2010}). \bibinfo{title}{Normal mode analysis and applications in biological physics}.
\newblock \bibinfo{journal}{Journal of Physics: Condensed Matter} \emph{\bibinfo{volume}{22}}. \DOIprefix\doi{10.1088/0953-8984/22/42/423202}.
\bibitem[{Tirion(1996)}]{Tirion1996}
\bibinfo{author}{Tirion, M.~M.} (\bibinfo{year}{1996}). \bibinfo{title}{Large amplitude elastic motions in proteins from a single-parameter, atomic analysis}.
\newblock \bibinfo{journal}{Phys. Rev. Lett.} \emph{\bibinfo{volume}{77}}, \bibinfo{pages}{1905--1908}. \DOIprefix\doi{10.1103/PhysRevLett.77.1905}.
\bibitem[{Bahar et~al.(2010)Bahar, Lezon, Yang and Eyal}]{bahar_global_2010}
\bibinfo{author}{Bahar, I.}, \bibinfo{author}{Lezon, T.~R.}, \bibinfo{author}{Yang, L.-W.}, and \bibinfo{author}{Eyal, E.} (\bibinfo{year}{2010}). \bibinfo{title}{Global {Dynamics} of {Proteins}: {Bridging} {Between} {Structure} and {Function}}.
\newblock \bibinfo{journal}{Annual Review of Biophysics} \emph{\bibinfo{volume}{39}}, \bibinfo{pages}{23--42}. \DOIprefix\doi{10.1146/annurev.biophys.093008.131258}.
\bibitem[{Yang et~al.(2009)Yang, Song and Jernigan}]{Yang2009}
\bibinfo{author}{Yang, L.}, \bibinfo{author}{Song, G.}, and \bibinfo{author}{Jernigan, R.~L.} (\bibinfo{year}{2009}). \bibinfo{title}{Protein elastic network models and the ranges of cooperativity}.
\newblock \bibinfo{journal}{Proceedings of the National Academy of Sciences} \emph{\bibinfo{volume}{106}}, \bibinfo{pages}{12347--12352}. \DOIprefix\doi{10.1073/pnas.0902159106}.
\bibitem[{Dobbins et~al.(2008)Dobbins, Lesk and Sternberg}]{Dobbins2008}
\bibinfo{author}{Dobbins, S.~E.}, \bibinfo{author}{Lesk, V.~I.}, and \bibinfo{author}{Sternberg, M. J.~E.} (\bibinfo{year}{2008}). \bibinfo{title}{Insights into protein flexibility: The relationship between normal modes and conformational change upon protein–protein docking}.
\newblock \bibinfo{journal}{Proceedings of the National Academy of Sciences} \emph{\bibinfo{volume}{105}}, \bibinfo{pages}{10390--10395}. \DOIprefix\doi{10.1073/pnas.0802496105}.
\bibitem[{Skjærven et~al.(2014)Skjærven, Yao, Scarabelli and Grant}]{nmaBio3D}
\bibinfo{author}{Skjærven, L.}, \bibinfo{author}{Yao, X.-Q.}, \bibinfo{author}{Scarabelli, G.}, and \bibinfo{author}{Grant, B.~J.} (\bibinfo{year}{2014}). \bibinfo{title}{Integrating protein structural dynamics and evolutionary analysis with bio3d}.
\newblock \bibinfo{journal}{BMC bioinformatics} \emph{\bibinfo{volume}{15}}, \bibinfo{pages}{399}. \DOIprefix\doi{10.1186/s12859-014-0399-6}.
\bibitem[{Lopéz-Blanco et~al.(2011)Lopéz-Blanco, Garzón and Chacón}]{LopezBlanco2011}
\bibinfo{author}{Lopéz-Blanco, J.~R.}, \bibinfo{author}{Garzón, J.~I.}, and \bibinfo{author}{Chacón, P.} (\bibinfo{year}{2011}). \bibinfo{title}{{iMod: multipurpose normal mode analysis in internal coordinates}}.
\newblock \bibinfo{journal}{Bioinformatics} \emph{\bibinfo{volume}{27}}, \bibinfo{pages}{2843--2850}. \DOIprefix\doi{10.1093/bioinformatics/btr497}.
\bibitem[{Bakan et~al.(2011)Bakan, Meireles and Bahar}]{Bakan2011}
\bibinfo{author}{Bakan, A.}, \bibinfo{author}{Meireles, L.~M.}, and \bibinfo{author}{Bahar, I.} (\bibinfo{year}{2011}). \bibinfo{title}{{ProDy: Protein Dynamics Inferred from Theory and Experiments}}.
\newblock \bibinfo{journal}{Bioinformatics} \emph{\bibinfo{volume}{27}}, \bibinfo{pages}{1575--1577}. \DOIprefix\doi{10.1093/bioinformatics/btr168}.
\bibitem[{Madsen et~al.(2024)Madsen, Mejias-Gomez, Pedersen, {Preben Morth}, Kristensen, Jenkins and Goletz}]{Madsen2024}
\bibinfo{author}{Madsen, A.~V.}, \bibinfo{author}{Mejias-Gomez, O.}, \bibinfo{author}{Pedersen, L.~E.}, \bibinfo{author}{{Preben Morth}, J.}, \bibinfo{author}{Kristensen, P.}, \bibinfo{author}{Jenkins, T.~P.}, and \bibinfo{author}{Goletz, S.} (\bibinfo{year}{2024}). \bibinfo{title}{Structural trends in antibody-antigen binding interfaces: a computational analysis of 1833 experimentally determined 3d structures}.
\newblock \bibinfo{journal}{Computational and Structural Biotechnology Journal} \emph{\bibinfo{volume}{23}}, \bibinfo{pages}{199--211}. \DOIprefix\doi{10.1016/j.csbj.2023.11.056}.
\bibitem[{Chothia and Lesk(1987)}]{Chothia1987}
\bibinfo{author}{Chothia, C.}, and \bibinfo{author}{Lesk, A.~M.} (\bibinfo{year}{1987}). \bibinfo{title}{Canonical structures for the hypervariable regions of immunoglobulins}.
\newblock \bibinfo{journal}{Journal of Molecular Biology} \emph{\bibinfo{volume}{196}}, \bibinfo{pages}{901--917}. \DOIprefix\doi{10.1016/0022-2836(87)90412-8}.
\bibitem[{Baldi and Pollastri(2003)}]{Baldi2003}
\bibinfo{author}{Baldi, P.}, and \bibinfo{author}{Pollastri, G.} (\bibinfo{year}{2003}). \bibinfo{title}{The principled design of large-scale recursive neural network architectures--dag-rnns and the protein structure prediction problem}.
\newblock \bibinfo{journal}{Journal of Machine Learning Research} \emph{\bibinfo{volume}{4}}. \DOIprefix\doi{10.1162/153244304773936054}.
\bibitem[{McInnes et~al.(2018)McInnes, Healy, Saul and Großberger}]{McInnes2018}
\bibinfo{author}{McInnes, L.}, \bibinfo{author}{Healy, J.}, \bibinfo{author}{Saul, N.}, and \bibinfo{author}{Großberger, L.} (\bibinfo{year}{2018}). \bibinfo{title}{Umap: Uniform manifold approximation and projection}.
\newblock \bibinfo{journal}{Journal of Open Source Software} \emph{\bibinfo{volume}{3}}, \bibinfo{pages}{861}. \DOIprefix\doi{10.21105/joss.00861}.
\bibitem[{Koenig et~al.(2017)Koenig, Lee, Walters, Janakiraman, Stinson, Patapoff and Fuh}]{Koenig2017}
\bibinfo{author}{Koenig, P.}, \bibinfo{author}{Lee, C.~V.}, \bibinfo{author}{Walters, B.~T.}, \bibinfo{author}{Janakiraman, V.}, \bibinfo{author}{Stinson, J.}, \bibinfo{author}{Patapoff, T.~W.}, and \bibinfo{author}{Fuh, G.} (\bibinfo{year}{2017}). \bibinfo{title}{Mutational landscape of antibody variable domains reveals a switch modulating the interdomain conformational dynamics and antigen binding}.
\newblock \bibinfo{journal}{Proceedings of the National Academy of Sciences} \emph{\bibinfo{volume}{114}}, \bibinfo{pages}{E486--E495}. \DOIprefix\doi{10.1073/pnas.1613231114}.
\bibitem[{Schiele et~al.(2015)Schiele, van Ryn, Litzenburger, Ritter, Seeliger and Nar}]{Schiele2015}
\bibinfo{author}{Schiele, F.}, \bibinfo{author}{van Ryn, J.}, \bibinfo{author}{Litzenburger, T.}, \bibinfo{author}{Ritter, M.}, \bibinfo{author}{Seeliger, D.}, and \bibinfo{author}{Nar, H.} (\bibinfo{year}{2015}). \bibinfo{title}{Structure-guided residence time optimization of a dabigatran reversal agent}.
\newblock \bibinfo{journal}{mAbs} \emph{\bibinfo{volume}{7}}, \bibinfo{pages}{871--880}. \DOIprefix\doi{10.1080/19420862.2015.1057364}.
\bibitem[{Monzon et~al.(2022)Monzon, Paysan-Lafosse, Wood and Bateman}]{Monzon2022}
\bibinfo{author}{Monzon, V.}, \bibinfo{author}{Paysan-Lafosse, T.}, \bibinfo{author}{Wood, V.}, and \bibinfo{author}{Bateman, A.} (\bibinfo{year}{2022}). \bibinfo{title}{{Reciprocal best structure hits: using AlphaFold models to discover distant homologues}}.
\newblock \bibinfo{journal}{Bioinformatics Advances} \emph{\bibinfo{volume}{2}}. \DOIprefix\doi{10.1093/bioadv/vbac072}.
\bibitem[{Wang et~al.(2022)Wang, Lin, Huang, He, Deng, Xu, Pei and Lai}]{Wang2022}
\bibinfo{author}{Wang, S.}, \bibinfo{author}{Lin, H.}, \bibinfo{author}{Huang, Z.}, \bibinfo{author}{He, Y.}, \bibinfo{author}{Deng, X.}, \bibinfo{author}{Xu, Y.}, \bibinfo{author}{Pei, J.}, and \bibinfo{author}{Lai, L.} (\bibinfo{year}{2022}). \bibinfo{title}{Cavityspace: A database of potential ligand binding sites in the human proteome}.
\newblock \bibinfo{journal}{Biomolecules} \emph{\bibinfo{volume}{12}}. \DOIprefix\doi{10.3390/biom12070967}.
\bibitem[{Jumper et~al.(2021)Jumper, Evans, Pritzel, Green, Figurnov, Ronneberger, Tunyasuvunakool, Bates, Žídek, Potapenko, Bridgland, Meyer, Kohl, Ballard, Cowie, Romera-Paredes, Nikolov, Jain, Adler and Hassabis}]{Jumper2021}
\bibinfo{author}{Jumper, J.}, \bibinfo{author}{Evans, R.}, \bibinfo{author}{Pritzel, A.}, \bibinfo{author}{Green, T.}, \bibinfo{author}{Figurnov, M.}, \bibinfo{author}{Ronneberger, O.}, \bibinfo{author}{Tunyasuvunakool, K.}, \bibinfo{author}{Bates, R.}, \bibinfo{author}{Žídek, A.}, \bibinfo{author}{Potapenko, A.}, \bibinfo{author}{Bridgland, A.}, \bibinfo{author}{Meyer, C.}, \bibinfo{author}{Kohl, S.}, \bibinfo{author}{Ballard, A.}, \bibinfo{author}{Cowie, A.}, \bibinfo{author}{Romera-Paredes, B.}, \bibinfo{author}{Nikolov, S.}, \bibinfo{author}{Jain, R.}, \bibinfo{author}{Adler, J.}, and \bibinfo{author}{Hassabis, D.} (\bibinfo{year}{2021}). \bibinfo{title}{Highly accurate protein structure prediction with alphafold}.
\newblock \bibinfo{journal}{Nature} \emph{\bibinfo{volume}{596}}, \bibinfo{pages}{1--11}. \DOIprefix\doi{10.1038/s41586-021-03819-2}.
\bibitem[{Evans et~al.(2022)Evans, O{\textquoteright}Neill, Pritzel, Antropova, Senior, Green, {\v Z}{\'\i}dek, Bates, Blackwell, Yim, Ronneberger, Bodenstein, Zielinski, Bridgland, Potapenko, Cowie, Tunyasuvunakool, Jain, Clancy, Kohli, Jumper and Hassabis}]{Evans2022}
\bibinfo{author}{Evans, R.}, \bibinfo{author}{O{\textquoteright}Neill, M.}, \bibinfo{author}{Pritzel, A.}, \bibinfo{author}{Antropova, N.}, \bibinfo{author}{Senior, A.}, \bibinfo{author}{Green, T.}, \bibinfo{author}{{\v Z}{\'\i}dek, A.}, \bibinfo{author}{Bates, R.}, \bibinfo{author}{Blackwell, S.}, \bibinfo{author}{Yim, J.}, \bibinfo{author}{Ronneberger, O.}, \bibinfo{author}{Bodenstein, S.}, \bibinfo{author}{Zielinski, M.}, \bibinfo{author}{Bridgland, A.}, \bibinfo{author}{Potapenko, A.}, \bibinfo{author}{Cowie, A.}, \bibinfo{author}{Tunyasuvunakool, K.}, \bibinfo{author}{Jain, R.}, \bibinfo{author}{Clancy, E.}, \bibinfo{author}{Kohli, P.}, \bibinfo{author}{Jumper, J.}, and \bibinfo{author}{Hassabis, D.} (\bibinfo{year}{2022}). \bibinfo{title}{Protein complex prediction with alphafold-multimer}.
\newblock \bibinfo{journal}{bioRxiv}. \DOIprefix\doi{10.1101/2021.10.04.463034}.
\bibitem[{Jr. et~al.(2001)Jr., Travers, Walport and Shlomchik}]{Janeway2001}
\bibinfo{author}{Jr., C. A.~J.}, \bibinfo{author}{Travers, P.}, \bibinfo{author}{Walport, M.}, and \bibinfo{author}{Shlomchik, M.~J.}
\newblock \bibinfo{title}{Immunobiology, 5th edition: The Immune System in Health and Disease}.
\newblock \bibinfo{publisher}{Garland Science} (\bibinfo{year}{2001}).
\bibitem[{Igawa et~al.(2011)Igawa, Tsunoda, Kuramochi, Sampei, Ishii and Hattori}]{Igawa2011}
\bibinfo{author}{Igawa, T.}, \bibinfo{author}{Tsunoda, H.}, \bibinfo{author}{Kuramochi, T.}, \bibinfo{author}{Sampei, Z.}, \bibinfo{author}{Ishii, S.}, and \bibinfo{author}{Hattori, K.} (\bibinfo{year}{2011}). \bibinfo{title}{Engineering the variable region of therapeutic igg antibodies}.
\newblock \bibinfo{journal}{mAbs} \emph{\bibinfo{volume}{3}}, \bibinfo{pages}{243--252}. \DOIprefix\doi{10.4161/mabs.3.3.15234}.
\bibitem[{Rader et~al.(2005)Rader, Chennubhotla, Yang and Bahar}]{Rader2005}
\bibinfo{author}{Rader, A.}, \bibinfo{author}{Chennubhotla, C.}, \bibinfo{author}{Yang, L.-W.}, and \bibinfo{author}{Bahar, I.} (\bibinfo{year}{2005}).
\newblock \bibinfo{title}{The Gaussian network model: Theory and applications}.
\newblock \bibinfo{publisher}{Chapman and Hall/CRC}, pp.  \bibinfo{pages}{41--64}.
\bibitem[{Amor et~al.(2016)Amor, Schaub, Yaliraki and Barahona}]{amor_prediction_2016}
\bibinfo{author}{Amor, B. R.~C.}, \bibinfo{author}{Schaub, M.~T.}, \bibinfo{author}{Yaliraki, S.~N.}, and \bibinfo{author}{Barahona, M.} (\bibinfo{year}{2016}). \bibinfo{title}{Prediction of allosteric sites and mediating interactions through bond-to-bond propensities}.
\newblock \bibinfo{journal}{Nature Communications} \emph{\bibinfo{volume}{7}}, \bibinfo{pages}{12477}. \DOIprefix\doi{10.1038/ncomms12477}.
\bibitem[{Tang and Kaneko(2020)}]{Tang2020}
\bibinfo{author}{Tang, Q.-Y.}, and \bibinfo{author}{Kaneko, K.} (\bibinfo{year}{2020}). \bibinfo{title}{Long-range correlation in protein dynamics: Confirmation by structural data and normal mode analysis}.
\newblock \bibinfo{journal}{PLOS Computational Biology} \emph{\bibinfo{volume}{16}}, \bibinfo{pages}{1--17}. \DOIprefix\doi{10.1371/journal.pcbi.1007670}.
\bibitem[{Barozet et~al.(2018)Barozet, Bianciotto, Siméon, Minoux and Cortés}]{Barozet2018}
\bibinfo{author}{Barozet, A.}, \bibinfo{author}{Bianciotto, M.}, \bibinfo{author}{Siméon, T.}, \bibinfo{author}{Minoux, H.}, and \bibinfo{author}{Cortés, J.} (\bibinfo{year}{2018}). \bibinfo{title}{Conformational changes in antibody fab fragments upon binding and their consequences on the performance of docking algorithms}.
\newblock \bibinfo{journal}{Immunology Letters} \emph{\bibinfo{volume}{200}}, \bibinfo{pages}{5--15}. \DOIprefix\doi{10.1016/j.imlet.2018.06.002}.
\bibitem[{Liu et~al.(2024)Liu, Denzler, Hood and Martin}]{Liu2024}
\bibinfo{author}{Liu, C.}, \bibinfo{author}{Denzler, L.~M.}, \bibinfo{author}{Hood, O.~E.}, and \bibinfo{author}{Martin, A.~C.} (\bibinfo{year}{2024}). \bibinfo{title}{Do antibody cdr loops change conformation upon binding?}
\newblock \bibinfo{journal}{mAbs} \emph{\bibinfo{volume}{16}}, \bibinfo{pages}{2322533}. \DOIprefix\doi{10.1080/19420862.2024.2322533}.
\bibitem[{Peach et~al.(2022)Peach, Saman, Yaliraki, Klug, Ying, Willison and Barahona}]{peach_unsupervised_2019}
\bibinfo{author}{Peach, R.~L.}, \bibinfo{author}{Saman, D.}, \bibinfo{author}{Yaliraki, S.~N.}, \bibinfo{author}{Klug, D.~R.}, \bibinfo{author}{Ying, L.}, \bibinfo{author}{Willison, K.~R.}, and \bibinfo{author}{Barahona, M.} (\bibinfo{year}{2022}).
\newblock \bibinfo{title}{Unsupervised {Graph}-{Based} {Learning} {Predicts} {Mutations} {That} {Alter} {Protein} {Dynamics}}.
\newblock \DOIprefix\doi{10.1101/847426}.
\bibitem[{Wu et~al.(2022)Wu, Str{\"o}mich and Yaliraki}]{Wu2022}
\bibinfo{author}{Wu, N.}, \bibinfo{author}{Str{\"o}mich, L.}, and \bibinfo{author}{Yaliraki, S.~N.} (\bibinfo{year}{2022}). \bibinfo{title}{Prediction of allosteric sites and signaling: Insights from benchmarking datasets}.
\newblock \bibinfo{journal}{Patterns} \emph{\bibinfo{volume}{3}}. \DOIprefix\doi{10.1016/j.patter.2021.100408}.
\bibitem[{Tubiana et~al.(2022)Tubiana, Schneidman-Duhovny and Wolfson}]{tubiana_scannet_2022}
\bibinfo{author}{Tubiana, J.}, \bibinfo{author}{Schneidman-Duhovny, D.}, and \bibinfo{author}{Wolfson, H.~J.} (\bibinfo{year}{2022}). \bibinfo{title}{{ScanNet}: an interpretable geometric deep learning model for structure-based protein binding site prediction}.
\newblock \bibinfo{journal}{Nature Methods} \emph{\bibinfo{volume}{19}}, \bibinfo{pages}{730--739}. \DOIprefix\doi{10.1038/s41592-022-01490-7}.
\bibitem[{Pak et~al.(2023)Pak, Markhieva, Novikova, Petrov, Vorobyev, Maksimova, Kondrashov and Ivankov}]{Pak2023}
\bibinfo{author}{Pak, M.~A.}, \bibinfo{author}{Markhieva, K.~A.}, \bibinfo{author}{Novikova, M.~S.}, \bibinfo{author}{Petrov, D.~S.}, \bibinfo{author}{Vorobyev, I.~S.}, \bibinfo{author}{Maksimova, E.~S.}, \bibinfo{author}{Kondrashov, F.~A.}, and \bibinfo{author}{Ivankov, D.~N.} (\bibinfo{year}{2023}). \bibinfo{title}{Using alphafold to predict the impact of single mutations on protein stability and function}.
\newblock \bibinfo{journal}{PLOS ONE} \emph{\bibinfo{volume}{18}}, \bibinfo{pages}{1--9}. \DOIprefix\doi{10.1371/journal.pone.0282689}.
\bibitem[{Li et~al.(2020)Li, Zhou, Dvornek, Gu, Ventola and Duncan}]{LiShapley2020}
\bibinfo{author}{Li, X.}, \bibinfo{author}{Zhou, Y.}, \bibinfo{author}{Dvornek, N.~C.}, \bibinfo{author}{Gu, Y.}, \bibinfo{author}{Ventola, P.}, and \bibinfo{author}{Duncan, J.~S.} (\bibinfo{year}{2020}).
\newblock \bibinfo{title}{Efficient Shapley Explanation for Features Importance Estimation Under Uncertainty}.
\newblock \bibinfo{publisher}{Springer International Publishing}, pp.  \bibinfo{pages}{792--801}.
\bibitem[{Li et~al.(2021)Li, Iyer, Prasath, Ni and Salomonis}]{Li2021}
\bibinfo{author}{Li, G.}, \bibinfo{author}{Iyer, B.}, \bibinfo{author}{Prasath, V. B.~S.}, \bibinfo{author}{Ni, Y.}, and \bibinfo{author}{Salomonis, N.} (\bibinfo{year}{2021}). \bibinfo{title}{{DeepImmuno: deep learning-empowered prediction and generation of immunogenic peptides for T-cell immunity}}.
\newblock \bibinfo{journal}{Briefings in Bioinformatics} \emph{\bibinfo{volume}{22}}. \DOIprefix\doi{10.1093/bib/bbab160}.
\bibitem[{Adams et~al.(2019)Adams, Kinney, Walczak and Mora}]{Adams2019}
\bibinfo{author}{Adams, R.~M.}, \bibinfo{author}{Kinney, J.~B.}, \bibinfo{author}{Walczak, A.~M.}, and \bibinfo{author}{Mora, T.} (\bibinfo{year}{2019}). \bibinfo{title}{Epistasis in a {{Fitness Landscape Defined}} by {{Antibody-Antigen Binding Free Energy}}}.
\newblock \bibinfo{journal}{Cell Systems} \emph{\bibinfo{volume}{8}}, \bibinfo{pages}{86--93}. \DOIprefix\doi{10.1016/j.cels.2018.12.004}.
\bibitem[{Dondelinger et~al.(2018)Dondelinger, Filée, Sauvage, Quinting, Muyldermans, Galleni and Vandevenne}]{Dondelinger2018}
\bibinfo{author}{Dondelinger, M.}, \bibinfo{author}{Filée, P.}, \bibinfo{author}{Sauvage, E.}, \bibinfo{author}{Quinting, B.}, \bibinfo{author}{Muyldermans, S.}, \bibinfo{author}{Galleni, M.}, and \bibinfo{author}{Vandevenne, M.~S.} (\bibinfo{year}{2018}). \bibinfo{title}{Understanding the significance and implications of antibody numbering and antigen-binding surface/residue definition}.
\newblock \bibinfo{journal}{Frontiers in Immunology} \emph{\bibinfo{volume}{9}}. \DOIprefix\doi{10.3389/fimmu.2018.02278}.
\bibitem[{Ruffolo et~al.(2022)Ruffolo, Sulam and Gray}]{ruffolo_antibody_2022}
\bibinfo{author}{Ruffolo, J.~A.}, \bibinfo{author}{Sulam, J.}, and \bibinfo{author}{Gray, J.~J.} (\bibinfo{year}{2022}). \bibinfo{title}{Antibody structure prediction using interpretable deep learning}.
\newblock \bibinfo{journal}{Patterns} \emph{\bibinfo{volume}{3}}, \bibinfo{pages}{100406}. \DOIprefix\doi{10.1016/j.patter.2021.100406}.
\bibitem[{Bauer et~al.(2019)Bauer, Pavlović and Bauerová-Hlinková}]{nmaBauer}
\bibinfo{author}{Bauer, J.~A.}, \bibinfo{author}{Pavlović, J.}, and \bibinfo{author}{Bauerová-Hlinková, V.} (\bibinfo{year}{2019}). \bibinfo{title}{Normal mode analysis as a routine part of a structural investigation}.
\newblock \bibinfo{journal}{Molecules} \emph{\bibinfo{volume}{24}}. \DOIprefix\doi{10.3390/molecules24183293}.
\bibitem[{Tama and Sanejouand(2001)}]{Tama2001}
\bibinfo{author}{Tama, F.}, and \bibinfo{author}{Sanejouand, Y.-H.} (\bibinfo{year}{2001}). \bibinfo{title}{{Conformational change of proteins arising from normal mode calculations}}.
\newblock \bibinfo{journal}{Protein Engineering, Design and Selection} \emph{\bibinfo{volume}{14}}, \bibinfo{pages}{1--6}. \DOIprefix\doi{10.1093/protein/14.1.1}.
\bibitem[{Dubanevics and McLeish(2022)}]{Dubanevics2022}
\bibinfo{author}{Dubanevics, I.}, and \bibinfo{author}{McLeish, T.~C.} (\bibinfo{year}{2022}). \bibinfo{title}{Optimising elastic network models for protein dynamics and allostery: Spatial and modal cut-offs and backbone stiffness}.
\newblock \bibinfo{journal}{Journal of Molecular Biology} \emph{\bibinfo{volume}{434}}, \bibinfo{pages}{167696}. \DOIprefix\doi{10.1016/j.jmb.2022.167696}.
\bibitem[{{Van Wynsberghe} and Cui(2006)}]{VanWynsberghe2006}
\bibinfo{author}{{Van Wynsberghe}, A.~W.}, and \bibinfo{author}{Cui, Q.} (\bibinfo{year}{2006}). \bibinfo{title}{Interpreting correlated motions using normal mode analysis}.
\newblock \bibinfo{journal}{Structure} \emph{\bibinfo{volume}{14}}, \bibinfo{pages}{1647--1653}. \DOIprefix\doi{10.1016/j.str.2006.09.003}.
\bibitem[{Hinsen(2005)}]{Hinsen2005}
\bibinfo{author}{Hinsen, K.}
\newblock \bibinfo{title}{{Normal mode theory and harmonic potential approximations}} (\bibinfo{year}{2005}).
\newblock In: \emph{\bibinfo{booktitle}{{Normal Mode Analysis}}}. Normal Mode Analysis: Theory and Applications to Biological and Chemical Systems \bibinfo{publisher}{Chapman \& Hall / CRC}, pp.  \bibinfo{pages}{1--16}.
\bibitem[{Cornell et~al.(1995)Cornell, Cieplak, Bayly, Gould, Merz, Ferguson, Spellmeyer, Fox, Caldwell and Kollman}]{Cornell1995}
\bibinfo{author}{Cornell, W.~D.}, \bibinfo{author}{Cieplak, P.}, \bibinfo{author}{Bayly, C.~I.}, \bibinfo{author}{Gould, I.~R.}, \bibinfo{author}{Merz, K.~M.}, \bibinfo{author}{Ferguson, D.~M.}, \bibinfo{author}{Spellmeyer, D.~C.}, \bibinfo{author}{Fox, T.}, \bibinfo{author}{Caldwell, J.~W.}, and \bibinfo{author}{Kollman, P.~A.} (\bibinfo{year}{1995}). \bibinfo{title}{A second generation force field for the simulation of proteins, nucleic acids, and organic molecules}.
\newblock \bibinfo{journal}{Journal of the American Chemical Society} \emph{\bibinfo{volume}{117}}, \bibinfo{pages}{5179--5197}. \DOIprefix\doi{10.1021/ja00124a002}.
\bibitem[{Paszke et~al.(2019)Paszke, Gross, Massa, Lerer, Bradbury, Chanan, Killeen, Lin, Gimelshein, Antiga, Desmaison, Kopf, Yang, DeVito, Raison, Tejani, Chilamkurthy, Steiner, Fang, Bai and Chintala}]{PyTorchPaper}
\bibinfo{author}{Paszke, A.}, \bibinfo{author}{Gross, S.}, \bibinfo{author}{Massa, F.}, \bibinfo{author}{Lerer, A.}, \bibinfo{author}{Bradbury, J.}, \bibinfo{author}{Chanan, G.}, \bibinfo{author}{Killeen, T.}, \bibinfo{author}{Lin, Z.}, \bibinfo{author}{Gimelshein, N.}, \bibinfo{author}{Antiga, L.}, \bibinfo{author}{Desmaison, A.}, \bibinfo{author}{Kopf, A.}, \bibinfo{author}{Yang, E.}, \bibinfo{author}{DeVito, Z.}, \bibinfo{author}{Raison, M.}, \bibinfo{author}{Tejani, A.}, \bibinfo{author}{Chilamkurthy, S.}, \bibinfo{author}{Steiner, B.}, \bibinfo{author}{Fang, L.}, \bibinfo{author}{Bai, J.}, and \bibinfo{author}{Chintala, S.}
\newblock \bibinfo{title}{Pytorch: An imperative style, high-performance deep learning library} (\bibinfo{year}{2019}).
\newblock In: \emph{\bibinfo{booktitle}{Advances in Neural Information Processing Systems 32}}. \bibinfo{publisher}{Curran Associates, Inc.}, pp.  \bibinfo{pages}{8024--8035}.
\bibitem[{Davila et~al.(2022)Davila, Xu, Li, Rozewicki, Wilamowski, Kotelnikov, Kozakov, Teraguchi and Standley}]{Davila2022}
\bibinfo{author}{Davila, A.}, \bibinfo{author}{Xu, Z.}, \bibinfo{author}{Li, S.}, \bibinfo{author}{Rozewicki, J.}, \bibinfo{author}{Wilamowski, J.}, \bibinfo{author}{Kotelnikov, S.}, \bibinfo{author}{Kozakov, D.}, \bibinfo{author}{Teraguchi, S.}, and \bibinfo{author}{Standley, D.~M.} (\bibinfo{year}{2022}). \bibinfo{title}{Abadapt: an adaptive approach to predicting antibody–antigen complex structures from sequence}.
\newblock \bibinfo{journal}{Bioinformatics Advances} \emph{\bibinfo{volume}{2}}. \DOIprefix\doi{10.1093/bioadv/vbac015}.
\bibitem[{Zhuang et~al.(2020)Zhuang, Tang, Ding, Tatikonda, Dvornek, Papademetris and Duncan}]{AdaBelief}
\bibinfo{author}{Zhuang, J.}, \bibinfo{author}{Tang, T.}, \bibinfo{author}{Ding, Y.}, \bibinfo{author}{Tatikonda, S.}, \bibinfo{author}{Dvornek, N.}, \bibinfo{author}{Papademetris, X.}, and \bibinfo{author}{Duncan, J.~S.} (\bibinfo{year}{2020}).
\newblock \bibinfo{title}{Adabelief optimizer: Adapting stepsizes by the belief in observed gradients}.
\newblock \DOIprefix\doi{10.48550/arxiv.2010.07468}.
\bibitem[{Zhuang et~al.(2021)Zhuang, Ding, Tang, Dvornek, Tatikonda and Duncan}]{zhuang2021acprop}
\bibinfo{author}{Zhuang, J.}, \bibinfo{author}{Ding, Y.}, \bibinfo{author}{Tang, T.}, \bibinfo{author}{Dvornek, N.}, \bibinfo{author}{Tatikonda, S.}, and \bibinfo{author}{Duncan, J.} (\bibinfo{year}{2021}). \bibinfo{title}{Momentum centering and asynchronous update for adaptive gradient methods}.
\newblock \bibinfo{journal}{Conference on Neural Information Processing Systems}.
\bibitem[{Ma et~al.(2019)Ma, Miao, Niu and Zhang}]{Ma2019}
\bibinfo{author}{Ma, R.}, \bibinfo{author}{Miao, J.}, \bibinfo{author}{Niu, L.}, and \bibinfo{author}{Zhang, P.} (\bibinfo{year}{2019}). \bibinfo{title}{Transformed $\ell_1$ regularization for learning sparse deep neural networks}.
\newblock \bibinfo{journal}{Neural Networks} \emph{\bibinfo{volume}{119}}, \bibinfo{pages}{286--298}. \DOIprefix\doi{10.1016/j.neunet.2019.08.015}.
\bibitem[{Akiba et~al.(2019)Akiba, Sano, Yanase, Ohta and Koyama}]{Akiba2019Optuna}
\bibinfo{author}{Akiba, T.}, \bibinfo{author}{Sano, S.}, \bibinfo{author}{Yanase, T.}, \bibinfo{author}{Ohta, T.}, and \bibinfo{author}{Koyama, M.} (\bibinfo{year}{2019}).
\newblock \bibinfo{title}{Optuna: A next-generation hyperparameter optimization framework}.
\newblock \DOIprefix\doi{10.48550/arXiv.1907.10902}.
\bibitem[{{Jasmin G{\"u}ven} et~al.(2023){Jasmin G{\"u}ven}, Molkenthin, M{\"u}hle and Mey}]{Guven2023}
\bibinfo{author}{{Jasmin G{\"u}ven}, J.}, \bibinfo{author}{Molkenthin, N.}, \bibinfo{author}{M{\"u}hle, S.}, and \bibinfo{author}{Mey, A.} (\bibinfo{year}{2023}). \bibinfo{title}{What geometrically constrained models can tell us about real-world protein contact maps}.
\newblock \bibinfo{journal}{Physical Biology} \emph{\bibinfo{volume}{20}}. \DOIprefix\doi{10.1088/1478-3975/acd543}.
\bibitem[{Mirdita et~al.(2022)Mirdita, Schütze, Moriwaki, Heo, Ovchinnikov and Steinegger}]{Mirdita2022}
\bibinfo{author}{Mirdita, M.}, \bibinfo{author}{Schütze, K.}, \bibinfo{author}{Moriwaki, Y.}, \bibinfo{author}{Heo, L.}, \bibinfo{author}{Ovchinnikov, S.}, and \bibinfo{author}{Steinegger, M.} (\bibinfo{year}{2022}). \bibinfo{title}{Colabfold: making protein folding accessible to all}.
\newblock \bibinfo{journal}{Nature Methods} \emph{\bibinfo{volume}{19}}, \bibinfo{pages}{679–682}. \DOIprefix\doi{10.1038/s41592-022-01488-1}.
\bibitem[{Steinegger and Söding(2017)}]{Steinegger2017}
\bibinfo{author}{Steinegger, M.}, and \bibinfo{author}{Söding, J.} (\bibinfo{year}{2017}). \bibinfo{title}{Mmseqs2 enables sensitive protein sequence searching for the analysis of massive data sets}.
\newblock \bibinfo{journal}{Nature Biotechnology} \emph{\bibinfo{volume}{35}}, \bibinfo{pages}{1026--1028}. \DOIprefix\doi{10.1038/nbt.3988}.
\bibitem[{Ruffolo et~al.(2023)Ruffolo, Chu, Mahajan and Gray}]{Ruffolo2023}
\bibinfo{author}{Ruffolo, J.}, \bibinfo{author}{Chu, L.-S.}, \bibinfo{author}{Mahajan, S.}, and \bibinfo{author}{Gray, J.} (\bibinfo{year}{2023}). \bibinfo{title}{Fast, accurate antibody structure prediction from deep learning on massive set of natural antibodies}.
\newblock \bibinfo{journal}{Nature Communications} \emph{\bibinfo{volume}{14}}, \bibinfo{pages}{2389}. \DOIprefix\doi{10.1038/s41467-023-38063-x}.
\bibitem[{Simmons et~al.(2023)Simmons, Watanabe, Oguin~III, Van~Itallie, Wiehe, Sempowski, Kuraoka, Kelsoe and McCarthy}]{7TRH}
\bibinfo{author}{Simmons, H.~C.}, \bibinfo{author}{Watanabe, A.}, \bibinfo{author}{Oguin~III, T.~H.}, \bibinfo{author}{Van~Itallie, E.~S.}, \bibinfo{author}{Wiehe, K.~J.}, \bibinfo{author}{Sempowski, G.~D.}, \bibinfo{author}{Kuraoka, M.}, \bibinfo{author}{Kelsoe, G.}, and \bibinfo{author}{McCarthy, K.~R.} (\bibinfo{year}{2023}). \bibinfo{title}{A new class of antibodies that overcomes a steric barrier to cross-group neutralization of influenza viruses}.
\newblock \bibinfo{journal}{PLOS Biology} \emph{\bibinfo{volume}{21}}, \bibinfo{pages}{1--19}. \DOIprefix\doi{10.1371/journal.pbio.3002415}.
\bibitem[{Rao et~al.(2023)Rao, Zhao, Tong, Guo, Peng, Liu, Li, Wu, Tong, Chai, Han, Wang, Jia, Li, Zhao, Li, Zhang, Zhang, Zou, Li, Wang, Gao, Wu, Dai and Gao}]{7Y3O}
\bibinfo{author}{Rao, X.}, \bibinfo{author}{Zhao, R.}, \bibinfo{author}{Tong, Z.}, \bibinfo{author}{Guo, S.}, \bibinfo{author}{Peng, W.}, \bibinfo{author}{Liu, K.}, \bibinfo{author}{Li, S.}, \bibinfo{author}{Wu, L.}, \bibinfo{author}{Tong, J.}, \bibinfo{author}{Chai, Y.}, \bibinfo{author}{Han, P.}, \bibinfo{author}{Wang, F.}, \bibinfo{author}{Jia, P.}, \bibinfo{author}{Li, Z.}, \bibinfo{author}{Zhao, X.}, \bibinfo{author}{Li, D.}, \bibinfo{author}{Zhang, R.}, \bibinfo{author}{Zhang, X.}, \bibinfo{author}{Zou, W.}, \bibinfo{author}{Li, W.}, \bibinfo{author}{Wang, Q.}, \bibinfo{author}{Gao, G.~F.}, \bibinfo{author}{Wu, Y.}, \bibinfo{author}{Dai, L.}, and \bibinfo{author}{Gao, F.} (\bibinfo{year}{2023}). \bibinfo{title}{Defining a de novo non-rbm antibody as rbd-8 and its synergistic rescue of immune-evaded antibodies to neutralize omicron sars-cov-2}.
\newblock \bibinfo{journal}{Proceedings of the National Academy of Sciences} \emph{\bibinfo{volume}{120}}, \bibinfo{pages}{e2314193120}. \DOIprefix\doi{10.1073/pnas.2314193120}.
\bibitem[{Hong et~al.(2023)Hong, Zhong, Liu, Wu, Zhang, Chen, Wei, Sun, Zhou, Zhang, Kang, Huang, Chen, Wang, Zhou, Chen, Feng, Yu, Li, Zeng, Zeng, Xu, Zheng, Chen, Zhang and Xia}]{7YP2}
\bibinfo{author}{Hong, J.}, \bibinfo{author}{Zhong, L.}, \bibinfo{author}{Liu, L.}, \bibinfo{author}{Wu, Q.}, \bibinfo{author}{Zhang, W.}, \bibinfo{author}{Chen, K.}, \bibinfo{author}{Wei, D.}, \bibinfo{author}{Sun, H.}, \bibinfo{author}{Zhou, X.}, \bibinfo{author}{Zhang, X.}, \bibinfo{author}{Kang, Y.-F.}, \bibinfo{author}{Huang, Y.}, \bibinfo{author}{Chen, J.}, \bibinfo{author}{Wang, G.}, \bibinfo{author}{Zhou, Y.}, \bibinfo{author}{Chen, Y.}, \bibinfo{author}{Feng, Q.-S.}, \bibinfo{author}{Yu, H.}, \bibinfo{author}{Li, S.}, \bibinfo{author}{Zeng, M.-S.}, \bibinfo{author}{Zeng, Y.-X.}, \bibinfo{author}{Xu, M.}, \bibinfo{author}{Zheng, Q.}, \bibinfo{author}{Chen, Y.}, \bibinfo{author}{Zhang, X.}, and \bibinfo{author}{Xia, N.} (\bibinfo{year}{2023}). \bibinfo{title}{Non-overlapping epitopes on the ghgl-gp42 complex for the rational design of a triple-antibody cocktail against ebv infection}.
\newblock \bibinfo{journal}{Cell Reports Medicine} \emph{\bibinfo{volume}{4}}. \DOIprefix\doi{10.1016/j.xcrm.2023.101296}.
\bibitem[{Zhang et~al.(2022)Zhang, Jiang, Gao, Zhang, Zhang, Zhou, Xu and Cai}]{8AG1}
\bibinfo{author}{Zhang, J.}, \bibinfo{author}{Jiang, X.}, \bibinfo{author}{Gao, H.}, \bibinfo{author}{Zhang, F.}, \bibinfo{author}{Zhang, X.}, \bibinfo{author}{Zhou, A.}, \bibinfo{author}{Xu, T.}, and \bibinfo{author}{Cai, H.} (\bibinfo{year}{2022}). \bibinfo{title}{Structural basis of a novel agonistic anti-ox40 antibody}.
\newblock \bibinfo{journal}{Biomolecules} \emph{\bibinfo{volume}{12}}. \DOIprefix\doi{10.3390/biom12091209}.
\bibitem[{Hargreaves et~al.(2023)Hargreaves, Carbajo, Bodnarchuk, Embrey, Rawlins, Packer, Degorce, Hird, Johannes, Chiarparin and Schade}]{8AV9}
\bibinfo{author}{Hargreaves, D.}, \bibinfo{author}{Carbajo, R.~J.}, \bibinfo{author}{Bodnarchuk, M.~S.}, \bibinfo{author}{Embrey, K.}, \bibinfo{author}{Rawlins, P.~B.}, \bibinfo{author}{Packer, M.}, \bibinfo{author}{Degorce, S.~L.}, \bibinfo{author}{Hird, A.~W.}, \bibinfo{author}{Johannes, J.~W.}, \bibinfo{author}{Chiarparin, E.}, and \bibinfo{author}{Schade, M.} (\bibinfo{year}{2023}). \bibinfo{title}{Design of rigid protein–protein interaction inhibitors enables targeting of undruggable mcl-1}.
\newblock \bibinfo{journal}{Proceedings of the National Academy of Sciences} \emph{\bibinfo{volume}{120}}. \DOIprefix\doi{10.1073/pnas.2221967120}.
\bibitem[{Davies et~al.(2023)Davies, Dedi, Jones, Kevorkian, McMillan, Ottone, Schulze, Scott-Tucker, Tewari, West, Wright and Rowley}]{8B7H}
\bibinfo{author}{Davies, G. C.~G.}, \bibinfo{author}{Dedi, N.}, \bibinfo{author}{Jones, P.~S.}, \bibinfo{author}{Kevorkian, L.}, \bibinfo{author}{McMillan, D.}, \bibinfo{author}{Ottone, C.}, \bibinfo{author}{Schulze, M.-S. E.~D.}, \bibinfo{author}{Scott-Tucker, A.}, \bibinfo{author}{Tewari, R.}, \bibinfo{author}{West, S.}, \bibinfo{author}{Wright, M.}, and \bibinfo{author}{Rowley, T.~F.} (\bibinfo{year}{2023}). \bibinfo{title}{Discovery of ginisortamab, a potent and novel anti-gremlin-1 antibody in clinical development for the treatment of cancer}.
\newblock \bibinfo{journal}{mAbs} \emph{\bibinfo{volume}{15}}. \DOIprefix\doi{10.1080/19420862.2023.2289681}.
\bibitem[{Hiemstra et~al.(2023)Hiemstra, Santegoets, Janmaat, {De Goeij}, {Ten Hagen}, {van Dooremalen}, Boross, {van den Brakel}, Bosgra, Andringa, {van Kessel-Welmers}, Verzijl, Hibbert, Frerichs, Mutis, {van de Donk}, Ahmadi, Satijn, Sasser and Breij}]{8BYU}
\bibinfo{author}{Hiemstra, I.~H.}, \bibinfo{author}{Santegoets, K.~C.}, \bibinfo{author}{Janmaat, M.~L.}, \bibinfo{author}{{De Goeij}, B.~E.}, \bibinfo{author}{{Ten Hagen}, W.}, \bibinfo{author}{{van Dooremalen}, S.}, \bibinfo{author}{Boross, P.}, \bibinfo{author}{{van den Brakel}, J.}, \bibinfo{author}{Bosgra, S.}, \bibinfo{author}{Andringa, G.}, \bibinfo{author}{{van Kessel-Welmers}, B.}, \bibinfo{author}{Verzijl, D.}, \bibinfo{author}{Hibbert, R.~G.}, \bibinfo{author}{Frerichs, K.~A.}, \bibinfo{author}{Mutis, T.}, \bibinfo{author}{{van de Donk}, N.~W.}, \bibinfo{author}{Ahmadi, T.}, \bibinfo{author}{Satijn, D.}, \bibinfo{author}{Sasser, A.~K.}, and \bibinfo{author}{Breij, E.~C.} (\bibinfo{year}{2023}). \bibinfo{title}{Preclinical anti-tumour activity of hexabody-cd38, a next-generation cd38 antibody with superior complement-dependent cytotoxic activity}.
\newblock \bibinfo{journal}{eBioMedicine} \emph{\bibinfo{volume}{93}}. \DOIprefix\doi{10.1016/j.ebiom.2023.104663}.
\bibitem[{Banach et~al.(2023)Banach, Pletnev, Olia, Xu, Zhang, Rawi, Bylund, Doria-Rose, Nguyen, Fahad, Lee, Lin, Liu, Louder, Madan, McKee, O'Dell, Sastry, Sch{\"o}n, Bui, Shen, Wolfe, Chuang, Mascola, Kwong and DeKosky}]{8ELI}
\bibinfo{author}{Banach, B.~B.}, \bibinfo{author}{Pletnev, S.}, \bibinfo{author}{Olia, A.~S.}, \bibinfo{author}{Xu, K.}, \bibinfo{author}{Zhang, B.}, \bibinfo{author}{Rawi, R.}, \bibinfo{author}{Bylund, T.}, \bibinfo{author}{Doria-Rose, N.~A.}, \bibinfo{author}{Nguyen, T.~D.}, \bibinfo{author}{Fahad, A.~S.}, \bibinfo{author}{Lee, M.}, \bibinfo{author}{Lin, B.~C.}, \bibinfo{author}{Liu, T.}, \bibinfo{author}{Louder, M.~K.}, \bibinfo{author}{Madan, B.}, \bibinfo{author}{McKee, K.}, \bibinfo{author}{O'Dell, S.}, \bibinfo{author}{Sastry, M.}, \bibinfo{author}{Sch{\"o}n, A.}, \bibinfo{author}{Bui, N.}, \bibinfo{author}{Shen, C.-H.}, \bibinfo{author}{Wolfe, J.~R.}, \bibinfo{author}{Chuang, G.-Y.}, \bibinfo{author}{Mascola, J.~R.}, \bibinfo{author}{Kwong, P.~D.}, and \bibinfo{author}{DeKosky, B.~J.} (\bibinfo{year}{2023}). \bibinfo{title}{Antibody-directed evolution reveals a mechanism for enhanced neutralization at the hiv-1 fusion peptide site}.
\newblock \bibinfo{journal}{Nature Communications} \emph{\bibinfo{volume}{14}}. \DOIprefix\doi{10.1038/s41467-023-42098-5}.
\bibitem[{Kapingidza et~al.(2023)Kapingidza, Marston, Harris, Wrapp, Winters, Mielke, Xiaozhi, Yin, Foulger, Parks, Barr, Newman, Sch{\"a}fer, Eaton, Flores, Harner, Catanzaro, Mallory, Mattocks, Beverly, Rhodes, Mansouri, Van~Itallie, Vure, Dunn, Keyes, Stanfield-Oakley, Woods, Petzold, Walter, Wiehe, Edwards, Montefiori, Ferrari, Baric, Cain, Saunders, Haynes and Azoitei}]{8F5I}
\bibinfo{author}{Kapingidza, A.~B.}, \bibinfo{author}{Marston, D.~J.}, \bibinfo{author}{Harris, C.}, \bibinfo{author}{Wrapp, D.}, \bibinfo{author}{Winters, K.}, \bibinfo{author}{Mielke, D.}, \bibinfo{author}{Xiaozhi, L.}, \bibinfo{author}{Yin, Q.}, \bibinfo{author}{Foulger, A.}, \bibinfo{author}{Parks, R.}, \bibinfo{author}{Barr, M.}, \bibinfo{author}{Newman, A.}, \bibinfo{author}{Sch{\"a}fer, A.}, \bibinfo{author}{Eaton, A.}, \bibinfo{author}{Flores, J.~M.}, \bibinfo{author}{Harner, A.}, \bibinfo{author}{Catanzaro, N.~J.}, \bibinfo{author}{Mallory, M.~L.}, \bibinfo{author}{Mattocks, M.~D.}, \bibinfo{author}{Beverly, C.}, \bibinfo{author}{Rhodes, B.}, \bibinfo{author}{Mansouri, K.}, \bibinfo{author}{Van~Itallie, E.}, \bibinfo{author}{Vure, P.}, \bibinfo{author}{Dunn, B.}, \bibinfo{author}{Keyes, T.}, \bibinfo{author}{Stanfield-Oakley, S.}, \bibinfo{author}{Woods, C.~W.}, \bibinfo{author}{Petzold, E.~A.}, \bibinfo{author}{Walter, E.~B.}, \bibinfo{author}{Wiehe, K.}, \bibinfo{author}{Edwards, R.~J.},
  \bibinfo{author}{Montefiori, D.~C.}, \bibinfo{author}{Ferrari, G.}, \bibinfo{author}{Baric, R.}, \bibinfo{author}{Cain, D.~W.}, \bibinfo{author}{Saunders, K.~O.}, \bibinfo{author}{Haynes, B.~F.}, and \bibinfo{author}{Azoitei, M.~L.} (\bibinfo{year}{2023}). \bibinfo{title}{Engineered immunogens to elicit antibodies against conserved coronavirus epitopes}.
\newblock \bibinfo{journal}{Nature Communications} \emph{\bibinfo{volume}{14}}. \DOIprefix\doi{10.1038/s41467-023-43638-9}.
\bibitem[{Cheung et~al.(2023)Cheung, Atwell, Bafetti, Cuenca, Froning, Hendle, Hickey, Ho, Huang, Lieu, Lim, Lippner, Obungu, Ward-Kavanagh, Weichert, Ware and Vendel}]{8F6O}
\bibinfo{author}{Cheung, T.~C.}, \bibinfo{author}{Atwell, S.}, \bibinfo{author}{Bafetti, L.}, \bibinfo{author}{Cuenca, P.~D.}, \bibinfo{author}{Froning, K.}, \bibinfo{author}{Hendle, J.}, \bibinfo{author}{Hickey, M.}, \bibinfo{author}{Ho, C.}, \bibinfo{author}{Huang, J.}, \bibinfo{author}{Lieu, R.}, \bibinfo{author}{Lim, S.}, \bibinfo{author}{Lippner, D.}, \bibinfo{author}{Obungu, V.}, \bibinfo{author}{Ward-Kavanagh, L.}, \bibinfo{author}{Weichert, K.}, \bibinfo{author}{Ware, C.~F.}, and \bibinfo{author}{Vendel, A.~C.} (\bibinfo{year}{2023}). \bibinfo{title}{Epitope topography of agonist antibodies to the checkpoint inhibitory receptor btla}.
\newblock \bibinfo{journal}{Structure} \emph{\bibinfo{volume}{31}}, \bibinfo{pages}{958--967.e3}. \DOIprefix\doi{10.1016/j.str.2023.05.011}.
\bibitem[{Thai et~al.(2023)Thai, Murugan, Binter, Burn~Aschner, Prieto, Kassardjian, Obraztsova, Kang, Flores-Garcia, Mathis-Torres, Li, Horn, Huntwork, Bolscher, de~Bruijni, Sauerwein, Dennison, Tomaras, Zavala, Kellam, Wardemann and Julien}]{8F9U}
\bibinfo{author}{Thai, E.}, \bibinfo{author}{Murugan, R.}, \bibinfo{author}{Binter, {\v{S}}.}, \bibinfo{author}{Burn~Aschner, C.}, \bibinfo{author}{Prieto, K.}, \bibinfo{author}{Kassardjian, A.}, \bibinfo{author}{Obraztsova, A.~S.}, \bibinfo{author}{Kang, R.~W.}, \bibinfo{author}{Flores-Garcia, Y.}, \bibinfo{author}{Mathis-Torres, S.}, \bibinfo{author}{Li, K.}, \bibinfo{author}{Horn, G.~Q.}, \bibinfo{author}{Huntwork, R.~H.}, \bibinfo{author}{Bolscher, J.~M.}, \bibinfo{author}{de~Bruijni, M.~H.}, \bibinfo{author}{Sauerwein, R.}, \bibinfo{author}{Dennison, S.~M.}, \bibinfo{author}{Tomaras, G.~D.}, \bibinfo{author}{Zavala, F.}, \bibinfo{author}{Kellam, P.}, \bibinfo{author}{Wardemann, H.}, and \bibinfo{author}{Julien, J.-P.} (\bibinfo{year}{2023}). \bibinfo{title}{Molecular determinants of cross-reactivity and potency by vh3-33 antibodies against the plasmodium falciparum circumsporozoite protein}.
\newblock \bibinfo{journal}{Cell Reports} \emph{\bibinfo{volume}{42}}. \DOIprefix\doi{10.1016/j.celrep.2023.113330}.
\bibitem[{Bedian et~al.(2023)Bedian, Biris, Omer, Chung, Fuller, Dagher, Chandran, Harwin, Kiselak, Violin, Nichols and Bista}]{8FGX}
\bibinfo{author}{Bedian, V.}, \bibinfo{author}{Biris, N.}, \bibinfo{author}{Omer, C.}, \bibinfo{author}{Chung, J.-K.}, \bibinfo{author}{Fuller, J.}, \bibinfo{author}{Dagher, R.}, \bibinfo{author}{Chandran, S.}, \bibinfo{author}{Harwin, P.}, \bibinfo{author}{Kiselak, T.}, \bibinfo{author}{Violin, J.}, \bibinfo{author}{Nichols, A.}, and \bibinfo{author}{Bista, P.} (\bibinfo{year}{2023}). \bibinfo{title}{Star-0215 is a novel, long-acting monoclonal antibody inhibitor of plasma kallikrein for the potential treatment of hereditary angioedema}.
\newblock \bibinfo{journal}{Journal of Pharmacology and Experimental Therapeutics} \emph{\bibinfo{volume}{387}}, \bibinfo{pages}{214--225}. \DOIprefix\doi{10.1124/jpet.123.001740}.
\bibitem[{Jenkins et~al.(2023)Jenkins, Park, Pederzoli-Ribeil, Eskiocak, Johnson, Guzman, McLaughlin, Moore-Lai, O{\textquoteright}Toole, Liu, Nicholson, Flesch, Qiu, Clackson, O{\textquoteright}Hagan, Rodeck, Karow, O{\textquoteright}Neil and Williams}]{8G8N}
\bibinfo{author}{Jenkins, K.~A.}, \bibinfo{author}{Park, M.}, \bibinfo{author}{Pederzoli-Ribeil, M.}, \bibinfo{author}{Eskiocak, U.}, \bibinfo{author}{Johnson, P.}, \bibinfo{author}{Guzman, W.}, \bibinfo{author}{McLaughlin, M.}, \bibinfo{author}{Moore-Lai, D.}, \bibinfo{author}{O{\textquoteright}Toole, C.}, \bibinfo{author}{Liu, Z.}, \bibinfo{author}{Nicholson, B.}, \bibinfo{author}{Flesch, V.}, \bibinfo{author}{Qiu, H.}, \bibinfo{author}{Clackson, T.}, \bibinfo{author}{O{\textquoteright}Hagan, R.~C.}, \bibinfo{author}{Rodeck, U.}, \bibinfo{author}{Karow, M.}, \bibinfo{author}{O{\textquoteright}Neil, J.}, and \bibinfo{author}{Williams, J.~C.} (\bibinfo{year}{2023}). \bibinfo{title}{Xtx101, a tumor-activated, fc-enhanced anti-ctla-4 monoclonal antibody, demonstrates tumor-growth inhibition and tumor-selective pharmacodynamics in mouse models of cancer}.
\newblock \bibinfo{journal}{Journal for ImmunoTherapy of Cancer} \emph{\bibinfo{volume}{11}}. \DOIprefix\doi{10.1136/jitc-2023-007785}.
\bibitem[{Moriyama et~al.(2023)Moriyama, Anraku, Taminishi, Adachi, Kuroda, Kita, Higuchi, Kirita, Kotaki, Tonouchi, Yumoto, Suzuki, Someya, Fukuhara, Kuroda, Yamamoto, Onodera, Fukushi, Maeda, Nakamura-Uchiyama, Hashiguchi, Hoshino, Maenaka and Takahashi}]{8HES}
\bibinfo{author}{Moriyama, S.}, \bibinfo{author}{Anraku, Y.}, \bibinfo{author}{Taminishi, S.}, \bibinfo{author}{Adachi, Y.}, \bibinfo{author}{Kuroda, D.}, \bibinfo{author}{Kita, S.}, \bibinfo{author}{Higuchi, Y.}, \bibinfo{author}{Kirita, Y.}, \bibinfo{author}{Kotaki, R.}, \bibinfo{author}{Tonouchi, K.}, \bibinfo{author}{Yumoto, K.}, \bibinfo{author}{Suzuki, T.}, \bibinfo{author}{Someya, T.}, \bibinfo{author}{Fukuhara, H.}, \bibinfo{author}{Kuroda, Y.}, \bibinfo{author}{Yamamoto, T.}, \bibinfo{author}{Onodera, T.}, \bibinfo{author}{Fukushi, S.}, \bibinfo{author}{Maeda, K.}, \bibinfo{author}{Nakamura-Uchiyama, F.}, \bibinfo{author}{Hashiguchi, T.}, \bibinfo{author}{Hoshino, A.}, \bibinfo{author}{Maenaka, K.}, and \bibinfo{author}{Takahashi, Y.} (\bibinfo{year}{2023}). \bibinfo{title}{Structural delineation and computational design of sars-cov-2-neutralizing antibodies against omicron subvariants}.
\newblock \bibinfo{journal}{Nature Communications} \emph{\bibinfo{volume}{14}}. \DOIprefix\doi{10.1038/s41467-023-39890-8}.
\bibitem[{Wang et~al.(2023)Wang, Li, Chen, Sun, Jin, Hu, Feng, Su, Ren, Hao, Wang, Zhu, Liu, Qi, Zhu and Shao}]{8HN6}
\bibinfo{author}{Wang, Z.}, \bibinfo{author}{Li, D.}, \bibinfo{author}{Chen, Y.}, \bibinfo{author}{Sun, Y.}, \bibinfo{author}{Jin, C.}, \bibinfo{author}{Hu, C.}, \bibinfo{author}{Feng, Y.}, \bibinfo{author}{Su, J.}, \bibinfo{author}{Ren, L.}, \bibinfo{author}{Hao, Y.}, \bibinfo{author}{Wang, S.}, \bibinfo{author}{Zhu, M.}, \bibinfo{author}{Liu, Y.}, \bibinfo{author}{Qi, J.}, \bibinfo{author}{Zhu, B.}, and \bibinfo{author}{Shao, Y.} (\bibinfo{year}{2023}). \bibinfo{title}{Characterization of rbd-specific cross-neutralizing antibodies responses against sars-cov-2 variants from covid-19 convalescents}.
\newblock \bibinfo{journal}{Frontiers in Immunology} \emph{\bibinfo{volume}{14}}. \DOIprefix\doi{10.3389/fimmu.2023.1160283}.
\bibitem[{Adachi et~al.(2023)Adachi, Kaneko, Kato and Nogi}]{8IPC}
\bibinfo{author}{Adachi, Y.}, \bibinfo{author}{Kaneko, M.~K.}, \bibinfo{author}{Kato, Y.}, and \bibinfo{author}{Nogi, T.} (\bibinfo{year}{2023}). \bibinfo{title}{Recombinant production of antibody antigen-binding fragments with an n-terminal human growth hormone tag in mammalian cells}.
\newblock \bibinfo{journal}{Protein Expression and Purification} \emph{\bibinfo{volume}{208-209}}. \DOIprefix\doi{10.1016/j.pep.2023.106289}.
\bibitem[{Xu et~al.(2023)Xu, Goel, Burkart, Burman, Chong, Barber, Geng, Zhai, Wang, Kumar, Menefee, Polizzi, Eide, Rauch, Rahman, Hamel, Fogassy, Klopp-Savino, Paz, Zhang, Cubitt, Nangle and Mercurio}]{8IVX}
\bibinfo{author}{Xu, Z.}, \bibinfo{author}{Goel, H.~L.}, \bibinfo{author}{Burkart, C.}, \bibinfo{author}{Burman, L.}, \bibinfo{author}{Chong, Y.~E.}, \bibinfo{author}{Barber, A.~G.}, \bibinfo{author}{Geng, Y.}, \bibinfo{author}{Zhai, L.}, \bibinfo{author}{Wang, M.}, \bibinfo{author}{Kumar, A.}, \bibinfo{author}{Menefee, A.}, \bibinfo{author}{Polizzi, C.}, \bibinfo{author}{Eide, L.}, \bibinfo{author}{Rauch, K.}, \bibinfo{author}{Rahman, J.}, \bibinfo{author}{Hamel, K.}, \bibinfo{author}{Fogassy, Z.}, \bibinfo{author}{Klopp-Savino, S.}, \bibinfo{author}{Paz, S.}, \bibinfo{author}{Zhang, M.}, \bibinfo{author}{Cubitt, A.}, \bibinfo{author}{Nangle, L.~A.}, and \bibinfo{author}{Mercurio, A.~M.} (\bibinfo{year}{2023}). \bibinfo{title}{Inhibition of vegf binding to neuropilin-2 enhances chemosensitivity and inhibits metastasis in triple-negative breast cancer}.
\newblock \bibinfo{journal}{Science Translational Medicine} \emph{\bibinfo{volume}{15}}. \DOIprefix\doi{10.1126/scitranslmed.adf1128}.
\bibitem[{Li et~al.(2024{\natexlab{a}})Li, Chong, Peng, Liu, Rao, Fu, Shu, Shen, Xiao, Liu, Li, Deng, Yan, Hu, Cao and Wang}]{8JLX}
\bibinfo{author}{Li, L.}, \bibinfo{author}{Chong, T.}, \bibinfo{author}{Peng, L.}, \bibinfo{author}{Liu, Y.}, \bibinfo{author}{Rao, G.}, \bibinfo{author}{Fu, Y.}, \bibinfo{author}{Shu, Y.}, \bibinfo{author}{Shen, J.}, \bibinfo{author}{Xiao, Q.}, \bibinfo{author}{Liu, J.}, \bibinfo{author}{Li, J.}, \bibinfo{author}{Deng, F.}, \bibinfo{author}{Yan, B.}, \bibinfo{author}{Hu, Z.}, \bibinfo{author}{Cao, S.}, and \bibinfo{author}{Wang, M.} (\bibinfo{year}{2024}{\natexlab{a}}). \bibinfo{title}{Neutralizing monoclonal antibodies against the gc fusion loop region of crimean–congo hemorrhagic fever virus}.
\newblock \bibinfo{journal}{PLOS Pathogens} \emph{\bibinfo{volume}{20}}, \bibinfo{pages}{1--22}. \DOIprefix\doi{10.1371/journal.ppat.1011948}.
\bibitem[{Liu et~al.(2023)Liu, Chiang, Xiong, Laurent, Griffiths, D{\"u}lfer, Deng, Sun, Yin, Li, Audoly, An, Sch{\"u}rpf, Li and Zhang}]{8PE9}
\bibinfo{author}{Liu, J.}, \bibinfo{author}{Chiang, H.-C.}, \bibinfo{author}{Xiong, W.}, \bibinfo{author}{Laurent, V.}, \bibinfo{author}{Griffiths, S.~C.}, \bibinfo{author}{D{\"u}lfer, J.}, \bibinfo{author}{Deng, H.}, \bibinfo{author}{Sun, X.}, \bibinfo{author}{Yin, Y.~W.}, \bibinfo{author}{Li, W.}, \bibinfo{author}{Audoly, L.~P.}, \bibinfo{author}{An, Z.}, \bibinfo{author}{Sch{\"u}rpf, T.}, \bibinfo{author}{Li, R.}, and \bibinfo{author}{Zhang, N.} (\bibinfo{year}{2023}). \bibinfo{title}{A highly selective humanized ddr1 mab reverses immune exclusion by disrupting collagen fiber alignment in breast cancer}.
\newblock \bibinfo{journal}{Journal for ImmunoTherapy of Cancer} \emph{\bibinfo{volume}{11}}. \DOIprefix\doi{10.1136/jitc-2023-006720}.
\bibitem[{Xiao et~al.(2023)Xiao, Wen, Chen, Shipman, Kostas, Reid, Warren, Tang, Luo, O’Donnell, Fridman, Chen, Vora, Zhang, Su and Eddins}]{8T9Z}
\bibinfo{author}{Xiao, X.}, \bibinfo{author}{Wen, Z.}, \bibinfo{author}{Chen, Q.}, \bibinfo{author}{Shipman, J.~M.}, \bibinfo{author}{Kostas, J.}, \bibinfo{author}{Reid, J.~C.}, \bibinfo{author}{Warren, C.}, \bibinfo{author}{Tang, A.}, \bibinfo{author}{Luo, B.}, \bibinfo{author}{O’Donnell, G.}, \bibinfo{author}{Fridman, A.}, \bibinfo{author}{Chen, Z.}, \bibinfo{author}{Vora, K.~A.}, \bibinfo{author}{Zhang, L.}, \bibinfo{author}{Su, H.-P.}, and \bibinfo{author}{Eddins, M.~J.} (\bibinfo{year}{2023}). \bibinfo{title}{Structural characterization of m8c10, a neutralizing antibody targeting a highly conserved prefusion-specific epitope on the metapneumovirus fusion trimerization interface}.
\newblock \bibinfo{journal}{Journal of Virology} \emph{\bibinfo{volume}{97}}. \DOIprefix\doi{10.1128/jvi.01052-23}.
\bibitem[{Li et~al.(2024{\natexlab{b}})Li, Ge, Guo and Tao}]{8X0T}
\bibinfo{author}{Li, N.}, \bibinfo{author}{Ge, Q.}, \bibinfo{author}{Guo, Q.}, and \bibinfo{author}{Tao, Y.} (\bibinfo{year}{2024}{\natexlab{b}}). \bibinfo{title}{Identification and functional validation of fzd8-specific antibodies}.
\newblock \bibinfo{journal}{International Journal of Biological Macromolecules} \emph{\bibinfo{volume}{254}}. \DOIprefix\doi{10.1016/j.ijbiomac.2023.127846}.
\bibitem[{Fleishman et~al.(2011)Fleishman, Leaver-Fay, Corn, Strauch, Khare, Koga, Ashworth, Murphy, Richter, Lemmon, Meiler and Baker}]{Fleishman2011}
\bibinfo{author}{Fleishman, S.~J.}, \bibinfo{author}{Leaver-Fay, A.}, \bibinfo{author}{Corn, J.~E.}, \bibinfo{author}{Strauch, E.-M.}, \bibinfo{author}{Khare, S.~D.}, \bibinfo{author}{Koga, N.}, \bibinfo{author}{Ashworth, J.}, \bibinfo{author}{Murphy, P.}, \bibinfo{author}{Richter, F.}, \bibinfo{author}{Lemmon, G.}, \bibinfo{author}{Meiler, J.}, and \bibinfo{author}{Baker, D.} (\bibinfo{year}{2011}). \bibinfo{title}{Rosettascripts: A scripting language interface to the rosetta macromolecular modeling suite}.
\newblock \bibinfo{journal}{PLOS ONE} \emph{\bibinfo{volume}{6}}, \bibinfo{pages}{1--10}. \DOIprefix\doi{10.1371/journal.pone.0020161}.
\bibitem[{Virtanen et~al.(2020)Virtanen, Gommers, Oliphant, Haberland, Reddy, Cournapeau, Burovski, Peterson, Weckesser, Bright, {van der Walt}, Brett, Wilson, Millman, Mayorov, Nelson, Jones, Kern, Larson, Carey, Polat, Feng, Moore, {VanderPlas}, Laxalde, Perktold, Cimrman, Henriksen, Quintero, Harris, Archibald, Ribeiro, Pedregosa, {van Mulbregt} and {SciPy 1.0 Contributors}}]{Virtanen2020}
\bibinfo{author}{Virtanen, P.}, \bibinfo{author}{Gommers, R.}, \bibinfo{author}{Oliphant, T.~E.}, \bibinfo{author}{Haberland, M.}, \bibinfo{author}{Reddy, T.}, \bibinfo{author}{Cournapeau, D.}, \bibinfo{author}{Burovski, E.}, \bibinfo{author}{Peterson, P.}, \bibinfo{author}{Weckesser, W.}, \bibinfo{author}{Bright, J.}, \bibinfo{author}{{van der Walt}, S.~J.}, \bibinfo{author}{Brett, M.}, \bibinfo{author}{Wilson, J.}, \bibinfo{author}{Millman, K.~J.}, \bibinfo{author}{Mayorov, N.}, \bibinfo{author}{Nelson, A. R.~J.}, \bibinfo{author}{Jones, E.}, \bibinfo{author}{Kern, R.}, \bibinfo{author}{Larson, E.}, \bibinfo{author}{Carey, C.~J.}, \bibinfo{author}{Polat, {\.I}.}, \bibinfo{author}{Feng, Y.}, \bibinfo{author}{Moore, E.~W.}, \bibinfo{author}{{VanderPlas}, J.}, \bibinfo{author}{Laxalde, D.}, \bibinfo{author}{Perktold, J.}, \bibinfo{author}{Cimrman, R.}, \bibinfo{author}{Henriksen, I.}, \bibinfo{author}{Quintero, E.~A.}, \bibinfo{author}{Harris, C.~R.}, \bibinfo{author}{Archibald, A.~M.},
  \bibinfo{author}{Ribeiro, A.~H.}, \bibinfo{author}{Pedregosa, F.}, \bibinfo{author}{{van Mulbregt}, P.}, and \bibinfo{author}{{SciPy 1.0 Contributors}} (\bibinfo{year}{2020}). \bibinfo{title}{{{SciPy} 1.0: Fundamental Algorithms for Scientific Computing in Python}}.
\newblock \bibinfo{journal}{Nature Methods} \emph{\bibinfo{volume}{17}}, \bibinfo{pages}{261--272}. \DOIprefix\doi{10.1038/s41592-019-0686-2}.

\end{thebibliography}


\newpage

\section*{Supplemental information}
\markboth{Supplemental information}{Supplemental information}

\setcounter{figure}{0}
\pagenumbering{gobble}
\renewcommand{\thesection}{S\arabic{section}}  
\renewcommand{\thetable}{S\arabic{table}}
\renewcommand{\thefigure}{S\arabic{figure}}

\begin{table}[!ht]
\begin{center}
    \begin{tabular}{| c | c | c |}
    \hline
    \thead{\textbf{Method}} & \thead{Based on} & \thead{R score} \\ \hline
    \small{Yang et al. (2023)} & \small{Structure} & $0.79$  \\ \hline
    \small{Kurumida et al. (2020)} & \small{Structure} & $0.69$ \\ \hline
    \small{Kang et al. (2021)} & \small{Sequence} & $0.66$ \\ \hline
    \small{Sirin et al. (2015)} & \small{Sequence} & $0.45$ \\
    \hline
    \end{tabular}
    \caption{\label{table:other-methods}\textbf{Comparison to existing methods, Related to Figure 2.} $R$ score reported for binding affinity prediction methods which, to the best of our knowledge, are the best performing in the literature.}
    \end{center}
\end{table} 

\begin{longtable}{|c|c|c|c|c|c|}
\hline
$n_\text{f}$ & $k$ & $p$ & $l_r$ & Average validation MSE & Standard deviation \\
\hline
\endfirsthead

\multicolumn{6}{c}%
\textit{\footnotesize{Continued from previous page}} \\
\hline
$n_\text{f}$ & $k$ & $p$ & $l_r$ & Average validation MSE & Standard deviation \\
\hline
\endhead

\hline \multicolumn{6}{r}\textit{\footnotesize{Continued on next page}} \\

\endfoot

\hline
\addlinespace[1.5ex]
\caption*{\textbf{Table S2: ANTIPASTI model selection, Related to Figure 2 and STAR Methods}. Average and standard deviation of the validation MSE across the $K=10$ folds for each combination of hyperparameters (see~\figref[]{SupplHyperparametric}). The notation for the different hyperparameters is the same as in Figure \ref{fig:SupplHyperparametric}.}
\label{table:stddev}

\endlastfoot

\begin{filecontents*}{stddevCV.csv}
n_filters,kernel_size,pooling,learning_rate,mean val_loss,standard_deviation
2,3,3,0.0001,2.98,0.14
2,3,3,5.00E-05,2.94,0.17
2,3,3,0.0005,2.85,0.14
2,3,3,0.001,2.77,0.19
2,3,2,0.0001,2.75,0.15
2,3,2,5.00E-05,2.89,0.14
2,3,2,0.0005,2.61,0.14
2,3,2,0.001,2.64,0.16
2,3,1,0.0001,2.38,0.14
2,3,1,5.00E-05,2.44,0.16
2,3,1,0.0005,3.06,0.21
2,3,1,0.001,4.26,0.29
2,4,3,0.0001,2.19,0.12
2,4,3,5.00E-05,2.06,0.14
2,4,3,0.0005,2.02,0.14
2,4,3,0.001,1.93,0.15
2,4,2,5.00E-05,1.93,0.14
2,4,2,0.0005,1.94,0.13
2,4,2,0.001,1.92,0.14
2,4,2,0.0001,1.94,0.15
2,4,1,0.0001,2.27,0.14
2,4,1,0.0005,2.55,0.14
2,4,1,0.001,3.11,0.18
2,4,1,5.00E-05,1.9,0.2
4,3,3,0.0001,2.07,0.14
4,3,3,5.00E-05,2.33,0.14
4,3,3,0.0005,2.04,0.15
4,3,3,0.001,2.16,0.15
4,3,2,0.0001,2.13,0.16
4,3,2,5.00E-05,2.28,0.14
4,3,2,0.0005,2.08,0.14
4,3,2,0.001,2.46,0.15
4,3,1,0.0001,2.39,0.16
4,3,1,5.00E-05,2.35,0.16
4,3,1,0.0005,2.79,0.17
4,3,1,0.001,3.1,0.18
8,3,3,0.0001,3.25,0.21
8,3,3,5.00E-05,3.29,0.25
8,3,3,0.0005,3.22,0.24
8,3,3,0.001,3.14,0.28
8,3,2,0.0001,2.74,0.19
8,3,2,5.00E-05,2.66,0.19
8,3,2,0.0005,2.82,0.18
8,3,2,0.001,2.87,0.21
8,3,1,0.0001,3.97,0.33
8,3,1,5.00E-05,4.01,0.29
8,3,1,0.0005,4.12,0.31
8,3,1,0.001,4.19,0.35
8,4,3,0.0001,3.18,0.18
8,4,3,5.00E-05,3.21,0.19
8,4,3,0.0005,3.13,0.18
8,4,3,0.001,3.07,0.19
8,4,2,5.00E-05,2.81,0.22
8,4,2,0.0005,2.68,0.22
8,4,2,0.001,2.88,0.21
8,4,2,0.0001,2.85,0.24
8,4,1,0.0001,3.9,0.28
8,4,1,0.0005,3.92,0.22
8,4,1,0.001,4.04,0.29
8,4,1,5.00E-05,3.61,0.28
8,5,1,5.00E-05,3.58,0.36
4,5,3,0.0001,2.24,0.18
4,5,3,5.00E-05,2.4,0.18
4,5,3,0.0005,2.13,0.15
4,5,3,0.001,2.17,0.19
4,5,2,0.0001,1.72,0.14
4,5,2,5.00E-05,1.91,0.16
4,5,2,0.0005,2.09,0.16
4,5,2,0.001,2.22,0.16
4,5,1,0.0001,1.88,0.21
4,5,1,0.0005,2.7,0.3
4,5,1,0.001,3.38,0.32
4,5,1,5.00E-05,1.64,0.26
2,5,3,0.0001,2.44,0.19
2,5,3,5.00E-05,2.12,0.17
2,5,3,0.0005,2.16,0.22
2,5,3,0.001,1.97,0.23
2,5,2,5.00E-05,2.05,0.16
2,5,2,0.0005,1.85,0.16
2,5,2,0.001,2.1,0.18
2,5,2,0.0001,1.91,0.16
2,5,1,0.0001,2.49,0.2
2,5,1,0.0005,2.63,0.24
2,5,1,0.001,3.01,0.24
2,5,1,5.00E-05,2.45,0.23
4,4,3,0.0001,1.91,0.16
4,4,3,5.00E-05,2.14,0.16
4,4,3,0.0005,1.87,0.14
4,4,3,0.001,1.94,0.15
4,4,2,5.00E-05,1.56,0.11
4,4,2,0.0005,1.88,0.12
4,4,2,0.001,2.02,0.11
4,4,1,0.0001,1.72,0.13
4,4,1,0.0005,2.83,0.12
4,4,1,0.001,3.51,0.12
4,4,1,5.00E-05,1.27,0.11
4,4,2,0.0001,1.25,0.12
\end{filecontents*}
\csvreader[late after line=\\]{stddevCV.csv}{}{{\csvcoli} & {\csvcolii} & {\csvcoliii} & {\csvcoliv} & {\csvcolv} & {\csvcolvi}}
\end{longtable}

\begin{figure}[htbp]
    \centering  \includegraphics[width=\linewidth, clip, trim=0cm 20cm 6cm 0cm]{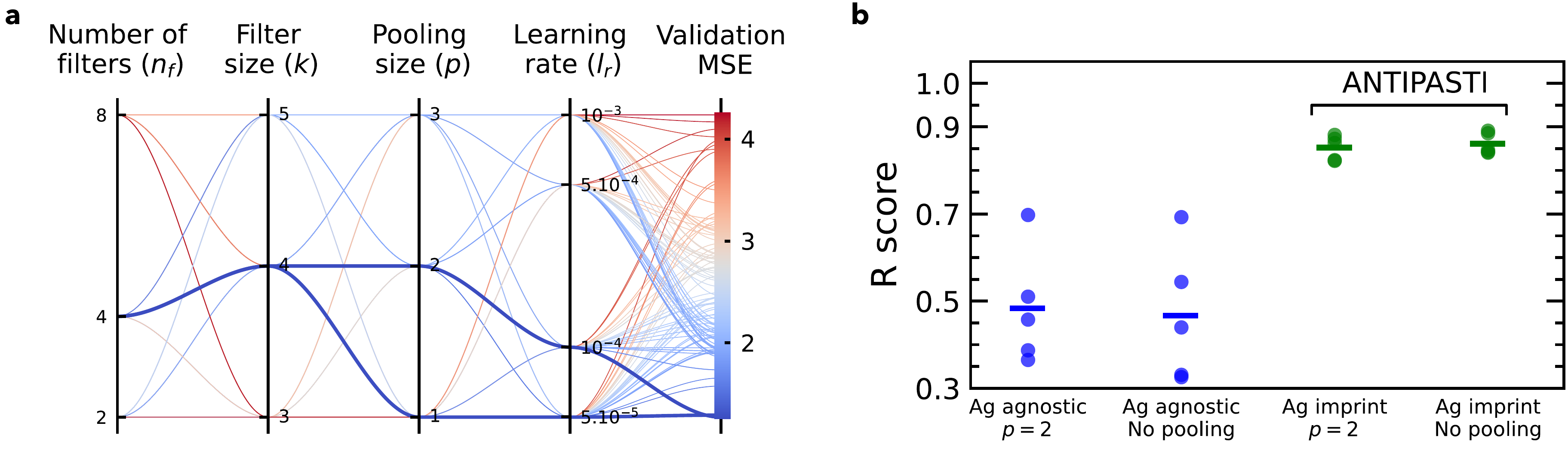}
  \caption{\textbf{ANTIPASTI model selection and performance, Related to Figure 2 and STAR Methods}. \textbf{(a)} Optuna and 10-fold cross-validation. The two hyperparameter sets with the lowest validation average MSE are represented by thicker lines and correspond to the two ANTIPASTI architectures whose performance is shown in Figures 2a–c (CNN with pooling $p=2$ and without pooling). The standard deviation of the validation MSE for each combination of hyperparameters is made available in Table S2. \textbf{(b)} Same plot as in Figure 2c, but comparing the two top-performing models with their antigen-agnostic analogues. }
    \label{fig:SupplHyperparametric}
\end{figure}

\begin{figure}[ht!]
    \centering
    \includegraphics[clip, trim=0.5cm 2.8cm 10.2cm 0.5cm, width=\linewidth]{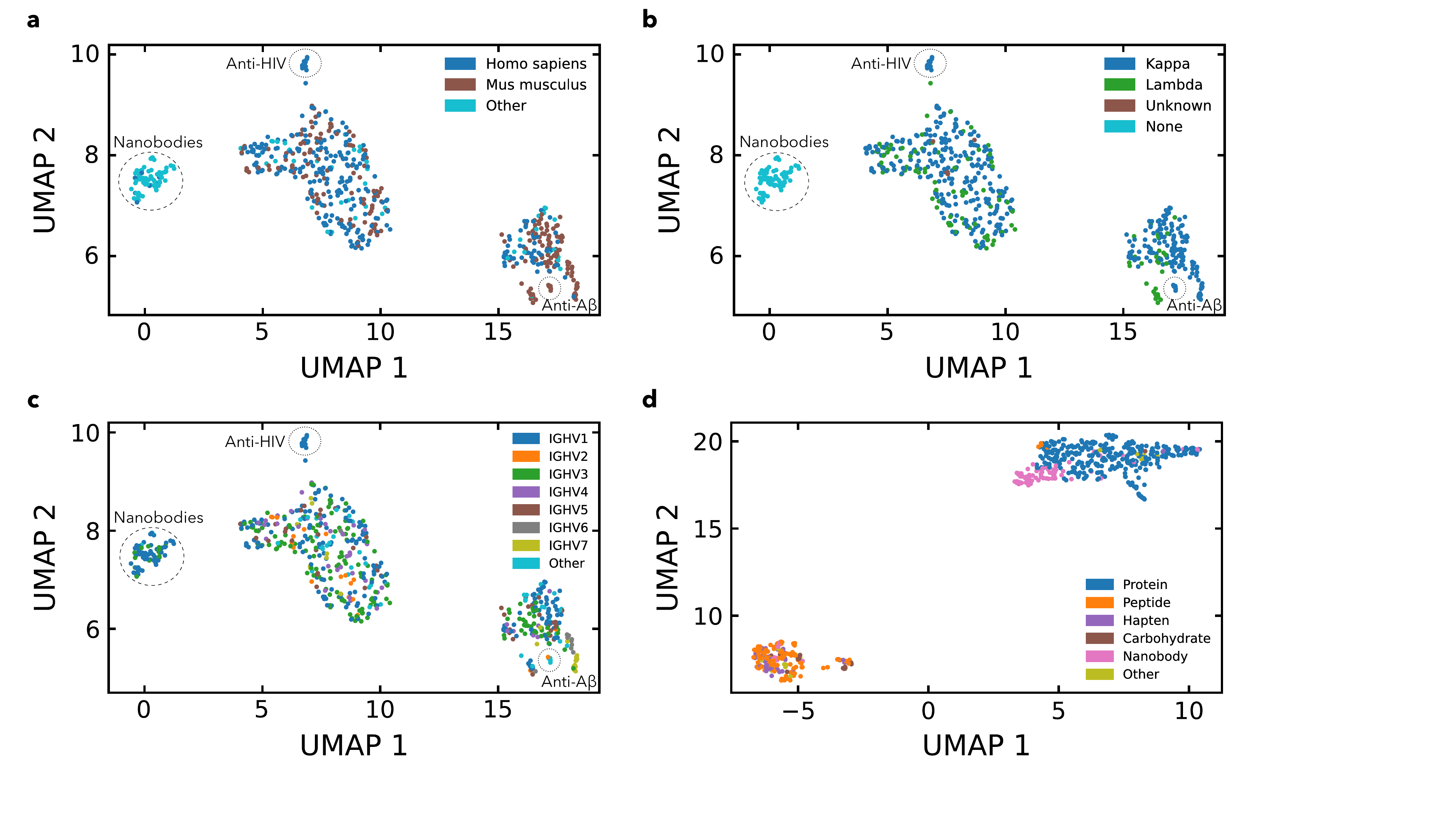}
    \includegraphics[clip, trim=0.5cm 2.8cm 10.2cm 0.5cm, width=\linewidth]{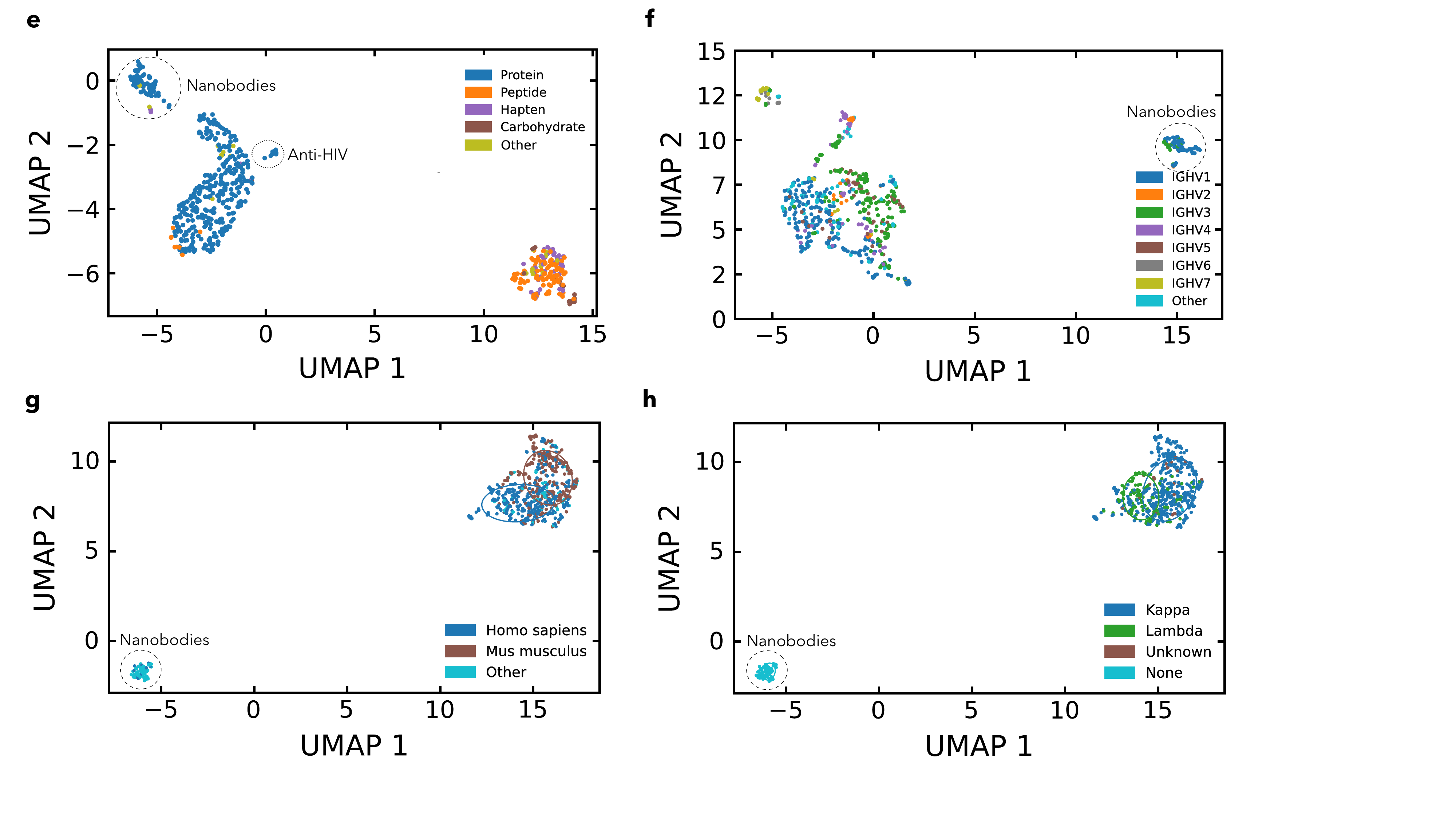}

\caption{\textbf{UMAP of the output layer representations, Related to Figure 3}. With antigen imprint, data points coloured according to:  (\textbf{a}) Antibody species. (\textbf{b}) Type of light chain. (\textbf{c}) Heavy chain V gene. (\textbf{d}) Binding target type, like in Figure 3b, but training ANTIPASTI only on the heavy chain residues. (\textbf{e}) Binding target type, like in Figure 3b, but computed using NM correlation maps at the atomistic level. (\textbf{f}, \textbf{g}, \textbf{h}) Antigen-agnostic case with data points using colour codes
that correspond to: heavy chain V gene, antibody species, and type of light chain respectively.}
\label{fig:UMAP}
\end{figure}

\begin{figure}[ht!]
\centering
  \includegraphics[clip, trim=0.5cm 0cm 2.5cm 0, width=0.96\linewidth]{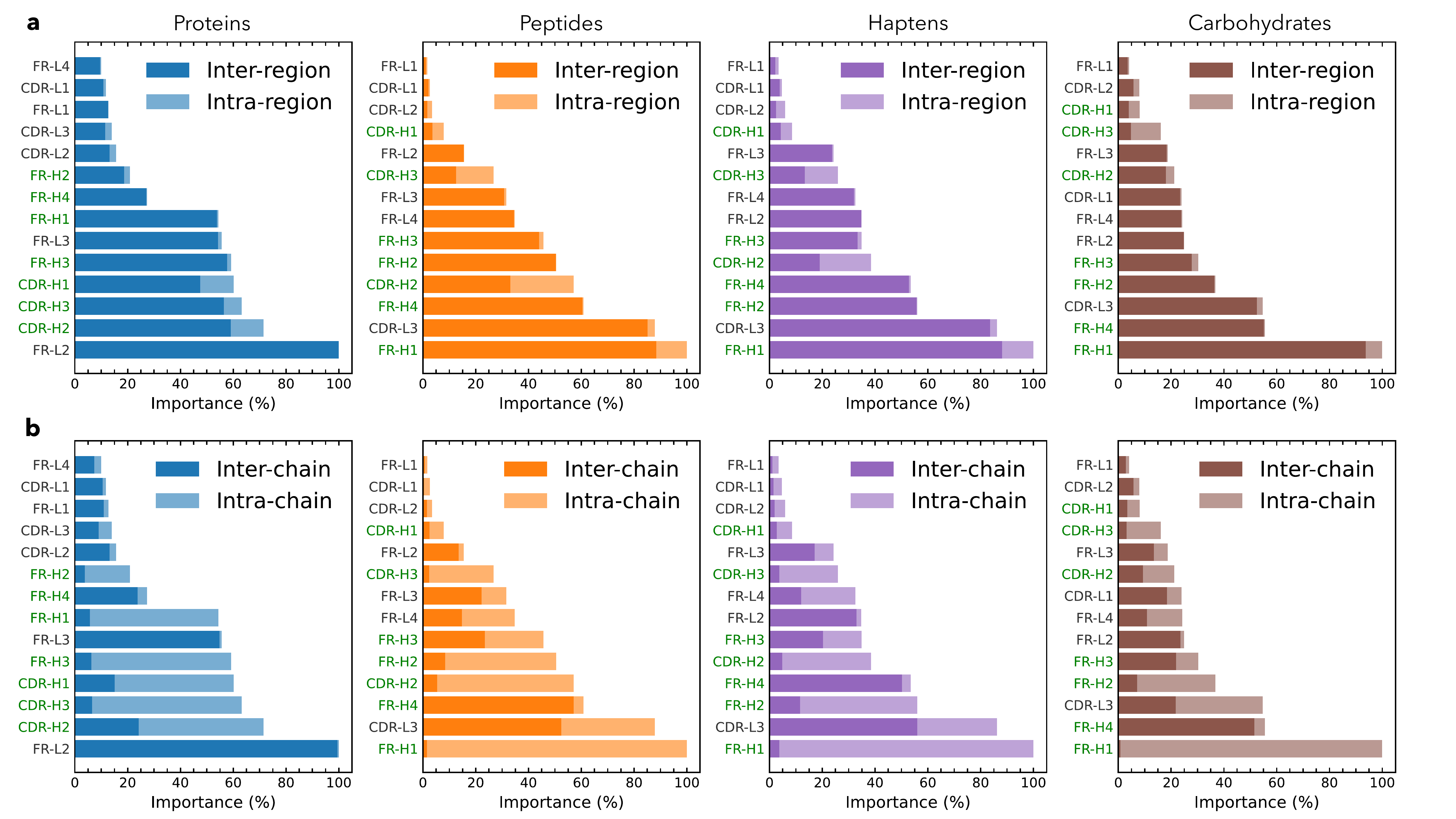}
    \includegraphics[clip, trim=0.5cm 0cm 2.5cm 0, width=0.96\linewidth]{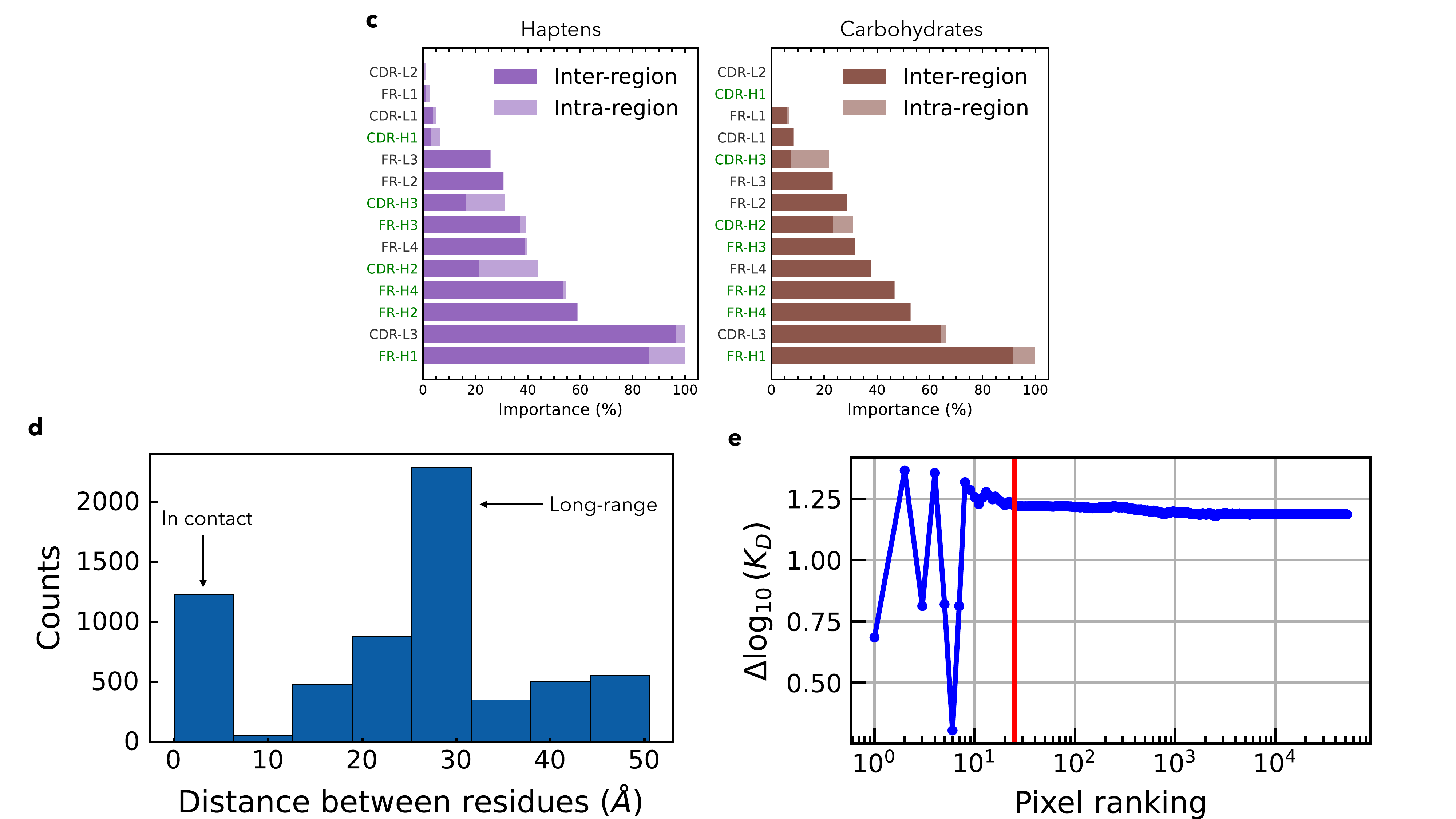}
\caption{\textbf{Region importance, Related to Figure 3.} Proportion of MSE deviation attributable to correlations between residues (\textbf{a}) within the same region
(intra-region) and that resulting from correlations with residues in other regions (inter-region), and (\textbf{b}) within the same chain
(intra-chain) and that resulting from correlations with residues in the other chain (inter-chain). (\textbf{c}) Same as Figure S3a (for haptens and carbohydrates) but computed using NM correlation maps at the atomistic level. Atomistic-level NM correlation maps have been computed using the \texttt{aanma} and \texttt{dccm} functions of Bio3D. (\textbf{d}) Distribution of residue $\alpha$-Carbon pairwise distances involved in the top $10$ affinity-relevant correlations for binding affinity across all $634$ antibody structures. (\textbf{e}) Difference in the predicted $\log_{10}(K_D)$, $\Delta\!\log_{10}(K_D)$, between a mutated antibody (\texttt{4yho}) and its original structure (\texttt{4yhi}) as function of the ranking in magnitude of changes
in affinity-relevant correlations between them (\textit{i.e.}, the pixels of the map in Figure 3f). For a given ranking $x$, only the top $x$ pixels are used to calculate $\Delta\!\log_{10}(K_D)$. The red vertical line is located at $x=25$.}
\label{fig:SupplImportance}
\end{figure}

\clearpage

\begin{figure}[ht!]
    \centering  \includegraphics[clip, trim=1.15cm 1.85cm 11cm 0.45cm,width=0.95\linewidth]{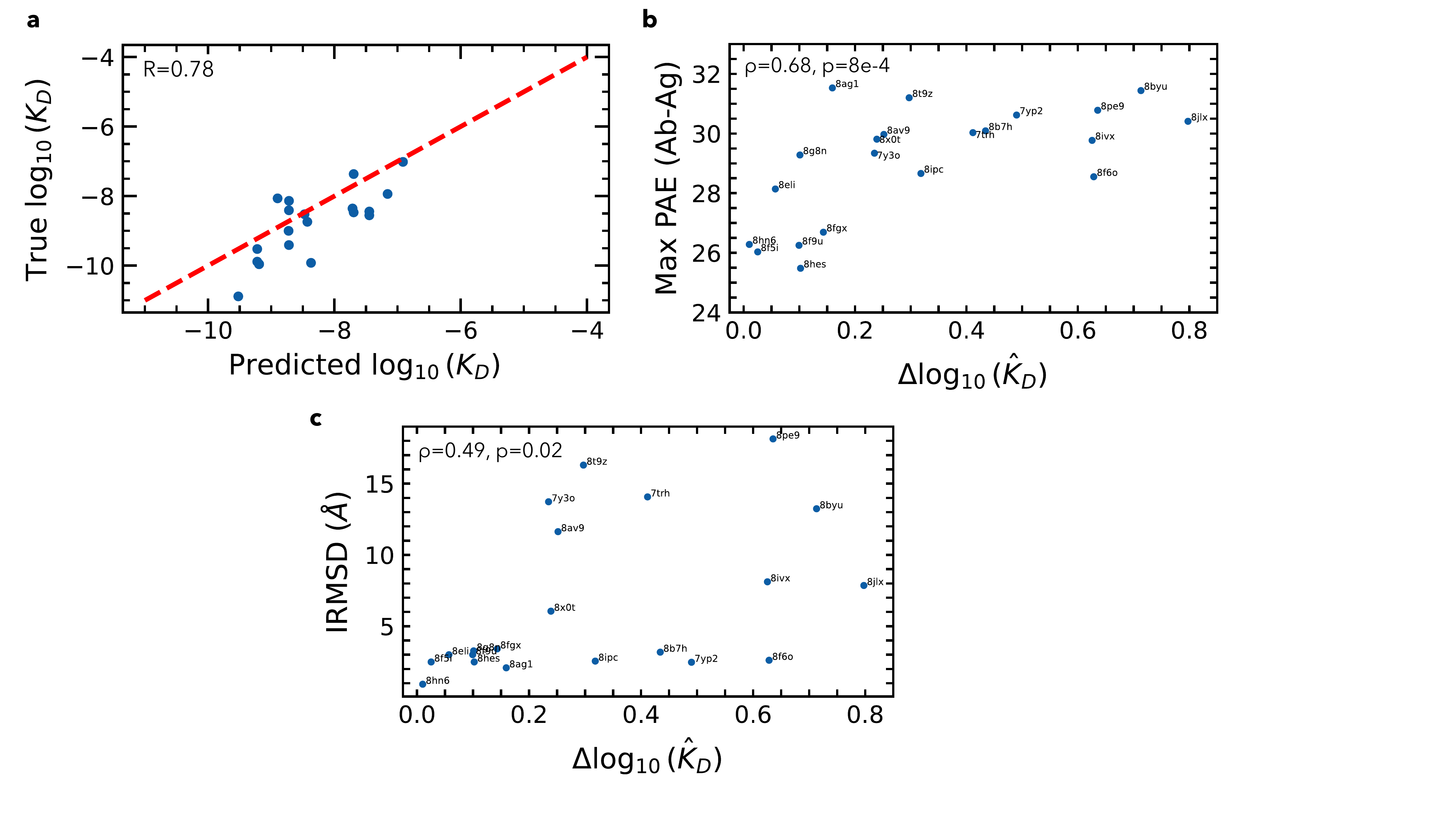}
    \caption{\textbf{AlphaFold-predicted structures, Related to Figure 4.} (\textbf{a}) ANTIPASTI predictions for a particular test set that
    yielded a correlation coefficient $R=0.78$. (\textbf{b}) Maximum predicted alignment error (PAE) for the antibody-antigen off-block only, as a function of
$\Delta\!\log_{10}(\hat{K}_D)$. (\textbf{c}) Interface Root Mean Square Deviation (IRMSD) between the original and AlphaFold-predicted structures, as a function of
$\Delta\!\log_{10}(\hat{K}_D)$.}
    \label{fig:SupplAlphaFold}
\end{figure}

\end{document}